\documentclass[12pt, draftclsnofoot, onecolumn]{IEEEtran}
\IEEEoverridecommandlockouts
\usepackage{caption} % added
\ifCLASSINFOpdf
\else
\fi
\ifCLASSOPTIONcaptionsoff
\usepackage[nomarkers]{endfloat}
\let\MYoriglatexcaption\caption
\renewcommand{\caption}[2][\relax]{\MYoriglatexcaption[#2]{#2}}
\fi
\usepackage{cite}
\usepackage{amsmath,amssymb,amsfonts}
\usepackage{graphicx}
\usepackage{epstopdf}
\usepackage{textcomp}
\usepackage{xcolor}
\def\BibTeX{{\rm B\kern-.05em{\sc i\kern-.025em b}\kern-.08em
		T\kern-.1667em\lower.7ex\hbox{E}\kern-.125emX}}
\usepackage{amsthm}
\usepackage{amssymb}
\usepackage{amsmath}

\newtheorem{proposition}{Proposition} 
\newtheorem{remark}{Remark}

\DeclareMathOperator{\diag}{diag}
\usepackage{amsmath,amsfonts,amsthm,bm} % Math packages
\usepackage{cite}
\usepackage{booktabs}
\usepackage{float}
\makeatletter

\newcommand{\Rmnum}[1]{\expandafter\@slowromancap\romannumeral #1@}
\makeatother
\DeclareMathOperator{\taninv}{tan^{-1}}
\DeclareMathOperator{\minimize}{minimize}
\DeclareMathOperator{\maximize}{maximize}
\DeclareMathOperator{\subto}{subject \hspace{0.125em} to}
\usepackage{color, soul} 
\usepackage{subfigure}
\usepackage{graphicx} % Required for the inclusion of images
\usepackage{graphics} % Required for the inclusion of images
\graphicspath{{figure/}}
\usepackage{amsmath} % Required for some math elements 
\usepackage{verbatim}%add comment
\usepackage{mathrsfs}
\usepackage{caption}
\usepackage{multirow}
\usepackage{float}
\usepackage{epstopdf}
\DeclareMathSizes{12}{10}{8}{6}
\usepackage{mleftright}
\mleftright
\newcommand{\dd}{\mathop{}\!\mathrm{d}}
\usepackage[linesnumbered,ruled,vlined]{algorithm2e}
\usepackage{algpseudocode}
\SetKwInOut{Input}{\textbf{Input :}}  
\SetKwInOut{Output}{\textbf{Output :}} 
\usepackage{array}
\renewcommand{\Re}{\operatorname{Re}}

\usepackage{caption}
\usepackage{color}
\definecolor{gray}{rgb}{0.6, 0.6, 0.6}
\definecolor{blue}{rgb}{0.0, 0.0, 1}
\newcommand{\tabincell}[2]{\begin{tabular}{@{}#1@{}}#2\end{tabular}}
\usepackage{makecell}

\usepackage{xpatch}
\makeatletter
\def\changeBibColor#1{%
	\in@{#1}{intro-IoT-2019, intro-IoT-EH-relay-2020,intro-UAV-2018,intro-UAV-2020, SER-multihop-2010,dm-af-ber-lou2017exact,dm-df-ser-1-liu2015energy, intro-5G-chen2020massive,ours-icassp,ref5-2-kadathlal2017generalised, intro-relay-new-2018,intro-relay-new-2019}%  list of colored bib items
	\ifin@\color{black}\else\normalcolor\fi
}

\xpatchcmd\@bibitem
{\item}
{\changeBibColor{#1}\item}
{}{\fail}

\xpatchcmd\@lbibitem
{\item}
{\changeBibColor{#2}\item}
{}{\fail}
\makeatother
\begin{document}
%	\title{Transmission under Non-coherent Cooperative Networks with Decode-and-Forward Relays: A Thorough Analysis}
%	\title{Detection with Optimized Transmission Scheme under Non-coherent Cooperative Networks with Decode-and-Forward Relays: A Thorough Analysis}
%	\title{Transmission under Non-coherent Relay Networks: An Optimized Power Allocation Scheme and  Performance Analysis}
	\title{Detection and Performance Analysis for Non-Coherent DF Relay Networks with Optimized Generalized Differential Modulation}
%	\title{ 
%		Detection and Performance Analysis for Non-Coherent Cooperative Networks with Decode-and-Forward Relays}
	\author{Yuxin~Lu,~\IEEEmembership{Student~Member,~IEEE,}
		and~Wai~Ho~Mow,~\IEEEmembership{Senior~Member,~IEEE}
		\thanks{This work was supported by the Hong Kong Research Grants Council under the GRF project no. 16207415.  
			Y. Lu and W. H. Mow are with the Department
			of Electronic and Computer
			Engineering, the Hong Kong University of Science and Technology, Hong
			Kong SAR (e-mail: \{ylubg, eewhmow\}@ust.hk).}}
	\maketitle
	
	% As a general rule, do not put math, special symbols or citatioNS
	
	%% Note that keywords are not normally used for peerreview papeRS.
	%\begin{IEEEkeywords}
	%	
	%\end{IEEEkeywords}
	
	% Note that keywords are not normally used for peerreview papeRS.
	%\begin{IEEEkeywords}
	%IEEEtran, journal, \LaTeX, paper, template.
	%\end{IEEEkeywords}
	
	% For peer review papeRS, you can put extra information on the cover
	% \fi
	%
	% For peerreview papeRS, this IEEEtran command iNSerts a page break and
	\IEEEpeerreviewmaketitle
	%%--------------------------------------------------------------------------------------------------------
	\begin{abstract}
		This paper studies the detection and performance analysis problems for a relay network with $N$ parallel decode-and-forward (DF) relays. Due to the distributed nature of this network, {\color{black} it is practically very challenging to fulfill the requirement of  instantaneous channel state information for coherent detection}. To bypass this requirement, we consider the use of {\color{black}non-coherent DF relaying based on} a generalized differential modulation (GDM) scheme, in which transmission power allocation over the $M$-ary phase shift keying symbols is exploited when performing differential encoding. In this paper, a novel detector at the destination {\color{black}of such a non-coherent DF relay network} is proposed. It is  an accurate approximation of the state-of-the-art detector, called the almost maximum likelihood detector (AMLD), but the detection complexity is considerably reduced from $\mathcal{O}(M^2N)$ to $\mathcal{O}(MN)$. {\color{black}By characterizing the dominant error terms, we derive an accurate approximate symbol error rate (SER) expression.} An optimized power allocation scheme for GDM is further designed based on this SER  expression. Our simulation {\color{black}demonstrates} that the proposed non-coherent scheme can perform close to the coherent counterpart {\color{black} as the block length increases}. {\color{black}Additionally, we prove that the diversity order of both the proposed detector and the AMLD is exactly $\lceil N/2 \rceil + 1$.} Extensive simulation results further verify the accuracy of our results in various scenarios.
	\end{abstract}
	
	%%--------------------------------------------------------------------------------------------------------
	\section{Introduction}	
	% logic: GDM includes power allocation scheme (treated as a minor contribution), put the deriveations in the appendix 
	Relay-assisted communication is an important technique to enhance the transmission reliability and achieve spatial diversity gains in future wireless systems \cite{intro-relay-5G, intro3-wang2014cellular}. Its basic idea is to introduce intermediate relay nodes to forward signals from the source to the destination. It is a particularly attractive technique for applications, such as ad hoc sensor networks, in which the use of multiple antennas is restricted by the size and cost limitations of the terminals. Among various relaying protocols, amplify-and-forward (AF) \cite{intro-AF-classic} and decode-and-forward (DF) \cite{intro-DF-classic} are the most popularly adopted and  widely studied.   DF relaying is an ongoing research topic and  has found many   important usages in many newly emerged applications, such as the Internet of Things (IoT), energy harvesting (EH) networks \cite{intro-IoT-2019, intro-IoT-EH-relay-2020}, and unmanned aerial vehicle (UAV) assisted communications \cite{intro-UAV-2018,intro-UAV-2020}. The devices in such systems may form a DF relay network  to facilitate  reliable communications among them, and they may harvest energy from the  received radio signals to meet the power-constrained operating condition \cite{dm-af-ber-lou2017exact,dm-df-ser-1-liu2015energy}.	 Currently, incorporating DF relaying into these new applications  involves a number of new challenges, such as low power consumption and high connectivity  \cite{intro-relay-new-2019,intro-5G-chen2020massive}. 
	It is thus important to investigate such issues on DF relaying in a timely manner so as to offer some insights and potentially facilitate such efforts. %  is drawing more and more attention.

	 Coherent detection at the destination of a relay network has been widely studied \cite{csi-relay-2004, cmrc-wang2007high,bhatnagar2011ml, kim2015low,group-wang2018near,ours-icassp}. It requires the prior knowledge of the instantaneous channel state information (CSI) of some network links.  {\color{black}However, to acquire the instantaneous CSI requires insertion of pilot symbols and frequent channel estimations, which may introduce heavy overhead in terms of bandwidth and  power consumption. This becomes a critical issue for future wireless systems facing the challenge of massive connectivity \cite{intro-5G-chen2020massive}.} Besides, the error performance is sensitive to CSI estimation errors. To eliminate the need for instantaneous CSI, differential modulation (DM) with non-coherent detection becomes an attractive alternative \cite{dm-AF-2006, ref5-3guan2011performance, ref2-bhatnagar2013pl, intro-dm-2015-DF,   dm-ref-df-1-liu2015energy,dm-relay-swipt-2016}. {\color{black} Unlike the widely studied coherent DF relay network,  its non-coherent counterpart has received much less attention. Meanwhile, since in the IoT and UAV relay networks the destination may be a complexity- and energy-constrained device (instead of the base station), low-complexity detection schemes are necessary in order to reduce the hardware complexity and power consumption at the destination.}
%	{\color{black} Moreover, most of the devices deployed in the IoT network are of small sizes and low-powered, which restraints the use of high power consumption coding and decoding, modulation and demodulation techniques. 	}
% ongoing researches
%	 	 , as studied for various scenarios in \cite{intro-dm-2015-DF, intro-dm-IOT, intro-dm-2018-sate, dm-ref-df-1-liu2015energy, intro-dm-2019-MIMO, intro-dm-2019-MIMO-DL, intro-dm-2019-PC, intro-dm-mMIMO, intro-dm-UWB, intro-dm-underwater}.
%Intermediate relay nodes are introduced to assist the source-destination communication, and two widely adopted relaying protocols are amplify-and-forward (AF) and decode-and-forward (DF) protocols. 	
	%
	
	%%--------------------------------------------------------------------------------------------------------
	%\subsection{Related Work and Challenges}
	A drawback of non-coherent detection lies in the performance loss as compared to its coherent counterpart. Specifically, {\color{black} it is well-known that} there is a $3$ dB performance loss in the case of a point-to-point communication channel, and a similar loss exists in both AF and DF relay networks using the conventional DM \cite{dm-ref7-1chen2006modulation}. To address this drawback, the generalized differential modulation (GDM) schemes are designed for AF relay networks in \cite{ref5-fang2009generalized,ref5-1-fang2014performance}. In GDM, a frame of transmitted symbols is divided into several blocks. In each block, the first symbol is called the reference symbol (RS), and the remaining symbols are called the normal symbols (NSs). 
	The RS of the current block is differentially encoded based on that of the previous block, while the NSs are differentially encoded based on  the RS of the current block. RSs and NSs are allocated different transmission power, and the performance can be improved by optimizing the associated power allocation scheme. Since both RSs and NSs convey information, {\color{black} the GDM is one of the most promising techniques that have the potential to improve the performance without reducing the transmission rate as compared to the conventional DM \cite{ref5-1-fang2014performance,ref5-2-kadathlal2017generalised}}.
{\color{black} To our best knowledge, it has not yet been considered for the DF relay networks. }
	For the AF relay networks, the power allocation schemes are obtained by maximizing the average output signal-to-noise ratio (SNR) (of the equivalent point-to-point channel), which is equivalent to minimizing the average symbol error rate (SER) \cite{ref5-fang2009generalized,ref5-1-fang2014performance}. However, this approach may be highly suboptimal for the DF relay networks due to the problem of erroneous relaying. 
% However, this approach may be inapplicable in DF due to the problem of erroneous relaying. In a extreme case when all relays detect wrongly, maximizing the output SNR would encourage the error propagation effect, and could degrade the overall SER performance. 	
	
	This paper considers a single-source single-destination network with $N$ parallel DF relays, assuming the availability of the average CSI at the destination. In this context,  non-coherent detection at the destination was studied in \cite{dm-ref7-1chen2006modulation,dm-ref7-zhu2010differential,ref1-bhatnagar2012decode} for the conventional DM\footnote{Hereafter, DM will be used to refer to the conventional DM as that used in \cite{dm-ref7-1chen2006modulation}.}. Two state-of-the-art detectors, i.e., the almost maximum likelihood detector (AMLD)\footnote{The so-called maximum-likelihood detector introduced in \cite{dm-ref7-1chen2006modulation} and \cite{dm-ref7-zhu2010differential} involves some approximations and is not truly optimal. Hence, it is referred to as the almost maximum likelihood detector here.} and the piecewise linear detector (PLD), were derived for non-coherent binary frequency shift keying (BFSK) in \cite{dm-ref7-1chen2006modulation}, and later for differential binary phase shift keying (DBPSK) in \cite{dm-ref7-zhu2010differential}. They were further extended to general $M$-ary differential PSK ($M$-DPSK) and differential $M$-QAM in \cite{ref1-bhatnagar2012decode}. 	
	The detection complexity of the AMLD is $\mathcal{O}(M^2N)$, while that of the PLD is $\mathcal{O}(MN)$. As shown in \cite{dm-ref7-1chen2006modulation,dm-ref7-zhu2010differential,ref1-bhatnagar2012decode}, the PLD is an accurate approximation of the AMLD. Some relay decoding error scenarios are ignored by the PLD to achieve the complexity reduction. 	
	Analytical SER and diversity order analyses were conducted in \cite{dm-ref7-1chen2006modulation,dm-ref7-zhu2010differential,ref1-bhatnagar2012decode} for the PLD, whereas no SER analysis results were given for the AMLD to our best knowledge. For ease of discussion, Table \ref{tab: results-diversity} summarizes the known performance analysis results in \cite{dm-ref7-1chen2006modulation,dm-ref7-zhu2010differential,ref1-bhatnagar2012decode}. 
		\begin{table}[htb]
		\caption{Known performance analysis results for the PLD in the non-coherent DF relay network with $N$ parallel relays and a direct link } % title of Table
		\centering % used for centering table
		\begin{tabular}{c | c |  c | c | c } % centered columns (4 columns)
			\toprule %inserts double horizontal lines
			$\#$ \textbf{of relays}  & \textbf{Modulation} & \tabincell{c}{\textbf{SER expression}} &  \textbf{Diversity order} $d$ & \textbf{Reference} \\ 
			\midrule % inserts single horizontal line
			 & \tabincell{c}{Non-coherent  BFSK} &  &  &
			 \cite{dm-ref7-1chen2006modulation}	 \\ 
			\cline{2-2}  \cline{5-5}
			% inserts single horizontal line
			\raisebox{0ex}{$N=1$}  & DBPSK & Available & \tabincell{c}{$N+1$  (full diversity)}  &
			\cite{dm-ref7-zhu2010differential}	 \\ [1ex]
			\cline{2-2}  \cline{5-5}
			% inserting body of the table
			& $M$-DPSK &   &  	&   \cite{ref1-bhatnagar2012decode} \\   	
			\midrule %inserts single line
			 & Non-coherent BFSK & &  \tabincell{c}{$ d = N/2+1$ for $N$ even; \\ $N/2+1 \leq d \leq (N+1)/2+1$ for $N$ odd \\ (assuming perfect relay-destination links)}&
			 \cite{dm-ref7-1chen2006modulation}   \\   [1ex]
			\cline{2-2}  \cline{4-5}
			\raisebox{2.5ex}{$N \geq 2$}	&	DBPSK & \raisebox{2.5ex}{Unavailable} &    $d > N/2+1$	 & \cite{dm-ref7-zhu2010differential} 	  \\ [1ex]
			\cline{2-2}  \cline{4-5} %inserts single line
			 & $M$-DPSK &  & \tabincell{c}{$N+1$  (assuming error-free relays)} &	\cite{ref1-bhatnagar2012decode}  \\  [1ex]
			\bottomrule
		\end{tabular}
		\label{tab: results-diversity} % is used to refer this table in the text
	\end{table}	
	It can be seen from the table that the SER expressions for the PLD were obtained only for the single relay case but not for the multiple relay case.  In addition, for the multiple relay case (when $N \geq 2$) with $M$-DPSK, the full diversity order result was obtained under the impractical assumption that all relays are error-free. The main challenge is due to the complexity of performance analysis, which grows exponentially with the number $N$ of relays \cite{dm-ref7-1chen2006modulation}. This paper aims to tackle the challenge and provide more thorough analyses on the SER and the diversity order.	

 In this paper, we propose a new detector, thoroughly analyze the error performance, and optimize the transmission power allocation for the non-coherent DF relay networks adopting the GDM scheme based on $M$-PSK. The main contributions are summarized as follows.  % We consider the usage of GDM with general $M$-ary DPSK. 
\begin{itemize}
		\item We propose an $\mathcal{O}(MN)$-complexity detector, called the near maximum likelihood detector (NMLD), by accurately approximating the detection metric of the $\mathcal{O}(M^2N)$-complexity AMLD. We also clarify the relation between the detection metrics of the proposed NMLD and the PLD. Our simulation results show that the NMLD performs similarly as the AMLD {\color{black}  but with a considerably reduced complexity in various relay channel scenarios. For example, when $M=8$ and $N=12$, the number of operations saved by the NMLD compared to the AMLD is up to $84.39 \%$.}
%		, both of which are approximations of the AMLD
		% especially when $M$ or $N$ increases. 
		\item {\color{black}We derive a new approximate SER expression that is applicable to any number of relays and for any value of $M$.  Such SER expression has not been developed in the literature,  to our best knowledge.}  Based on this SER expression, we further design an optimized transmission power allocation scheme for GDM. As verified by simulation, the SER expression is rather accurate {\color{black} and the proposed scheme can perform close to the coherent counterpart as the block length increases. For example, in the single relay case, for a block length of $256$ DQPSK symbols at SER $10^{-5}$, the performance gap in SNR is within $0.5$ dB}. 
%		a relatively strong $S-R$ link is needed usually because the direct $S-D$ link is relatively weak.
		\item {\color{black} By analyzing the high-SNR behavior of  our SER expression, we study how the number of erroneous relays affects the diversity order, and prove that the diversity order is exactly $\Big\lceil \frac{N}{2} \Big\rceil + 1$ for both the NMLD and the AMLD. Such diversity order results are novel in that they are applicable to any value of $N$ (and $M$). Our simulation results demonstrate that the diversity order results are accurate in various relay channel scenarios. }  
\end{itemize}
	%%--------------------------------------------------------------------------------------------------------
	%%--------------------------------------------------------------------------------------------------------
	\subsection{Organization and Notation}
	
	%{\color{black}check this subsection last}
	
	The rest of this paper is organized as follows. Sections \Rmnum{2}-\Rmnum{4} are for the single relay network. In Section \Rmnum{2}, we introduce the GDM transmission scheme followed by a brief review of the AMLD and the PLD for DM. In Section \Rmnum{3}, our NMLD is proposed. The SER analysis for NMLD is presented in Section \Rmnum{4}, followed by the design of an optimized power allocation scheme. The proposed NMLD is extended to the multiple relay network in Section \Rmnum{5}, where its performance analysis is also performed. Our simulation results with discussions are presented in Section \Rmnum{6}, followed by conclusions and future work in Section \Rmnum{7}.
	
	\textit{Notation}: $\Re \{\cdot\}$ denotes the real part of a number or an expression.
	Bold upper and lower case letters denote matrices and vectors, respectively.  $\mathbf{1}_{n}$ and $\mathbf{0}_{n}$ stand for all-$1$ and all-$0$ column vectors of length $n$, respectively. $\mathbf{a}[i]$ denotes the $i$-th element of a vector $\mathbf{a}$. $\diag(\mathbf{a})$ denotes a diagonal matrix whose diagonal entries starting in the upper left corner are $a_1, \dots, a_n$.  $a^*$ denotes the complex conjugate of $a$. The sum $\sum_{a}^{b}(\cdot)$ is $0$ if $a > b$. $\lceil \cdot \rceil$ denotes the ceil operator.
	%%--------------------------------------------------------------------------------------------------------
	\section{\color{black} System Model and Existing Works}
	\label{sec:sys-model}
%	{\color{black} In this section, we present the GDM scheme with transmission power allocation over symbols, and introduce the AMLD to facilitate the development of our detector in the next section.} 
{\color{black}
The system model of a multiple relay network is shown in Fig. \ref{Fig:system-model-general}, where $N$ parallel relays $R_1, \dots, R_N$ form $N$ parallel branches in addition to the direct source to destination ($S-D$) link. All relays adopt the DF protocol, and are operated in the half-duplex mode. It is assumed that the relays have no CSI knowledge, and the destination only has the average CSI, i.e., the average link SNRs.  % representing the transmitting and receiving node pair
 $\Theta \triangleq \{1,2,\dots,N\}$ denotes the relay index set. For the $I-J$ link, $(I,J) \in \{ (s,d), (s, r_{n}), (r_{n}, d) \}$, $n \in \Theta$, $n_{I, J}$ is used to denote the complex additive white Gaussian noise (AWGN) with zero mean and variance $N_{I, J}$, i.e., $n_{I, J} \sim \mathcal{CN}(0,N_{I, J})$, and $h_{I, J}$ denotes the complex fading coefficient modeled as a zero-mean complex Gaussian random variable. $\gamma_{{I}, {J}} \triangleq \frac{|h_{{I}, {J}}|^2 }{N_{{I}, {J}}}$ is defined as the instantaneous link SNR, and $\bar{\gamma}_{{I}, {J}} = \mathbb{E}_{ |h_{{I}, {J}}| } [\gamma_{{I}, {J}}]$ is the average link SNR.  

\begin{figure}[t]
	\centering
	\includegraphics[width=3.2in]{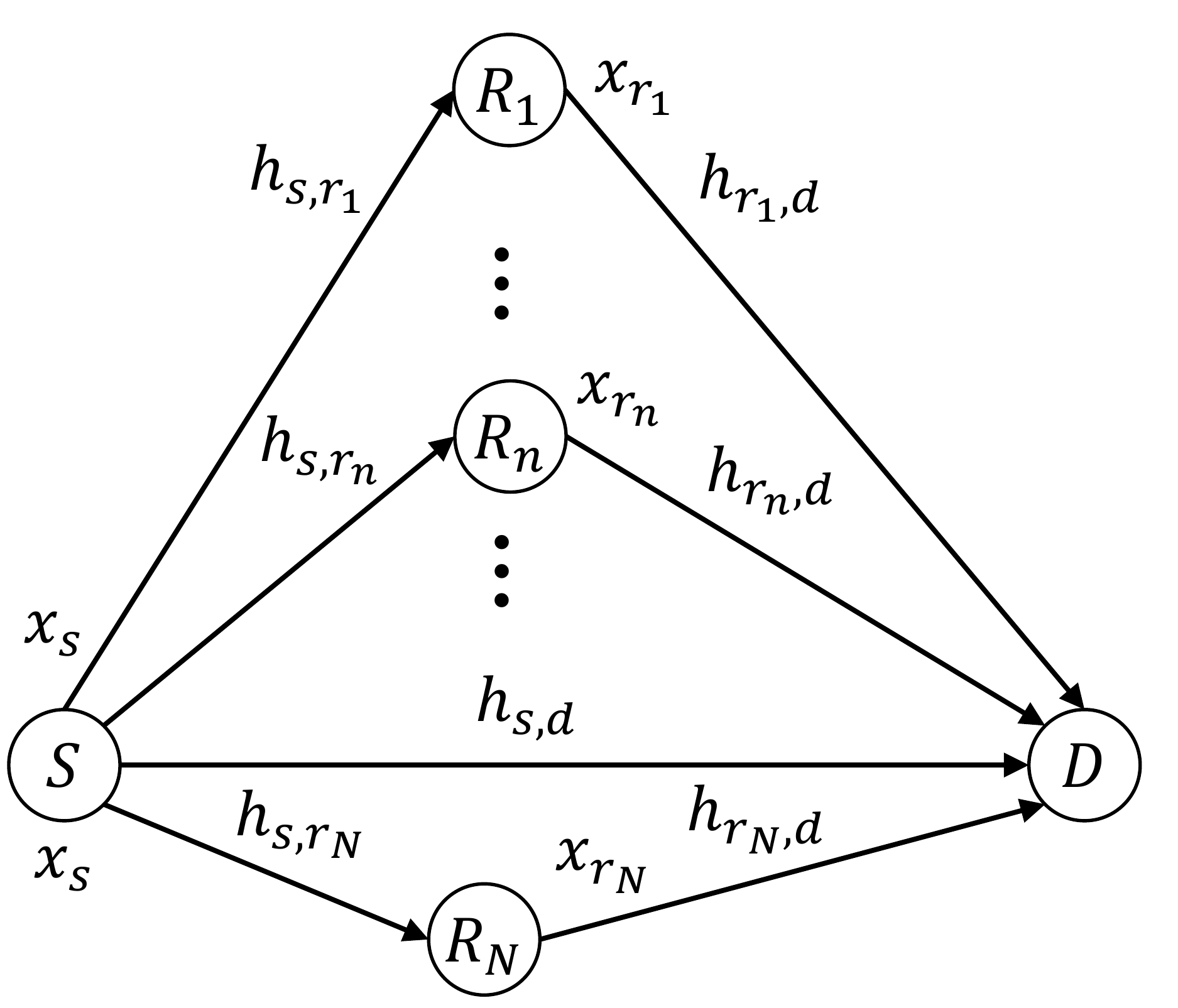}
	\caption{The sysem model of a relay network with $N$ parallel relays ($R_1, \dots, R_N$) and a source to destination ($S-D$) pair.}
	\label{Fig:system-model-general}
\end{figure}
	Let the source transmit a frame of $K$ information symbols with an average transmitting power $P_s$. At symbol time interval $k, k= 1,2,\dots, K$, the source selects a symbol, denoted as $x_s[k]$, from the $M$-PSK alphabet, defined as $\mathcal{X} \triangleq \{x_t=e^{j 2 \pi (t-1) /M}, t=1,2,\dots,M\}$, with equal probability. $x_s[k]$ is then differentially encoded as $u_s[k]$ for transmission.
	 Each symbol transmission takes $N+1$ time slots. In the first time slot, the source broadcasts its differentially encoded signal to all relays and the destination.  In the $(n+1)$-th time slot, $n \in \Theta$, $R_n$ differentially encodes its detected $x_{r_{n}}$ for transmission. }
	% $u_s[0]$ is the initialization symbol. 
	% The initialization symbol is power-constrained with $|u_s[0]|^2=\rho_{\rm{R}}$, and we set $u_s[0]=\sqrt{\rho_{\rm{R}}}$ for simplicity.  
	% The frame length is $K$ including the initialization symbol.  
	%%--------------------------------------------------------------------------------------------------------
	\subsection{Generalized Differential Modulation and Transmission}
	\label{sec:GDM}
	%%--------------------------------------------------------------------------------------------------------

\begin{figure}[t]
			\centering
			\includegraphics[width=4.2in]{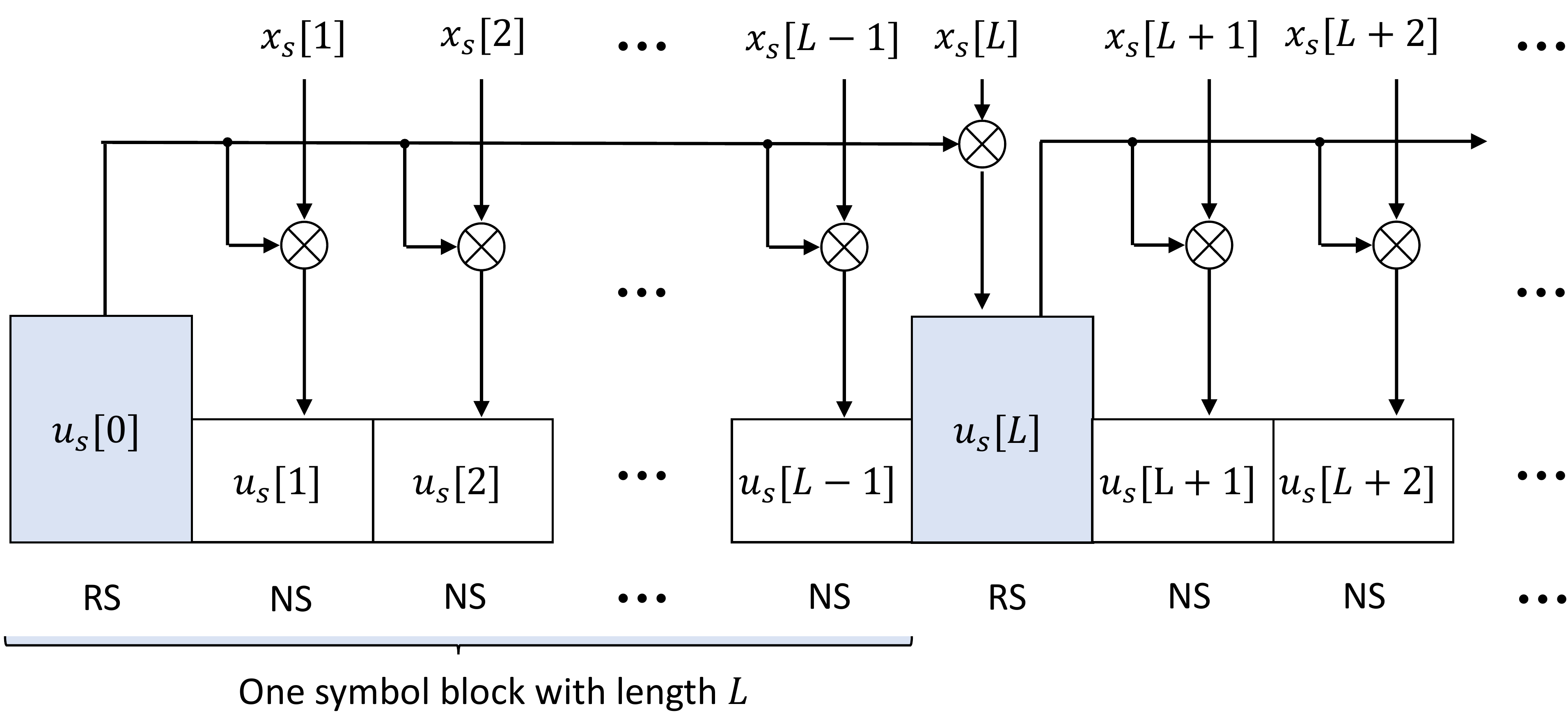}
			\caption{Diagram of the GDM scheme at the source. Different heights of the blocks denote the different power allocated to the RSs and NSs. Blocks labeled DE represent the differential encoding (DE) operations. A frame of $K+1$ symbols (including the initial symbol $u_s[0]$) is divided into several blocks each with length $L$.  }
			\label{fig:GDM-example}
		\end{figure}

	% % one frame is divided into $B$ blocks each with length $L$, and the frame length $K=BL$ % One block consists of $L$ symbols. 
	Let us first consider a classical three-node cooperative network, where one relay ($R$) helps the source ($S$) to communicate with the destination ($D$). Fig.~\ref{fig:GDM-example} shows a diagram of the GDM scheme, in which one frame of $K+1$ symbols (including the initial symbol $u_s[0]$) is divided into several blocks each with length $L$.  
		In each block, there are two types of symbols that are allocated different transmission power. The first symbol in each block, denoted as $u_s[bL], b=0,1,\dots$, is referred to as the reference symbol (RS) (including the initialization symbol $u_s[0]$), and the other symbols are referred to as the normal symbols (NSs). The RS and NSs are transmitted with the power values of $\rho_{\rm{T}}$, $\rm{T} \in \{\rm{R},\rm{N}\}$. The RS in the $b$-th block is differentially encoded based on the RS in the $(b-1)$-th block as
	\begin{IEEEeqnarray}{rCl}
			u_s[bL] = & u_s[(b-1)L]x_s[bL], \quad b=1,2,\dots, (K+1)/L-1, \label{eq:DE-RS}
	\end{IEEEeqnarray}
	where for simplicity, we set $u_s[0]=\sqrt{\rho_{\rm{R}}}$, and assume that $(K+1)/L$ is an integer and that $\rho_{\rm{R}}$ and $\rho_{\rm{N}}$ are known by the receivers, i.e., the relay and destination. The NSs in the $b$-th block are differentially encoded based on the RS in the same block as
	\begin{IEEEeqnarray}{rCl}
		u_s[bL+l] = & \sqrt{\rho_{\rm{N}}/\rho_{\rm{R}}} u_s[bL]x_s[bL+l], \quad b=0,1,\dots, (K+1)/L-1,\quad  l=1,2,\dots,L-1. \label{eq:DE-NS}
	\end{IEEEeqnarray}
	Based on \eqref{eq:DE-RS} and \eqref{eq:DE-NS}, the average power constraint at the source should be satisfied as
	\begin{IEEEeqnarray}{rCl} \label{eq:Ps-const}
		\rho_{\rm{R}}+(L-1)\rho_{\rm{N}}=LP_s.
	\end{IEEEeqnarray}
	
%Here we describe the channel models. 
	%%--------------------------------------------------------------------
 In the first time slot, the source broadcasts its signal $u_s[k], k=0,1,\dots, K$, and the respectively received signals at the relay and the destination are
	\begin{IEEEeqnarray}{rCl}  \label{eq:y-slot1}
		y_{s,J} [k]= h_{s,J}[k] u_s[k]+n_{s,J} [k],  \quad J \in \{r,d\} .
	\end{IEEEeqnarray}
 In the second time slot, the relay performs detection to get $x_r[k]$, differentially encodes $x_r[k]$ as $u_r[k]$ following the same approach as that of the source, and transmits $u_r[k]$ to the destination. Let the relay transmitting power be $P_s$, and the destination receives 
	\begin{IEEEeqnarray}{rCl} \label{eq:y-slot2}
		y_{r,d} [k]= h_{r,d}[k]u_r[k]+n_{r,d} [k].
	\end{IEEEeqnarray}
	% similarly to that in \eqref{eq:DE-RS}-\eqref{eq:DE-NS}
	%\begin{IEEEeqnarray}{rCl}
	%	y_{r,d} [k]= h_{r,d}u_r[k]+n_{r,d} [k].
	%\end{IEEEeqnarray} % quasi-static
	For slow Rayleigh fading channels, we follow \cite{ref5-fang2009generalized} to assume each $h_{I, J}$, $({I}, {J}) \in \{(s,r),(s,d), (r,d)\}$, remains invariant within one frame duration, while varies independently from one frame to another (quasi-static Rayleigh fading). Then we have for the two consecutive blocks that $h_{I, J}[bL] = h_{I, J}[(b-1)L] $, $b=1,2,\dots, (K+1)/L-1$. Based on \eqref{eq:DE-RS}-\eqref{eq:y-slot2}, the received signals can be written equivalently as
	\begin{IEEEeqnarray}{rCl}
		y_{{I}, {J}}[bL] & = &  y_{{I}, {J}}[(b-1)L] x_{I}[bL] + n_{{I}, {J}}[bL]-  x_{I}[bL] n_{{I}, {J}}[(b-1)L],   \label{eq:y-RS}\\
		y_{{I}, {J}}[bL+l] & = & \sqrt{\rho_{\rm{N}}/\rho_{\rm{R}}} y_{{I}, {J}}[bL] x_{I}[bL+l] + n_{{I}, {J}}[bL+l]- \sqrt{\rho_{\rm{N}}/\rho_{\rm{R}}}  x_{I}[bL+l] n_{{I}, {J}}[bL].  \label{eq:y-NS}
	\end{IEEEeqnarray}
	Equations \eqref{eq:y-RS} and \eqref{eq:y-NS} can be written in a unified form as
	\begin{IEEEeqnarray}{rCl}
		y_{{I}, {J}}[k_1] & = & \sqrt{\rho_{\rm{T}}/\rho_{\rm{R}}} y_{{I}, {J}}[k_2] x_{I}[k_1] + n'_{{I}, {J}},  \label{eq:receive-p2p}
	\end{IEEEeqnarray}
%	\rm{T} \in \{\rm{R},\rm{N}\},
	where for the RSs, $k_1=bL$, $k_2=(b-1)L$ and $\rm{T}=\rm{R}$, and for the NSs, $k_1=bL+l$, $k_2=bL$ and $\rm{T}=\rm{N}$. The noises $n'_{{I}, {J}}=n_{{I}, {J}}[k_1]- \sqrt{\rho_{\rm{T}}/\rho_{\rm{R}}}  x_{I}[k_1] n_{{I}, {J}}[k_2] \sim \mathcal{CN}(0, (1+\rho_{\rm{T}}/\rho_{\rm{R}})N_{{I}, {J}})$. 
	\subsection{Existing Detectors for the Conventional Differential Modulation }
	The DM scheme can be regarded as a special case of the GDM scheme when the block length $L=1$ and all symbols are allocated the same power as $\rho_{\rm{R}} = \rho_{\rm{N}} = P_s$. For $k= 1,2,\dots, K$, we have
	$
	u_I[k] = u_I[k-1] x_I[k]
	$ with the power constraint $|u_I[0]|^2 = P_s$, and $y_{{I}, {J}}[k] =  y_{{I}, {J}}[k-1] x_{I}[k] + n'_{{I}, {J}}$ with $n'_{{I}, {J}} \sim \mathcal{CN}(0, 2 N_{{I}, {J}})$, $({I}, {J}) \in \{(s,r),(s,d), (r,d)\}$. The state-of-the-art detectors are the AMLD and the PLD \cite{ref1-bhatnagar2012decode}. The optimal maximum likelihood detector for DM is performed as $\max_{x_s \in \mathcal{X}} f(y_{s,d}[k]|x_s, y_{s,d}[k-1])
	\sum_{x_r \in \mathcal{X}} \Pr(x_r|x_s) f(y_{r,d}[k]|x_r,y_{r,d}[k-1]),$ which finds the transmitted source symbol $x_s \in \mathcal{X} $ that maximizes the conditional joint probability density of the received signals. By approximating the transition probability term $\Pr(x_r|x_s)$ using the average SER at the relay, denoted as $\epsilon$, the AMLD metric is given as
	\begin{IEEEeqnarray}{rCl} 
		f
		\left( y_{s,d}[k]|x_s, y_{s,d}[k-1] \right) \Big[ (1-\epsilon) f\left(y_{r,d}[k]|x_s,y_{r,d}[k-1]\right)+  \frac{\epsilon}{M-1} \sum_{\substack{x_r \in \mathcal{X} x_r \neq x_s}} f(y_{r,d}[k]|x_r,y_{r,d}[k-1]) \Big]. \IEEEeqnarraynumspace \label{eq:AML-metric}
	\end{IEEEeqnarray}
	The AMLD has $\mathcal{O}(M^2)$ complexity. Based on \eqref{eq:AML-metric}, the $\mathcal{O}(M)$-complexity PLD can be developed by omitting some probability terms and applying the piece-wise linear
	approximation. The detection metric of the PLD will be discussed in detail in Section \ref{sec:multi-NMLD}, where we analyze the relation between the PLD and the proposed NMLD.
	
%	{\color{black} 	Based on \eqref{eq:receive-p2p} and \eqref{eq:AML-metric}, the AMLD metric for GDM can be written straightforwardly as 
%		\begin{IEEEeqnarray}{rCl} 
%			& & f
%			\left( y_{s,d}[k_1]|x_s, y_{s,d}[k_2] \right) \Bigg[ (1-\epsilon) f\left(y_{r,d}[k_1]|x_s,y_{r,d}[k_2]\right)+  \epsilon/(M-1) \sum_{\substack{x_r \in \mathcal{X} \\ x_r \neq x_s}} f(y_{r,d}[k_1]|x_r,y_{r,d}[k_2]) \Bigg] \nonumber 
%		\end{IEEEeqnarray}
%		\begin{IEEEeqnarray}{rCl} 
%			& = & \exp \left( - |y_{s,d}[k_1]-\sqrt{\rho_{\rm{T}}/\rho_{\rm{R}}} y_{s,d}[k_2]x_s|^2/N_{s,d} \right) \left[  (1-\epsilon) \exp \left( - |y_{r,d}[k_1]-\sqrt{\rho_{\rm{T}}/\rho_{\rm{R}}} y_{r,d}[k_2]x_s|^2/N_{r,d} \right) \right. \nonumber \\
%			& &  + \epsilon/(M-1)  \sum_{\substack{x_r \in \mathcal{X}, x_r \neq x_s}} \exp \left( - |y_{r,d}[k_1]-\sqrt{\rho_{\rm{T}}/\rho_{\rm{R}}} y_{r,d}[k_2]x_r|^2/N_{r,d} \right) \Big]. \label{eq:AML-metric-GDM}
%	\end{IEEEeqnarray}}	
	%%--------------------------------------------------------------------
	%%--------------------------------------------------------------------------------------------------------
	\section{Single Relay Detection} % at the Destination
	In this section, we derive an expression of the average SER at the relay (the value of $\epsilon$) and introduce our $\mathcal{O}(M)$-complexity near maximum likelihood detector (NMLD).  
	\subsection{Average SER at the Relay}
	\label{sec:SER-relay}
%	\begin{figure}[H]
%		\centering
%		\includegraphics[width=3.2in]{SER-p2p-4816-v2}
%		\caption{Analytical expressions and simulated results of the SER for the quasi-static Rayleigh fading $S-R$ link (the fading coefficient remains fixed for the duration of one frame, while varies independently from one frame to another) using GDM based on $M$-DPSK, where the frame length $K=320$, the block length $L=16$ and $P_s=1$.}
%		\label{fig:SER-point-to-point-PSK-GDM}
%	\end{figure} 

	The optimal detection at the relay can be performed as $x_r[k_1] = \arg\min_{x \in \mathcal{X}} |y_{s,r}[k_1]-\sqrt{\rho_{\rm{T}}/\rho_{\rm{R}}} \\ y_{s,r}[k_2] x|^2$ according to \eqref{eq:receive-p2p}.  For the $I-J$ link, by ignoring the higher order noise terms \cite{ref1-bhatnagar2012decode}, the average receive SNR can be obtained as 
	\begin{IEEEeqnarray}{rCl} 
	\mathbb{E}_{|h_{{I}, {J}}|} \left[ \frac{(\rho_{\rm{T}}/\rho_{\rm{R}}) |y_{{I}, {J}}[k_2]|^2}{(1+\rho_{\rm{T}}/\rho_{\rm{R}}) N_{{I}, {J}}} \right]
	\approx    \phi_{\rm{T}}  \bar{\gamma}_{{I}, {J}}, \quad \rm{T} \in \{\rm{R},\rm{N}\}, \label{eq:approx-high-noise} 
	% & = \rho_{R} |h_{{I}, {J}}|^2.
	\end{IEEEeqnarray}
	where $\phi_{\rm{T}}  \triangleq \frac{1}{1/\rho_{\rm{T}} + 1/\rho_{\rm{R}}}$. Based on \eqref{eq:approx-high-noise}, an expression of $\epsilon$ can be obtained (see  \eqref{eq:ser-relay} and its derivation in Appendix \ref{app:SER-SRlink}). 

%	{\color{black} Since this expression  is obtained using the approximation  in \eqref{eq:approx-high-noise}, we can verify the accuracy of \eqref{eq:approx-high-noise} by verifying the accuracy of this expression. To achieve this, we have  compared the expression with a baseline given in \cite[eq. (33)]{dm-ref-df-1-liu2015energy} and the Monte-Carlo simulation. It was observed that the proposed expression is accurate  for different modulation sizes over a range of SNRs.}
%	To achieve this, we compare our expression of $\epsilon$ with a baseline given in \cite[eq. (33)]{dm-ref-df-1-liu2015energy} and the Monte-Carlo simulation
%	 in Fig.~\ref{fig:SER-point-to-point-PSK-GDM}, where the frame length $K=320$, the block length $L=16$ and $P_s=1$. The results show that the proposed expression of $\epsilon$ is accurate for different modulation sizes and at all SNRs considered, verifying that \eqref{eq:approx-high-noise} is accurate. 	
	
	%%--------------------------------------------------------------------
	\subsection{NMLD at the Destination}
	\label{sec:NMLD-single}
	%\hl{to be modified based on the first draft}
%	 for $\rm{T} \in \{\rm{R},\rm{N}\}$,

	Based on \eqref{eq:receive-p2p} and \eqref{eq:AML-metric}, the AMLD metric for GDM can be written straightforwardly as 
	\begin{IEEEeqnarray}{rCl} 
		& & f
		\left( y_{s,d}[k_1]|x_s, y_{s,d}[k_2] \right) \Bigg[ (1-\epsilon) f\left(y_{r,d}[k_1]|x_s,y_{r,d}[k_2]\right)+  \epsilon/(M-1) \sum_{\substack{x_r \in \mathcal{X} \\ x_r \neq x_s}} f(y_{r,d}[k_1]|x_r,y_{r,d}[k_2]) \Bigg] \nonumber 
	\end{IEEEeqnarray}
	\begin{IEEEeqnarray}{rCl} 
		& = & \exp \left( - |y_{s,d}[k_1]-\sqrt{\rho_{\rm{T}}/\rho_{\rm{R}}} y_{s,d}[k_2]x_s|^2/N_{s,d} \right) \left[  (1-\epsilon) \exp \left( - |y_{r,d}[k_1]-\sqrt{\rho_{\rm{T}}/\rho_{\rm{R}}} y_{r,d}[k_2]x_s|^2/N_{r,d} \right) \right. \nonumber \\
		& &  + \epsilon/(M-1)  \sum_{\substack{x_r \in \mathcal{X}, x_r \neq x_s}} \exp \left( - |y_{r,d}[k_1]-\sqrt{\rho_{\rm{T}}/\rho_{\rm{R}}} y_{r,d}[k_2]x_r|^2/N_{r,d} \right) \Big]. \label{eq:AML-metric-GDM}
	\end{IEEEeqnarray}
By applying the widely-used max-sum approximation (see also \cite{kim2015low,qian2017near}) to replace the summation operations of the exponentials in \eqref{eq:AML-metric-GDM} with the maximum operations, we obtain the detection rule of the NMLD as 
	\begin{IEEEeqnarray}{rCl} 
		\hat{x}_s 
		& = & \arg\max_{x_s \in \mathcal{X}} \left\{ \exp \left( - |y_{s,d}[k_1]-\sqrt{\rho_{\rm{T}}/\rho_{\rm{R}}} y_{s,d}[k_2]x_s|^2/N_{s,d} \right) \max \left\{  (1-\epsilon) \exp \left( - |y_{r,d}[k_1]-\sqrt{\rho_{\rm{T}}/\rho_{\rm{R}}} \right.\right. \right. \nonumber \\
		& & \left. \left. \left.  y_{r,d}[k_2]x_s|^2/N_{r,d} \right), \epsilon/(M-1) \max_{\substack{x_r \in \mathcal{X} , x_r \neq x_s}} \exp \left( - |y_{r,d}[k_1]-\sqrt{\rho_{\rm{T}}/\rho_{\rm{R}}} y_{r,d}[k_2]x_r|^2/N_{r,d} \right)
		\right\} \right\} \nonumber \\
		& = & \arg\min_{x_s \in \mathcal{X}} \left\{
		\frac{1}{N_{s,d}} |y_{s,d} [k_1]- \sqrt{\rho_{\rm{T}}/\rho_{\rm{R}}} y_{s,d} [k_2]x_s|^2 + \min \left\{ \frac{1}{N_{r,d}}|y_{r,d} [k_1]- \sqrt{\rho_{\rm{T}}/\rho_{\rm{R}}} y_{r,d} [k_2]x_s|^2,  \right. \right.  \nonumber \\
		& & \left.\left.  \min_{\substack{x_r \in \mathcal{X} , x_r \neq x_s}} \frac{1}{N_{r,d}} |y_{r,d} [k_1]-\sqrt{\rho_{\rm{T}}/\rho_{\rm{R}}} y_{r,d} [k_2]x_{r}|^2 + \eta\right\}		
		\right\} \label{eq:NMLD-metric-pre}\\
		& = & \arg\min_{x_s \in \mathcal{X}} \left\{
		\frac{1}{N_{s,d}} |y_{s,d} [k_1]- \sqrt{\rho_{\rm{T}}/\rho_{\rm{R}}} y_{s,d} [k_2]x_s|^2 + \min \left\{ \frac{1}{N_{r,d}}|y_{r,d} [k_1]- \sqrt{\rho_{\rm{T}}/\rho_{\rm{R}}} y_{r,d} [k_2]x_s|^2,  \right. \right.  \nonumber  \\
		& & \left.\left.  \min_{\substack{x_r \in \mathcal{X}}} \frac{1}{N_{r,d}}|y_{r,d} [k_1]-\sqrt{\rho_{\rm{T}}/\rho_{\rm{R}}} y_{r,d} [k_2]x_{r}|^2 + \eta\right\}		
		\right\} , \label{eq:NMLD-metric}
	\end{IEEEeqnarray}
	where $\eta \triangleq \left(1+\rho_{\rm{T}}/\rho_{\rm{R}}\right)\log\frac{(1-\epsilon)(M-1)}{\epsilon}$. From \eqref{eq:NMLD-metric-pre} to \eqref{eq:NMLD-metric}, we remove the constraint of $ x_r \neq x_s$. This is because $\eta > 0$ holds when $\epsilon <1/2$ and $M \geq 2$, which is considered to be true.
	{ \color{black}
	The NMLD with $\mathcal{O}(M)$  complexity can be implemented according to \eqref{eq:NMLD-metric}, as described next.  First, we solve for $C_1=  \min_{x_{r} \in \mathcal{X}} \frac{1}{N_{r,d}} |y_{r,d} [k_1]-\sqrt{\rho_{\rm{T}}/\rho_{\rm{R}}}  y_{r,d} [k_2]x_{r}|^2 + \eta$, and the obtained $C_1$ value becomes a fixed threshold. This step has complexity $\mathcal{O}(M)$ since $|\mathcal{X}|=M$, where $|\mathcal{X}|$ denotes the cardinality of $\mathcal{X}$. Then, for each $x_s \in \mathcal{X}$, we compare the value of $\frac{1}{N_{r,d}}|y_{r,d} [k_1]- \sqrt{\rho_{\rm{T}}/\rho_{\rm{R}}} y_{r,d} [k_2]x_s|^2$ with $C_1$, and calculate the detection metrics to choose the symbol $x_s$ with the smallest metric value as $\hat{x}_s$. This step also has  complexity $\mathcal{O}(M)$. Overall, the complexity of the NMLD is $\mathcal{O}(M)$. }
	\section{SER Performance and Optimized Power Allocation Scheme}
	In this section, we derive an approximate SER expression, and based on which an optimized power allocation scheme is obtained. {\color{black} The solutions to all optimization problems under consideration will be given in explicit forms.}%  
	
	% where $\tau (x_s, x_r) = 0$ if $x_r=x_s$, and $\tau (x_s, x_r) =  1$ if $x_r \neq x_s$. 
%	We respectively characterize the SER performances for the two scenarios of perfect and erroneous relaying. Perfect (erroneous) relaying means that the relay detects correctly (wrongly). 
	The overall error probability can be written as 
	$\Pr [\text{error}]  =  (1-\epsilon) \Pr [\text{error}|\text{error-free relaying}] + \epsilon \Pr [\text{error}|\text{erroneous relaying}]$.
%	\begin{IEEEeqnarray}{rCl}
%		\Pr [\text{error}] & =  (1-\epsilon) \Pr [\text{error}|\text{error-free relaying}] + \epsilon \Pr [\text{error}|\text{erroneous relaying}] . \nonumber 
%%		\nonumber 
%%		\\
%%		& < 	\mathcal{P}^{\rm{C}} + 	\mathcal{P}^{\rm{E}} \label{eq:union-bound} ,
%	\end{IEEEeqnarray}
$\Pr [\text{error}|\text{error-free relaying}]$ and $\Pr [\text{error}|\text{erroneous relaying}]$ characterize the conditional SER performances for the scenarios of error-free relaying and erroneous relaying, respectively. Error-free (erroneous) relaying means that the relay detects correctly (wrongly). The key idea is to determine the dominant pairwise error probability (PEP) terms for the above two scenarios. To facilitate later analysis, the following primary results are presented. 

%The link metric associated with the $I-d$ link, $I \in \{s, r \}$, is $|y_{I,d}[k_1]- \sqrt{\rho_{\rm{T}}/\rho_{\rm{R}}} y_{I,d}[k_2] x_r |^2$, and the difference of 	the detection metric is denote $\omega_{I,d} (t_1, t_2)$ as the difference between 
For notational convenience, we define $\lambda (t_1, t_2)  \triangleq \frac{1}{N_{s,d}} |y_{s,d}[k_1]-\sqrt{\rho_{\rm{T}}/\rho_{\rm{R}}} y_{s,d}[k_2] t_1 |^2 + \frac{1}{N_{r,d}} |y_{r,d}[k_1]-\sqrt{\rho_{\rm{T}}/\rho_{\rm{R}}} y_{r,d}[k_2] t_2 |^2$, and write the detection metric in \eqref{eq:NMLD-metric-pre} equivalently as 
%$\min \Big\{\min_{x_s \in \mathcal{X} } \lambda (x_s, x_s)  ,  \\ \min_{ (x_s, x_r) \in \mathcal{X}^2, x_r \neq x_s} \lambda (x_s, x_r)  + \eta \Big\} = \min_{x_s \in \mathcal{X} } \min_{x_r \in \mathcal{X}, x_r \neq x_s } \left\{ \lambda (x_s, x_s),  \lambda (x_s, x_r) +  \eta  \right\}  $, 
\begin{IEEEeqnarray}{rCl}
	& & \min \left\{\min_{x_s \in \mathcal{X} } \lambda (x_s, x_s)  , \min_{ (x_s, x_r) \in \mathcal{X}^2, x_r \neq x_s} \lambda (x_s, x_r)  + \eta \right\} \nonumber \\
	%	= &  \min_{(x_s, x_r) \in \mathcal{X}^2 } \left\{  \lambda (x_s, x_r) + \tau (x_s, x_r)  \eta  \right\}  \\
	& = & \min_{x_s \in \mathcal{X} } \min_{x_r \in \mathcal{X}, x_r \neq x_s } \left\{ \lambda (x_s, x_s),  \lambda (x_s, x_r) +  \eta  \right\}  ,	 \nonumber
	% \label{eq:NMLD-metric-v2} .
	%	\\ 
	%	= & \min_{x_s \in \mathcal{X} } \left\{ \lambda (x_s, x_s)  , \min_{ x_r \in \mathcal{X}, x_r \neq x_s} \lambda (x_s, x_r)  + \eta \right\} 
\end{IEEEeqnarray}
where the detection metric associated with a symbol pair $( x_s, x_r )$ is $\lambda (x_s, x_s)$ if $x_s = x_r$ and is $\lambda (x_s, x_r) +  \eta$ otherwise. For performance analysis convenience and without loss of generality, the transmitted symbol pair from the source and the relay is denoted as $(x_1,x_r) \in \mathcal{X}^2$. We consider the case when $x_1$ is wrongly detected to any other symbol $x_v \in \mathcal{X}$ at the destination, and denote the competing symbol pair as $(x_v, x_u) \in \mathcal{X}^2 $, $x_v \neq x_1$. 

For $(t_1, t_2) \in \mathcal{X}^2$ and $ I \in \{s, r \}$, we define $		\omega_{I,d} (t_1, t_2)  \triangleq  \frac{1}{ 2 \sqrt{\rho_{\rm{T}}/\rho_{\rm{R}}} N_{I,d}}
\Big(|y_{I,d}[k_1]- \sqrt{\rho_{\rm{T}}/\rho_{\rm{R}}} y_{I,d}[k_2] t_1 |^2- \\ |y_{I,d}[k_1]- \sqrt{\rho_{\rm{T}}/\rho_{\rm{R}}} y_{I,d}[k_2] t_2 |^2
\Big)$, %	
%\begin{IEEEeqnarray}{rCl}
%		\omega_{I,d} (t_1, t_2) & \triangleq & \frac{1}{ 2 \sqrt{\rho_{\rm{T}}/\rho_{\rm{R}}} N_{I,d}}
%		\left(|y_{I,d}[k_1]- \sqrt{\rho_{\rm{T}}/\rho_{\rm{R}}} y_{I,d}[k_2] t_1 |^2-|y_{I,d}[k_1]- \sqrt{\rho_{\rm{T}}/\rho_{\rm{R}}} y_{I,d}[k_2] t_2 |^2
%		\right) ,  \nonumber 
%		& = & \frac{1}{ N_{I,d}} \Re \{ y_{I,d}^*[k_1] y_{I,d}[k_2] (t_2-t_1)\}, \quad I \in 
%		\{s, r \}, \nonumber
		%	\sim & \mathcal{N} (u_{I,d}(t_1, t_2), W_{I,d} (t_1, t_2)),
%	\end{IEEEeqnarray} 
and $\omega_{I,d} (t_1, t_2)$ denotes the difference between their associated $I-d$ link metrics. Then according to \eqref{eq:NMLD-metric-pre}, we have
\begin{align} \label{eq:diff-lam}
	\lambda (x_1, x_r) - \lambda (x_v, x_u) = 2 \sqrt{\rho_{\rm{T}}/\rho_{\rm{R}}} \big( \omega_{s,d} (x_1, x_v) + \omega_{r,d} (x_r, x_u) \big).
\end{align} 
%$\lambda (x_1, x_r) - \lambda (x_v, x_u) = 2 \sqrt{\rho_{\rm{T}}/\rho_{\rm{R}}} \big( \omega_{s,d} (x_1, x_v) + \omega_{r,d} (x_r, x_u) \big) $, 
%with
%$	\omega_{I,d} (t_1, t_2) \triangleq \frac{1}{ 2 \sqrt{\rho_{\rm{T}}/\rho_{\rm{R}}} N_{I,d}}
%\left(|y_{I,d}[k_1]- \sqrt{\rho_{\rm{T}}/\rho_{\rm{R}}} y_{I,d}[k_2] t_1 |^2-|y_{I,d}[k_1]- \sqrt{\rho_{\rm{T}}/\rho_{\rm{R}}} y_{I,d}[k_2] t_2 |^2
%\right)
%=  \frac{1}{ N_{I,d}} \\ \Re \{ y_{I,d}^*[k_1] y_{I,d}[k_2] (t_2-t_1)\},  I \in 
%\{s, r \}$,
%
	%The justification is
	%\begin{IEEEeqnarray}{rCl}
	%y_{I,d}^*[k_1] y_{I,d}[k_2] 
	%%= &  (h_{I,d}^*u_{I}^*[k_1]+n_{I,d}^*[k_1]) y_{I,d}[k_2] \\
	%%= & h_{I,d}^*u_{I}^*[k_1]y_{I,d}[k_2] + n_{I,d}^*[k_1] y_{I,d}[k_2] \\ 
	%%= & h_{I,d}^*\sqrt{\rho_{\rm{T}}/\rho_{\rm{R}}} u_{I}^*[k_2]x_{I}^* (h_{I,d}u_{I}[k_2]+n_{I,d}[k_2]) + n_{I,d}^*[k_1] y_{I,d}[k_2] \\
	%%= & \sqrt{\rho_{\rm{T}}/\rho_{\rm{R}}} |h_{I,d}|^2 |u_{I}[k_2]|^2 x_{I}^* + h_{I,d}^*\sqrt{\rho_{\rm{T}}/\rho_{\rm{R}}} u_{I}^*[k_2]x_{I}^*n_{I,d}[k_2] + y_{I,d}[k_2] n_{I,d}^*[k_1]  \\
	%= & \sqrt{\rho_{\rm{T}}\rho_{\rm{R}}} |h_{I,d}|^2	x_{I}^* + h_{I,d}^*\sqrt{\rho_{\rm{T}}/\rho_{\rm{R}}} u_{I}^*[k_2]x_{I}^*n_{I,d}[k_2] + y_{I,d}[k_2] n_{I,d}^*[k_1], 
	%\end{IEEEeqnarray}
	%then by substitution, we have
	By substituting the expressions of $y_{I,d}[k_1]$ and $ y_{I,d}[k_2]$ (based on the system equations given in Section \ref{sec:GDM}) it yields 
	\begin{IEEEeqnarray}{rCl}
		\omega_{I,d} (t_1, t_2) & = 	& \frac{1}{N_{I,d}}\Re \{ \sqrt{\rho_{\rm{T}}\rho_{\rm{R}}} |h_{I,d}|^2 x_{I}^*(t_2-t_1) + h_{I,d}^*\sqrt{\rho_{\rm{T}}/\rho_{\rm{R}}} u_{I}^*[k_2]x_{I}^*(t_2-t_1)n_{I,d}[k_2] + \nonumber \\
		& &  y_{I,d}[k_2](t_2-t_1) n_{I,d}^*[k_1] \}, \label{eq:wId}
	\end{IEEEeqnarray}
	where $h_{I,d}[k_1] = h_{I,d}[k_2] = h_{I,d}$. 
	Similar to  \eqref{eq:approx-high-noise}, by ignoring the higher order noise terms in \eqref{eq:wId}, it can be shown after some calculations that $\omega_{I,d} (t_1, t_2) \sim \mathcal{N} (u_{I,d}(t_1, t_2), W_{I,d} (t_1, t_2))$ with 
	\begin{IEEEeqnarray}{rCl}
		u_{I,d}(t_1, t_2) & = & \sqrt{\rho_{\rm{T}}\rho_{\rm{R}}}  \Re \{
		x_{I}^*(t_2-t_1)
		\}  \gamma_{I,d}, \label{eq:mean}\\
		W_{I,d}(t_1, t_2) & =  & \frac{1}{2}\rho_{\rm{R}}(1 + \rho_{\rm{T}}/\rho_{\rm{R}} )|t_2-t_1|^2 \gamma_{I,d}. \label{eq:var}
	\end{IEEEeqnarray}
%	\begin{align}
%	u_{I,d}(t_1, t_2)  = & \sqrt{\rho_{\rm{T}}\rho_{\rm{R}}}  \Re \{
%	x_{I}^*(t_2-t_1)
%	\}  \gamma_{I,d}, \label{eq:mean}\\
%	W_{I,d}(t_1, t_2)  =  & \frac{1}{2}\rho_{\rm{R}}(1 + \rho_{\rm{T}}/\rho_{\rm{R}} )|t_2-t_1|^2 \gamma_{I,d}. \label{eq:var}
%\end{align}
	%--------------------------------
	\subsection{SER Performance for Error-Free Relaying}
	% \textcolor{red}{Till here}
	The scenario of error-free relaying happens with probability $1-\epsilon$, in which we have $(x_1,x_r)=(x_1,x_1)$. The conditional error probability for detecting $(x_1,x_1)$ as $(x_v, x_u)$ can be written as 
		
	\begin{align}
		 & \Pr[ \lambda ( x_1, x_1 ) > \min_{x_u \neq x_v} \left\{ \lambda ( x_v, x_v ),  \lambda( x_v, x_u ) + \eta \right\} ]  \nonumber \\
		 = &  \Pr [\lambda( x_v, x_v ) <  \lambda( x_v, x_u ) + \eta,\lambda( x_1, x_1 ) > \lambda( x_v, x_v )]   + \nonumber \\
		 &  \Pr [\lambda( x_v, x_v ) >  \lambda( x_v, x_u ) + \eta, \lambda( x_1, x_1 ) > \lambda( x_v, x_u ) + \eta ],  \quad x_u \neq x_v \nonumber  \\
%	\end{align}
%		\begin{IEEEeqnarray}{rCl}
		 \overset{(a)}{=} & 
		\Pr [\omega_{r,d}(x_v, x_u) < \frac{\eta}{2 \sqrt{\rho_{\rm{T}}/\rho_{\rm{R}}}}, \omega_{s,d} (x_1, x_v)+ \omega_{r,d} (x_1, x_v) >0]	+  \nonumber \\
		 & \Pr \Bigg[\omega_{r,d}(x_v, x_u) > \frac{\eta}{2 \sqrt{\rho_{\rm{T}}/\rho_{\rm{R}}}}, \omega_{s,d} (x_1, x_v) + \omega_{r,d} (x_1, x_u) > \frac{\eta}{2 \sqrt{\rho_{\rm{T}}/\rho_{\rm{R}}}} \Bigg] , \quad x_u \neq x_v, \nonumber
	\end{align}
	and upper bounded by
	\begin{IEEEeqnarray}{rCl}
		&  & \Pr [\omega_{s,d} (x_1, x_v) + \omega_{r,d} (x_1, x_v)>0 ]	+ \Pr \Bigg[ \omega_{s,d} (x_1, x_v) + \omega_{r,d} (x_1, x_u) > \frac{\eta}{2 \sqrt{\rho_{\rm{T}}/\rho_{\rm{R}}}}\Bigg], \quad x_u \neq x_v , \IEEEeqnarraynumspace \label{eq:relay-correct-case1}
	\end{IEEEeqnarray}	
	where $(a)$ is obtained according to \eqref{eq:diff-lam}.
	%  instead of considering all feasible pairs of $(x_v, x_u)$,
	We assume that $x_1$ is wrongly detected to its two nearest neighbors, i.e.,  $x_v \in \{x_2, x_M\}$, at the destination, which is well justified for the scenario of error-free relaying. Then the problem of obtaining the dominant PEP terms
	in \eqref{eq:relay-correct-case1} is formulated as	 
	\begin{IEEEeqnarray}{rCl} 
		\raisebox{-.5\normalbaselineskip}[0pt][0pt]{%
			({\bf P1})
		} \quad	 \underset{x_v, x_u}{\maximize} & \quad & 
		\left\{  \Pr [\omega_{s,d} (x_1, x_v) + \omega_{r,d} (x_1, x_v)>0 ] , 
		\Pr \Bigg[ \omega_{s,d} (x_1, x_v) + \omega_{r,d} (x_1, x_u) > \frac{\eta}{2 \sqrt{\rho_{\rm{T}}/\rho_{\rm{R}}}}\Bigg]
		\right\} \nonumber \\
		\subto & \quad & x_v \in \{x_2, x_M\}, x_u \in \mathcal{X}, x_u \neq x_v. \nonumber
	\end{IEEEeqnarray}
	It can be observed that the exact solution to problem $({\bf P1})$ varies with channel coefficients $h_{s,d}$ and $h_{r,d}$, and our approach is to take all possible solutions and obtain a solution set, considering all values of $h_{s,d}$ and $h_{r,d}$. There are two probability terms in the objective function of $({\bf P1})$, and we subsequently maximize them. For the first term, according to \eqref{eq:mean} and \eqref{eq:var}, we ignore the higher order noise and maximize
%	the approximate distribution given in 
	%\begin{IEEEeqnarray}{rCl}
	%	\omega_{s,d} (x_1, x_v) + \omega_{r,d} (x_1, x_v) \sim & \mathcal{N} (u_{s,d}(x_1, x_v) + u_{r, d}(x_1, x_v), W_{s,d}(x_1, x_v) + W_{r, d}(x_1, x_v)) . 
	%\end{IEEEeqnarray}
	%Then we should maximize
	\begin{IEEEeqnarray}{rCl}
		\Pr [\omega_{s,d} (x_1, x_v) + \omega_{r,d} (x_1, x_v) >0 ] 
		\approx 	Q \left(	
		\sqrt{ \phi_{\rm{T}} \Re \{
			1-x_1^*x_v
			\} (\gamma_{s,d}+\gamma_{r,d})}
		\right). \label{eq:case1}
	\end{IEEEeqnarray} % \phi_{\rm{T}}
	This is equivalent to 
	$\underset{x_v}{\minimize} \
	\Re \{
	1-x_1^*x_v
	\} 
	$, of which the solutions are obtained when $
	x_v \in \{x_2, x_M\}.
	$ For the second term, similarly, we maximize
	\begin{IEEEeqnarray}{rCl}
		& \Pr \left[ \omega_{s,d} (x_1, x_v) + \omega_{r,d} (x_1, x_u) > \frac{\eta}{2 \sqrt{\rho_{\rm{T}}/\rho_{\rm{R}}}}  \right] \nonumber \\
		\approx  & 	Q \left(
		\frac{\frac{\eta}{2 \sqrt{\rho_{\rm{T}}/\rho_{\rm{R}}}}-\sqrt{\rho_{\rm{T}}\rho_{\rm{R}}}  (\Re \{
			x_1^*(x_v-x_1)
			\} \gamma_{s,d}+\Re \{
			x_1^*(x_u-x_1)
			\} \gamma_{r,d})}{\sqrt{\frac{1}{2}\rho_{\rm{R}}(1 + \rho_{\rm{T}}/\rho_{\rm{R}} )(|x_v-x_1|^2 \gamma_{s,d} + |x_u-x_1|^2 \gamma_{r,d})}}
		\right). \label{eq:case2}
	\end{IEEEeqnarray}
	%which is equivalent to 
	%\begin{IEEEeqnarray}{rCl}
	%%  \underset{x_v,x_u}{\min}	\quad & 
	%%\sqrt{\rho_{\rm{T}}/\rho_{\rm{R}}}  \rho_{\rm{R}} \left( \Re \{1-x_1^*x_v
	%%\} \gamma_{s,d}+\Re \{
	%%1-x_1^*x_u
	%%\} \gamma_{r,d} \right)^{\frac{1}{2}}  + \frac{\eta}{2 \sqrt{\rho_{\rm{T}}/\rho_{\rm{R}}} } \left(  \Re \{1-x_1^*x_v
	%%\} \gamma_{s,d}+\Re \{
	%%1-x_1^*x_u
	%%\} \gamma_{r,d} \right)^{-\frac{1}{2}} \\
	% \underset{x_v,x_u}{\min}	\quad & 
	% \sqrt{\rho_{\rm{T}}/\rho_{\rm{R}}}  \rho_{\rm{R}} \sqrt{\Re \{1-x_1^*x_v
	% 	\} \gamma_{s,d}+\Re \{
	% 	1-x_1^*x_u
	% 	\} \gamma_{r,d}}  + \frac{\eta}{2 \sqrt{\rho_{\rm{T}}/\rho_{\rm{R}}} } \frac{1}{\sqrt{\Re \{1-x_1^*x_v
	% 		\} \gamma_{s,d}+\Re \{
	% 		1-x_1^*x_u
	% 		\} \gamma_{r,d}}} .
	%%\subto \quad & (x_v,x_u) \in \mathcal{X}^2, x_v \neq x_1, x_u \neq x_v.
	%\end{IEEEeqnarray}
	It is observed that only when $x_u=x_1$, the value of \eqref{eq:case2} could be larger than that of \eqref{eq:case1} with $x_v \in \{x_2, x_M\}$. Then the solutions for maximizing the second term are obtained when
	$
	x_v \in \{x_2, x_M\}, x_u=x_1.
	$
	
	Overall, the solution set to problem $({\bf P1})$ is obtained when $(x_v, x_u) \in \{(x_2, x_2), (x_M, x_M), (x_2, x_1), \\ (x_M, x_1) \}$.
	%\begin{IEEEeqnarray}{rCl}
	%	x_v \in \{x_2, x_M\}, x_u=x_v \\
	%	x_v \in \{x_2, x_M\}, x_u=x_1
	%\end{IEEEeqnarray}
	By summing over all elements in the solution set based on \eqref{eq:case1} and \eqref{eq:case2}, an approximate SER expression can be obtained as
%	after some mathematical calculations, 
%	$\left\{  \Pr [\omega_{s,d} (x_1, x_v) + \omega_{r,d} (x_1, x_v)>0 ] , 
%	\Pr [ \omega_{s,d} (x_1, x_v) + \omega_{r,d} (x_1, x_u) > \frac{\eta}{2 \sqrt{\rho_{\rm{T}}/\rho_{\rm{R}}}}]
%	\right\}$
	\begin{IEEEeqnarray}{rCl} 	
	& & (1-\epsilon) \sum_{(x_v, x_u ) \in \{ (x_2, x_2), (x_M, x_M) \}}	\Pr [\omega_{s,d} (x_1, x_v) + \omega_{r,d} (x_1, x_v)>0 ]  + \nonumber \\
	& & (1-\epsilon) \sum_{(x_v, x_u ) \in \{ (x_2, x_1), (x_M, x_1) \}}	 \Pr \Bigg[ \omega_{s,d} (x_1, x_v) + \omega_{r,d} (x_1, x_u) > \frac{\eta}{2 \sqrt{\rho_{\rm{T}}/\rho_{\rm{R}}}} \Bigg] \nonumber \\
	&	\approx	& 2(1-\epsilon) Q 
		\bigg( \sin \left( \frac{\pi}{M}\right)  \sqrt{ 2 \phi_{\rm{T}} (\gamma_{s,d}+\gamma_{r,d})} 	\bigg) + \nonumber \\
	&	&   2(1-\epsilon)
		Q 	\left( \sin \left( \frac{\pi}{M}\right)
		\sqrt{  2 \phi_{\rm{T}}   \gamma_{s,d}} + \frac{\eta }{2 (1+\rho_{\rm{T}}/\rho_{\rm{R}}) \sin \left( \frac{\pi}{M}\right)  \sqrt{  2 \phi_{\rm{T}}   \gamma_{s,d}} } 
		\right) \label{eq:PC}, \quad \rm{T} \in \{\rm{R},\rm{N}\}. 
	\end{IEEEeqnarray} 
%	Note that for $M=2$ (BPSK), the terms $2(1-\epsilon)$  in \eqref{eq:PC} should be replaced by $(1-\epsilon)$.
	% \phi_{\rm{T}}
	%\textcolor{green}{where} $\Re \{1-x_1^*x_v\}=2 \sin^2 \left( \frac{\pi}{M}\right) \triangleq c_1_1 $ when $x_v \in \{x_2, x_M\}$.
	%-------------------------------------
	\subsection{SER Performance for Erroneous Relaying}
	% \textcolor{red}{Till here}
	The scenario of erroneous  relaying happens with probability $\epsilon$, in which we have $x_r \neq x_1$. The conditional error probability for detecting $(x_1,x_r)$ as $(x_v, x_u) $ can be written as 
	\begin{IEEEeqnarray}{rCl}
		& & \Pr \big[
		\lambda ( x_1, x_r ) + \eta > \min_{x_u \neq x_v} \{  \lambda ( x_v, x_v ) ,  \lambda ( x_v, x_u ) + \eta\}
		\big] \nonumber \\
		& = & \Pr [ \lambda ( x_v, x_v ) >  \lambda ( x_v, x_u ) + \eta,  \lambda (x_1, x_r) + \eta >  \lambda ( x_v, x_u ) + \eta] + \nonumber \\
		& & \Pr [  \lambda ( x_v, x_v ) <  \lambda ( x_v, x_u ) + \eta,  \lambda (x_1, x_r) + \eta >  \lambda ( x_v, x_v )], \quad x_u \neq x_v \nonumber \\ 
		& \overset{(b)}{=} &  \Pr\Bigg[ \omega_{r,d}(x_v, x_u) > \frac{\eta}{2 \sqrt{\rho_{\rm{T}}/\rho_{\rm{R}}}}, \omega_{s,d}( x_1, x_v ) + \omega_{r,d}(x_r, x_u) > 0 \Bigg] + \nonumber \\
		& & 
		\Pr\Bigg[\omega_{r,d}(x_v, x_u) < \frac{\eta}{2 \sqrt{\rho_{\rm{T}}/\rho_{\rm{R}}}}, \omega_{s,d}( x_1, x_v ) + \omega_{r,d}(x_r, x_v) > -\frac{\eta}{2 \sqrt{\rho_{\rm{T}}/\rho_{\rm{R}}}}\Bigg], \quad x_u \neq x_v, \nonumber
	\end{IEEEeqnarray}
	 and upper bounded by
	\begin{IEEEeqnarray}{rCl}
		\Pr [\omega_{s,d}( x_1, x_v ) + \omega_{r,d}(x_r, x_u) > 0] +  \Pr \Bigg[\omega_{s,d}( x_1, x_v ) + \omega_{r,d}(x_r, x_v) > -\frac{\eta}{2 \sqrt{\rho_{\rm{T}}/\rho_{\rm{R}}}}\Bigg], \quad x_u \neq x_v, \IEEEeqnarraynumspace \label{eq:relay-wrong-case2} 
	\end{IEEEeqnarray}
	where $(b)$ is obtained according to \eqref{eq:diff-lam}. The dominant terms in \eqref{eq:relay-wrong-case2} can be found by solving the two problems $({\bf P2.1})$ and $({\bf P2.2})$ as	
	\begin{IEEEeqnarray}{rCl}
		\raisebox{-.5\normalbaselineskip}[0pt][0pt]{%
			({\bf P2.1})
		} \quad \underset{x_v,x_u}{\maximize} 
		& \quad	& \Pr [\omega_{s,d} (x_1, x_v) + \omega_{r,d}(x_r, x_u) > 0 ]    \nonumber \\
		\subto & \quad & (x_v,x_u) \in \mathcal{X}^2, x_v \neq x_1, x_u \neq x_v.
		\nonumber 
	\end{IEEEeqnarray}	
	\begin{IEEEeqnarray}{rCl}
		\raisebox{-.5\normalbaselineskip}[0pt][0pt]{%
			({\bf P2.2})
		} 
		\quad \underset{x_v}{\maximize}  & \quad	& \Pr \Bigg[\omega_{s,d} (x_1, x_v) + \omega_{r,d}(x_r, x_v) > -\frac{\eta}{2 \sqrt{\rho_{\rm{T}}/\rho_{\rm{R}}}} \Bigg]   \nonumber \\
		\subto & \quad & x_v \in \mathcal{X}, x_v \neq x_1. \nonumber
	\end{IEEEeqnarray}
	Different from problem $({\bf P1})$ in the scenario of error-free relaying, we formulate the two separate problems $({\bf P2.1})$ and $({\bf P2.2})$ to include more dominant terms. This is in consideration of the accuracy of the SER expression at low SNRs. For problem $({\bf P2.1})$, according to \eqref{eq:mean} and \eqref{eq:var}, we have
	\begin{IEEEeqnarray}{rCl} \label{eq:relay-wrong-1-approx}
		\Pr [\omega_{s,d} (x_1, x_v) + \omega_{r,d}(x_r, x_u) > 0] 
		\approx  Q \left(
		\frac{-\sqrt{\rho_{\rm{T}}\rho_{\rm{R}}} (\gamma_{s,d} \Re \{x_1^*(x_v-x_1)\} + \gamma_{r,d} \Re \{ x_r^*(x_u-x_r)\} )}{\sqrt{\frac{1}{2}\rho_{\rm{R}}(1 + \rho_{\rm{T}}/\rho_{\rm{R}} )(\gamma_{s,d}|x_v-x_1|^2 +\gamma_{r,d}|x_u-x_r|^2)}}
		\right). \IEEEeqnarraynumspace
	\end{IEEEeqnarray}
	Based on \eqref{eq:relay-wrong-1-approx}, we have that problem $({\bf P2.1})$ can be written as $\underset{x_u, x_v}{\minimize} \ \gamma_{s,d} \Re \{1-x_1^*x_v\} + \gamma_{r,d} \Re \{ 1-x_r^*x_u\} $, of which the solutions are obtained when
	$
	x_v \in \{x_2, x_M\}, x_u=x_r.
	$ Similarly for problem $({\bf P2.2})$, the objective function becomes
	\begin{IEEEeqnarray}{rCl} \label{eq:relay-wrong-2-approx}
%		& & \Pr \Bigg[\omega_{s,d} (x_1, x_v) + \omega_{r,d}(x_r, x_v) > -\frac{\eta}{2 \sqrt{\rho_{\rm{T}}/\rho_{\rm{R}}}} \Bigg] \nonumber \\  
%		& \approx  &	
		Q \left(
		\frac{-\frac{\eta}{2 \sqrt{\rho_{\rm{T}}/\rho_{\rm{R}}}}-\sqrt{\rho_{\rm{T}}\rho_{\rm{R}}} (\gamma_{s,d} \Re \{x_1^*(x_v-x_1)\} + \gamma_{r,d} \Re \{ x_r^*(x_v-x_r)\} )}{\sqrt{\frac{1}{2}\rho_{\rm{R}}(1 + \rho_{\rm{T}}/\rho_{\rm{R}} )(\gamma_{s,d}|x_v-x_1|^2 +\gamma_{r,d}|x_v-x_r|^2)}}
		\right).
	\end{IEEEeqnarray}
	%then we should
	%\begin{IEEEeqnarray}{rCl}
	%  \underset{x_v,x_u}{\min}	\quad & 
	%  \sqrt{\rho_{\rm{T}}\rho_{\rm{R}}} \sqrt{\Re \{1-x_1^*x_v
	%  	\} \gamma_{s,d}+\Re \{
	%  	1-x_r^*x_v
	%  	\} \gamma_{r,d} } -  \frac{\eta}{2 \sqrt{\rho_{\rm{T}}/\rho_{\rm{R}}} } \frac{1}{\sqrt{\Re \{1-x_1^*x_v
	%  		\} \gamma_{s,d}+\Re \{
	%  		1-x_r^*x_v
	%  		\} \gamma_{r,d} }} \\
	%  \subto \quad & (x_v,x_u) \in \mathcal{X}^2, x_v \neq x_1, x_u \neq x_v .
	%\end{IEEEeqnarray}
	It is straightforward to prove that the right side of \eqref{eq:relay-wrong-2-approx} is monotonically decreasing in the term $\gamma_{s,d} \Re \{1-x_1^*x_v\} + \gamma_{r,d} \Re \{ 1-x_r^*x_v\}$, provided that the values of $\Re \{1-x_1^*x_v\}$ and $\Re \{1-x_r^*x_v\}$ are both non-negative, which holds true for $M$-DPSK. Therefore problem $({\bf P2.2})$ can be written as $\underset{x_v}{\minimize} \     \gamma_{s,d} \Re \{1-x_1^*x_v\} + \gamma_{r,d} \Re \{ 1-x_r^*x_v\}  $. To tighten the SER expression, $x_r$ is included as an additional optimization variable, and the objective becomes $\underset{x_v, x_r}{\minimize} \     \gamma_{s,d} \Re \{1-x_1^*x_v\} + \gamma_{r,d} \Re \{ 1-x_r^*x_v\} $ (with added constraints $x_r \in \mathcal{X}$, $x_r \neq x_1$), and then the solutions are obtained when $x_v=x_r \in \{x_2, x_M\}$.

	Overall, the solution sets to problems $({\bf P2.1})$ and $({\bf P2.2})$ are obtained when $(x_v, x_u) \in \{(x_2, x_r), \\ (x_M, x_r) \}$ and $ x_v = x_r \in \{ x_2, x_M \}$, respectively. By summing over all elements in the solution sets based on \eqref{eq:relay-wrong-1-approx} and \eqref{eq:relay-wrong-2-approx}, an SER expression is obtained as
		\begin{IEEEeqnarray}{rCl} 	
	& &\frac{\epsilon}{M-1}   \sum_{(x_v, x_u ) \in \{ (x_2, x_r), (x_M, x_r) \}}	\Pr [\omega_{s,d} (x_1, x_v) + \omega_{r,d}(x_r, x_u) > 0]   + \nonumber \\
	& &\frac{\epsilon}{M-1}   \sum_{x_v = x_r \in \{ x_2, x_M \}}		\Pr \Bigg[\omega_{s,d} (x_1, x_v) + \omega_{r,d}(x_r, x_v) > -\frac{\eta}{2 \sqrt{\rho_{\rm{T}}/\rho_{\rm{R}}}} \Bigg]  \nonumber \\
	& \approx	& 2 \epsilon 
	Q \left( \sin \left( \frac{\pi}{M}\right)
	\sqrt{ 2 \phi_{\rm{T}}    \gamma_{s,d} } 
	\right) + \nonumber \\
	& & \frac{2 \epsilon}{M-1}  Q \left(  \sin \left( \frac{\pi}{M}\right)
	\sqrt{  2 \phi_{\rm{T}}   \gamma_{s,d}} - \frac{\eta }{2 (1+\rho_{\rm{T}}/\rho_{\rm{R}}) \sin \left( \frac{\pi}{M}\right)  \sqrt{ 2 \phi_{\rm{T}}  \gamma_{s,d}} } 
	\right) \label{eq:PE}, \quad \rm{T} \in \{\rm{R},\rm{N}\} . 
	\end{IEEEeqnarray} 
%	\begin{IEEEeqnarray}{rCl} \label{eq:PE}
%		\mathcal{P}^{\rm{E}}(\sqrt{\rho_{\rm{T}}/\rho_{\rm{R}}},\gamma_{s,d}, \gamma_{r,d} ) \triangleq & 2 \epsilon
%		Q \left( \sin \left( \frac{\pi}{M}\right)
%		\sqrt{ 2 \phi_{\rm{T}}    \gamma_{s,d} } 
%		\right) + \nonumber \\
%		& \frac{2 \epsilon}{M-1}  Q \left(  \sin \left( \frac{\pi}{M}\right)
%		\sqrt{  2 \phi_{\rm{T}}   \gamma_{s,d}} - \frac{\eta }{2 (1+\rho_{\rm{T}}/\rho_{\rm{R}}) \sin \left( \frac{\pi}{M}\right)  \sqrt{ 2 \phi_{\rm{T}}  \gamma_{s,d}} } 
%		\right). 
%	\end{IEEEeqnarray}

%As explained earlier, in consideration of the tightness of the SER bound at low SNRs, the first $Q$-function term in \eqref{eq:PE} is preserved, although it is larger than the second term. 
Finally, \eqref{eq:PE} together with \eqref{eq:PC} constitute the overall SER expression. 	Note that for $M=2$ (the BPSK case), the terms $2(1-\epsilon)$  in \eqref{eq:PC} and $2\epsilon$ in \eqref{eq:PE} should be replaced by $(1-\epsilon)$ and $\epsilon$, respectively. 	{\color{black}	For Rayleigh fading channels, the average SER can be calculated by averaging over the channel coefficients $h_{s,d}$ and $h_{r,d}$. The expression will be derived in Section \ref{sec:multi}.}

	%%-----------------------------------------
	%%----------------------------------------- 
	\subsection{Optimized Power Allocation Scheme}
%	\hl{rewrite, carefully}
	\label{sec:power-scheme}
	As shown in \eqref{eq:approx-high-noise}, the average receive SNR $\phi_{\rm{T}}\bar{\gamma}_{I,J}$, $(I,J) \in \{ (s,d), (s,r), (r,d) \}$, is proportional to  $\phi_{\rm{T}}$. Therefore, in the scenario of error-free relaying, maximizing $\phi_{\rm{T}}$ is equivalent to minimizing the SER. Next, we optimize the SER performance for the scenario of erroneous relaying, by minimizing  the SER expression given in \eqref{eq:PE}. To show the advantages of GDM, the NSs usually take more portion than the RSs \cite{ref5-fang2009generalized,ref5-1-fang2014performance}, i.e., $L \gg 1$.  Therefore we focus on the NSs with $\rm{T}=\rm{N}$, and then we have $	\phi_{\rm{N}}   = \frac{1}{1/\rho_{\rm{N}} + 1/\rho_{\rm{R}}}$. 
	
	There are two summation terms in \eqref{eq:PE}, and we analyze their monotonicity properties with respect to $\phi_{\rm{N}} $ subsequently. It is straightforward to show that $\epsilon$ is monotonically decreasing with $\phi_{\rm{N}} $ by taking derivative, and we have that the first term in \eqref{eq:PE} is monotonically decreasing in $\phi_{\rm{N}} $. 
%	\begin{IEEEeqnarray}{rCl}
%		& \tilde{\mathcal{P}}^{\rm{E}}(\phi_{\rm{N}}) = \sin \left( \frac{\pi}{M}\right)
%		\sqrt{  2 \phi_{\rm{N}}   \gamma_{s,d}} - \frac{ \log \frac{(1-\epsilon)(M-1)}{\epsilon} }{2  \sin \left( \frac{\pi}{M}\right)  \sqrt{ 2 \phi_{\rm{N}}  \gamma_{s,d}} }. \label{eq:PE-app}
%	\end{IEEEeqnarray}
%$\tilde{\mathcal{P}}^{\rm{E}}(\phi_{\rm{N}})$ is a function of $\phi_{\rm{N}}$. 
We prove in Appendix \ref{app-mono-PE} that the second term is also monotonically decreasing in $\phi_{\rm{N}} $, provided that the condition $ (\rho_{\rm{R}}, \rho_{\rm{N}}) \in \mathbb{S} $ is satisfied, with
		\begin{IEEEeqnarray}{rCl}
		\mathbb{S} & = & \left\{ (\rho_{\rm{R}}, \rho_{\rm{N}}) \bigg|  \phi_{\rm{N}}   = \frac{1}{1/\rho_{\rm{N}} + 1/\rho_{\rm{R}}}, \frac{ \log \frac{(1-\epsilon)(M-1)}{\epsilon}  }{4  \sin^2 ( \frac{\pi}{M}) \phi_{\rm{N}} \gamma_{s,d} } \to  \infty  \right\}. \label{eq:feasi-S}
		\end{IEEEeqnarray} In this context, both summation terms in \eqref{eq:PE} are monotonically decreasing in $\phi_{\rm{N}} $, and minimizing \eqref{eq:PE} can be achieved by maximizing $\phi_{\rm{N}}$.
		
	 Finally, the objective becomes maximizing $\phi_{\rm{N}} =  \frac{1}{1/\rho_{\rm{N}} + 1/\rho_{\rm{R}}} $, and the problem is formulated as 
		% , and minimizing \eqref{eq:PE} can be achieved by maximizing $\phi_{\rm{N}} $
	\begin{IEEEeqnarray}{rCl} % \label{eq:relay-correct}
		\IEEEyesnumber
		\IEEEyessubnumber*
		\raisebox{-2.0\normalbaselineskip}[0pt][0pt]{%
			({\bf P3})
		} \quad	 \underset{\rho_{\rm{R}}, \rho_{\rm{N}}}{\maximize} \quad & 
		 \frac{1}{1/\rho_{\rm{N}} + 1/\rho_{\rm{R}}} \nonumber \\
		\subto \quad & \rho_{\rm{R}}+(L-1)\rho_{\rm{N}}=LP_s, \label{eq:P3_subeq_1}  \\
		& \rho_{\rm{N}} > 0, \rho_{\rm{R}} > 0, \label{eq:P3_subeq_2} \\
		&  (\rho_{\rm{R}}, \rho_{\rm{N}}) \in \mathbb{S}  , \label{eq:P3_subeq_3} 
%		&  4 \sin^2 \left( \frac{\pi}{M}\right) \frac{1}{1/\rho_{\rm{N}} + 1/\rho_{\rm{R}}}  \gamma_{s,d} \ll  \log \frac{(1-\epsilon)(M-1)}{\epsilon} . \\
%		& \sin \left( \frac{\pi}{M}\right)
%		\sqrt{  \frac{2 \gamma_{s,d} }{1/\rho_{\rm{N}} + 1/\rho_{\rm{R}}}  } \ll  \frac{ \log \frac{(1-\epsilon)(M-1)}{\epsilon} }{2  \sin \left( \frac{\pi}{M}\right)
%		\sqrt{  \frac{2 \gamma_{s,d}}{1/\rho_{\rm{N}} + 1/\rho_{\rm{R}}}  } } & \text{\eqref{eq:P3_subeq_3}} \nonumber
	\end{IEEEeqnarray}   
	where \eqref{eq:P3_subeq_1} is due to the average power constraint at the source (given in \eqref{eq:Ps-const}). The constraint of \eqref{eq:P3_subeq_3} is too complicated to be addressed directly, considering that both $\phi_{N}$ and $\epsilon$ are functions of $\rho_{\rm{R}}$ and $\rho_{\rm{N}}$ (see \eqref{eq:feasi-S} and \eqref{eq:ser-relay}). As an alternative, we first ignore \eqref{eq:P3_subeq_3} to obtain a solution, and then verify \eqref{eq:P3_subeq_3} by substituting the solution. Without constraint \eqref{eq:P3_subeq_3}, problem $({\bf P3})$ degrades to that in \cite[eq. (25)]{ref5-fang2009generalized} for AF relaying. It can be solved in several ways, such as using the Lagrange multiplier method, and the solution is obtained as
	\begin{IEEEeqnarray}{rCl} \label{eq:opt-rho}
	(	\rho_{\rm{R}}^{\star}, \rho_{\rm{N}}^{\star}) = \left( \frac{LP_s}{1+\sqrt{L-1}},
	\frac{LP_s}{L-1+\sqrt{L-1}}
	\right).
	\end{IEEEeqnarray}
%	\begin{IEEEeqnarray}{rCl}
%		\rho_{\rm{R}}^{\star} = \frac{LP_s}{1+\sqrt{L-1}}, \quad \rho_{\rm{N}}^{\star} = \frac{LP_s}{L-1+\sqrt{L-1}}.
%	\end{IEEEeqnarray}
%	$
%	\rho_{\rm{R}}^{\star} = \frac{LP_s}{1+\sqrt{L-1}}$ and $\rho_{\rm{N}}^{\star} = \frac{LP_s}{L-1+\sqrt{L-1}}
%	$. 
%$(\rho_{\rm{R}}^{\star}, \rho_{\rm{N}}^{\star}) \in \mathbb{S}^{\star} = \{ (\rho_{\rm{R}}, \rho_{\rm{N}}) | \epsilon =  \epsilon(\rho_{\rm{R}}, \rho_{\rm{N}}), \sin \left( \frac{\pi}{M}\right)
%\sqrt{  \frac{2 L P_s \gamma_{s,d} }{L+2\sqrt{L-1}}  } \ll  \frac{ \log \frac{(1-\epsilon)(M-1)}{\epsilon} }{2  \sin \left( \frac{\pi}{M}\right)
%	\sqrt{  \frac{2 L P_s \gamma_{s,d} }{L+2\sqrt{L-1}} } }, \frac{ \log \frac{(1-\epsilon)(M-1)}{\epsilon} }{4  \sin^2 \left( \frac{\pi}{M}\right)
%	\frac{ L P_s \gamma_{s,d} }{L+2\sqrt{L-1}} } \to \infty \}$ 
By substituting \eqref{eq:opt-rho} into \eqref{eq:P3_subeq_3} it yields the condition that $\frac{ \log \frac{(1-\epsilon^{\star})(M-1)}{\epsilon^{\star}} }{2 z_{L,M}^2 \gamma_{s,d} } \to \infty $ with $z_{L,M} = \sin \left( \frac{\pi}{M}\right) \sqrt{  \frac{2 L P_s  }{L+2\sqrt{L-1}}  }$, and $\epsilon^{\star}$ is calculated from \eqref{eq:ser-relay} using  $(\rho_{\rm{R}}, \rho_{\rm{N}}) = (\rho_{\rm{R}}^{\star}, \rho_{\rm{N}}^{\star})$. To gain a better understanding of this condition, we give an interpretation in terms of the SNRs of the $S-D$ and $S-R$ links. Since $1/2 \leq \frac{ L   }{L+2\sqrt{L-1}} < 1  $ when $L \geq 2$, $z_{L,M}$ is guaranteed to be a bounded positive value for all $M>1$. In addition, since $\epsilon$ denotes the average SER of the $S-R$ link, we have $ \frac{1}{\epsilon^{\star}} \propto \bar{\gamma}_{s,r} $. Based on the above, we have $ \frac{ \log \frac{(1-\epsilon^{\star})(M-1)}{\epsilon^{\star}} }{2 z_{L,M}^2 \gamma_{s,d} } \to \infty  \iff \frac{ \log \frac{(1-\epsilon^{\star})}{\epsilon^{\star}} }{\gamma_{s,d} } \to \infty  \iff \frac{ \log \bar{\gamma}_{s,r} }{\gamma_{s,d}} \to \infty  $. It is now clear that the condition can be satisfied when the average SNR of the $S-R$ link $\bar{\gamma}_{s,r}$ is much larger than the instantaneous SNR of the $S-D$ link $\gamma_{s,d}$. This is generally practical for the DF relay network, in which the help from a relatively strong $S-R$ link is needed usually because the direct $S-D$ link is relatively weak.

Let us focus on the solution in \eqref{eq:opt-rho} and get insights on how this optimized power allocation scheme works. We first note that only the values of $P_s$  and $L$ are needed to obtain $\rho_{\rm{R}}^{\star}$ and $\rho_{\rm{N}}^{\star}$, while no prior information of the channels are required.  $\rho_{\rm{R}}^{\star}/\rho_{\rm{N}}^{\star}  =  \sqrt{L-1}$  only depends on the block length $L$. For $L=2$, we have $\rho_{\rm{R}}^{\star}=\rho_{\rm{N}}^{\star}=P_s$, which means all symbols are allocated the same power as $P_s$. For $L>2$, we have $\rho_{\rm{R}}^{\star}/\rho_{\rm{N}}^{\star} > 1$, which reveals that to show the advantages of GDM, more power should be allocated to the RSs. It is obvious that the value of $L$ largely affects the performance of the optimized power allocation scheme, and this effect will be discussed by simulation in Section \ref{sec:simu}.
	% Approximate expressions for the optimal values of $\rho_{\rm{N}}$ and $\rho_{\rm{R}}$ are given. 
	%%--------------------------------------------------------------------------------------------------------
	\section{Multiple Relay Detection and Performance Analysis} 
		\label{sec:multi}
%	In this section, the proposed NMLD is extended to the  DF multiple relay networks. 
%	{\color{black} First, the multi-branch relay network is considered. Then, the general single-source single-destination   relay network is considered.}
	%\textcolor{red}{till here}
	
	\subsection{\color{black} Multiple Relay Detection}
	\label{sec:multi-NMLD}
%	The block model of a multiple relay network is shown in Fig. \ref{Fig:system-model-general}, where $N$ parallel relays $R_1, \dots, R_N$ form $N$ cooperative branches besides the direct $S-D$ link. The destination has the average CSI of all links. The GDM scheme can be applied, and each relay processes and transmits symbols as $R$ does in the single relay case. We define $\Theta \triangleq \{1,2,\dots,N\}$ as the relay index set, and the noise has distribution $\mathcal{CN} (0,N_{I, J})$, $(I,J) \in \{ (s,d), (s, r_{n}), (r_{n}, d) \}$, $n \in \Theta$. 
%	\begin{figure}[t]
%		\centering
%		\includegraphics[width=3.2in]{system-model-general.pdf}
%		\caption{Block model of a relay network with $N$ parallel relays ($R_1, \dots, R_N$) and a source-destination ($S-D$) pair.}
%		\label{Fig:system-model-general}
%	\end{figure}
%		
%	Each symbol transmission takes $N+1$ time slots. In the first time slot, the source broadcasts its differentially encoded signal to all relays and the destination. In the $(n+1)$-th time slot, $n \in \Theta$, $R_n$ differentially encodes its detected $x_{r_{n}}$ for transmission. 
	{\color{black} Let us recall that the multi-relay system model has been stated in Section \ref{sec:sys-model}. The GDM scheme is applied, and each relay processes and transmits symbols as $R$ does in the single relay case. } At the destination, the optimal maximum likelihood detection metric is given as
	$ \max_{x_s \in \mathcal{X}} f(y_{s,d} [k_1]|x_s, y_{s,d} [k_2]) \prod_{n \in \Theta} \left(
	\sum_{x_{r_{n}} \in \mathcal{X}} \Pr(x_{r_{n}}|x_s) f(y_{r_{n},d} [k_1]|x_{r_{n}}, y_{r_{n},d} [k_2]) \right).$
	By approximating the transmission probability term $\Pr(x_{r_{n}}|x_s)$ using the average SER of the corresponding $S-R_n$ link, denoted as $\epsilon_n$, and applying the widely-used max-sum approximation, the detection rule of NMLD is obtained as
	\begin{IEEEeqnarray}{rCl} 
	\hat{x}_s 
	& = & \arg\min_{x_s \in \mathcal{X}} \left\{
	\frac{1}{N_{s,d}} |y_{s,d} [k_1]- \sqrt{\rho_{\rm{T}}/\rho_{\rm{R}}} y_{s,d} [k_2]x_s|^2 + \sum_{n=1}^{N} \min \left\{ \frac{1}{N_{r_{n},d}}|y_{r_{n},d} [k_1]- \sqrt{\rho_{\rm{T}}/\rho_{\rm{R}}} y_{r_{n},d} [k_2]x_s|^2,  \right. \right.  \nonumber  \\
	& & \left.\left.  \min_{\substack{x_{r_{n}} \in \mathcal{X}}} \frac{1}{N_{r_{n},d}}|y_{r_{n},d} [k_1]-\sqrt{\rho_{\rm{T}}/\rho_{\rm{R}}} y_{r_{n},d} [k_2]x_{r_{n}}|^2 + \eta_n\right\}		
	\right\} , \quad  \rm{T} \in \{\rm{R},\rm{N}\}, \label{eq:NMLD-general}
	\end{IEEEeqnarray}
	where $\eta_n \triangleq (1+\rho_{\rm{T}}/\rho_{\rm{R}}) \log \frac{(1-\epsilon_n)(M-1)}{\epsilon_n}$.  {\color{black} Our NMLD is performed based on \eqref{eq:NMLD-general}, and  the detection procedure is described in \textbf{Algorithm} \ref{alg:NMLD-multi}. It can be seen from \textbf{Algorithm} \ref{alg:NMLD-multi}  that lines  $2$-$4$  have $\mathcal{O}(MN)$  complexity, and so are lines $5$-$6$ (since the cardinality $|\mathcal{X}| = M$) and lines $7$-$10$. Line $11$ has   $\mathcal{O}(M)$ complexity. Therefore, the overall complexity is $\mathcal{O}(MN)$. The single relay case corresponds to the special case when $N=1$. Table \ref{table:metric-complexity} shows the number of operations required for the AMLD, the PLD, and the proposed NMLD for detecting one $M$-DPSK symbol. We can see that the AMLD and PLD have $\mathcal{O}(M^2N)$ and $\mathcal{O}(MN)$ complexities, respectively, while the proposed  NMLD has $\mathcal{O}(MN)$ complexity. The total number of operations per symbol detection required for the AMLD and  NMLD are $(20M^2+2M)N+19M-1$ and $(24M-1)N+20M-1$, respectively. An interesting observation is that asymptotically, the overall complexity saving by the NMLD (compared to the AMLD) is up to
		$\lim\limits_{N \to \infty} \frac{(20M^2+2M)N+19M-1 - ((24M-1)N+20M-1)}{(20M^2+2M)N+19M-1} = 100 \% - \frac{24M-1}{20M^2+2M}$, which  increases with the modulation size $M$, and approaches $100 \%$ for large enough $M$.} 
	% The run times will be compared in Section \ref{sec:simu}
	
	% clearly the overall complexity is $\mathcal{O}(MN)$. \footnote{\color{black} This is a direct implementation of \eqref{eq:NMLD-general}. The procedure can be further simplified using methods such as }
	
		\begin{algorithm}[t]
		\small
		{	\color{black}
			\caption{Proposed NMLD for the multiple relay network based on \eqref{eq:NMLD-general}}
			\label{alg:NMLD-multi}
			\KwIn {\ \ Received $y_{s,d}[k_1]$, $y_{s,d}[k_2]$,  $y_{r_{n},d}[k_1]$,  and $y_{r_{n},d}[k_2]$; Average SERs $\epsilon_n$; \\ 
			\ \ \ \ 	\ \ \ \  	\ \ \  Power allocation factors $\rho_{\rm{R}}$ and $\rho_{\rm{N}}$; Alphabet $\mathcal{X}$}
			\KwOut {Detected source symbol $\hat{x}_s$ }
%			\tcc{Step 1} 
		$\mathcal{X} = \{x_t=e^{j 2 \pi (t-1) /M}, \ t=1,2,\dots,M\}$\;
		\For{$n = 1,2,\dots,N$}
		{\For{$ t=1,2,\dots,M $}{
				Calculate and store $\mathbf{F}(n,x_t) \leftarrow \frac{1}{N_{r_{n},d}}|y_{r_{n},d} [k_1]-\sqrt{\rho_{\rm{T}}/\rho_{\rm{R}}} y_{r_{n},d} [k_2]x_{t}|^2 $\;
			}
		
		}			
			\For{$n = 1,2,\dots,N$}
			{
				$C_{n}  \leftarrow \min_{x_{t} \in \mathcal{X} } \mathbf{F}(n,x_t) + \eta_n$\;
			}
%		\tcc{Step 2} 
		\For{$ t=1,2,\dots,M $}
		{
			\For{$n = 1,2,\dots,N$}
			{ $\tilde{C}_n(x_t) \leftarrow \min \{ \mathbf{F}(n,x_t), C_{n} \} $\;				
			}
		$C(x_t)  \leftarrow \frac{1}{N_{s,d}} |y_{s,d} [k_1]- \sqrt{\rho_{\rm{T}}/\rho_{\rm{R}}} y_{s,d} [k_2]x_t|^2  + \sum_{n=1}^N \tilde{C}_n(x_t)$\;
		}
		$\hat{x}_s  \leftarrow \arg\min_{\substack{x_{t} \in \mathcal{X}}} C(x_t) $\;
		}
	\end{algorithm}	

%		\begin{algorithm}[t]
%	\small
%	{	\color{black}
%		\caption{Proposed NMLD for the multi-branch relay network}
%		\label{alg:NMLD-multi}
%		\KwIn {\ \ Received $y_{s,d}[k_1]$, $y_{s,d}[k_2]$,  $y_{r_{n},d}[k_1]$,  and $y_{r_{n},d}[k_2]$; Average SERs $\epsilon_n$; \\ 
%			\ \ \ \ 	\ \ \ \  	\ \ \  Power allocation factors $\rho_{\rm{R}}$ and $\rho_{\rm{N}}$; Alphabet $\mathcal{X}$}
%		\KwOut {Detected source symbol $\hat{x}_s$ }
%		%			\tcc{Step 1} 
%		\For{$n = 1,2,\dots,N$}
%		{   \For{$x_{r_{n}} \in \mathcal{X}$}
%			{ $\tilde{C}_n(x_{r_{n}}) \leftarrow \frac{1}{N_{r_{n},d}}|y_{r_{n},d} [k_1]-\sqrt{\rho_{\rm{T}}/\rho_{\rm{R}}} y_{r_{n},d} [k_2]x_{r_{n}}|^2 + \eta_n $\;				
%			}
%			$C_{n}  \leftarrow \min_{\substack{x_{r_{n}} \in \mathcal{X}}} \tilde{C}_n(x_{r_{n}}) $\;
%		}
%		%		\tcc{Step 2} 
%		\For{$x_{s} \in \mathcal{X}$}
%		{
%			\For{$n = 1,2,\dots,N$}
%			{ $\bar{C}_n(x_{s}) \leftarrow \min \{ \frac{1}{N_{r_{n},d}}|y_{r_{n},d} [k_1]-\sqrt{\rho_{\rm{T}}/\rho_{\rm{R}}} y_{r_{n},d} [k_2]x_{s}|^2, C_{n} \} $\;				
%			}
%			$C(x_s)  \leftarrow \frac{1}{N_{s,d}} |y_{s,d} [k_1]- \sqrt{\rho_{\rm{T}}/\rho_{\rm{R}}} y_{s,d} [k_2]x_s|^2  + \sum_{n=1}^N \bar{C}_n(x_{s})$\;
%		}
%		$\hat{x}_s  \leftarrow \arg\min_{\substack{x_{s} \in \mathcal{X}}} C(x_s) $\;
%	}
%\end{algorithm}	

\begin{table}[h] 
	{ \color{black}
	\centering
	\captionsetup{font={small}}
	\caption{ \color{black} Number of operations per symbol detection required by AMLD, PLD, and the proposed NMLD for a non-coherent parallel DF relay network with $N$ relays using $M$-DPSK GDM }
	\label{table:metric-complexity}
	\centering
	\scalebox{0.94}{
		\begin{tabular}{|c|c|c|c|c|}
			\hline
			\diaghead{\theadfont Diag ColumnmnHead}%
			{\textbf{Detector}}{\textbf{Operation}}&
			\thead{\textbf{Addition}}&\thead{
			\textbf{Multiplication}}&\thead{\textbf{Max/Min}}&\thead{ \textbf{Exponential/Logarithm}} \\ 
			\hline
			AMLD & $M(9MN-N+7)$ & $M(11MN+2N+11)$ &  $M-1$ & $NM$ \\ % $5M^2-M-4$ & $10M^2+7M-17$
			\hline
			PLD & $(M-1)(18N+17)$ & $(M-1)(23N+23)$ &  $(M-1)(2N+1)$  & $0$  \\
			\hline
			Proposed NMLD & $M(11N+8)$ & $M(11N+11)$ &  $(M-1)(N+1)+NM$ & $0$  \\
			\hline
	\end{tabular} }
}
\end{table}

%	Our NMLD is performed based on \eqref{eq:NMLD-general} as: First, for each $n =1,\cdots, N$, we solve for $\min_{x_{r_{n}} \in \mathcal{X}} \\ \frac{1}{  N_{r_{n},d}}|y_{r_{n},d} [k_1]-\sqrt{\rho_{\rm{T}}/\rho_{\rm{R}}} y_{r_{n},d} [k_2]x_{r_{n}}|^2+ \eta_n$, and this term becomes a fixed threshold associated with the $n$-th branch. Then for each $x_s \in \mathcal{X}$, we compare the value of $\frac{1}{  N_{r_{n},d}}|y_{r_{n},d} [k_1]- \sqrt{\rho_{\rm{T}}/\rho_{\rm{R}}} y_{r_{n},d} [k_2]x_s|^2$ with the threshold associated with the $n$-th branch for all $n \in \{ 1,\cdots, N \}$, and calculate the overall metric for $x_s$. Finally, the symbol $x_s$ with the smallest metric value is selected as $\hat{x}_s$. The overall complexity of the NMLD is $\mathcal{O}(MN)$. 	
	For the rest of the analysis in this section, $L=1$ is assumed to simplify our presentation even if it is trivial to
	extend to the case with $L>1$. It is denoted in \eqref{eq:NMLD-general} that $k_1=k$, $k_2=k-1$, $\sqrt{\rho_{\rm{T}}/\rho_{\rm{R}}}=1$ and $\eta_n = 2 \log \frac{(1-\epsilon_n)(M-1)}{\epsilon_n}$. 
	
	The NMLD and the PLD are both accurate approximations of the AMLD, and we analyze their relation in Appendix \ref{app:NMLD and PLD}. Specifically, by proving that the PLD metric can be obtained by approximating the NMLD metric in \eqref{eq:NMLD-general}, it is shown that the PLD involves additional approximations compared to the NMLD, for $M>2$.
	
	%%------------------------------------------------------
	\subsection{Pairwise Error Probability}
	Similarly to the single relay case, the key idea to obtain an SER expression is to determine the dominant PEP terms. 
	The following analysis is provided for calculating the PEP between any two different symbol sets $\mathbf{x}^{\rm{C}} = [x_1, x_{r_{1}}, \dots, x_{r_{N}}]$ and $\mathbf{x}^{\rm{E}} = [x_v, x_{u_{1}}, \dots, x_{u_{N}}]$.
	 
	Similarly to the single relay case, for all relay-destination links, we define $\omega_{r_{n}, d}(t_1, t_2) \triangleq  \frac{1}{ 2 N_{r_{n},d}}
	\left( |y_{r_{n}, d}[k]-y_{r_{n}, d}[k-1]t_1|^2 - |y_{r_{n}, d}[k]-y_{r_{n}, d}[k-1]t_2|^2 \right)$, $n \in \Theta$. Their distributions are obtained as $\omega_{r_{n}, d}(t_1, t_2) 
	 \sim  \mathcal{N} (u_{r_{n}, d}(t_1, t_2), W_{r_{n}, d}(t_1, t_2))$, 
%	\begin{IEEEeqnarray}{rCl}
%		\omega_{r_{n}, d}(t_1, t_2) 
%		& \sim & \mathcal{N} (u_{r_{n}, d}(t_1, t_2), W_{r_{n}, d}(t_1, t_2)), \nonumber
%	\end{IEEEeqnarray}
	where 
	\begin{IEEEeqnarray}{rCl}
		u_{r_{n}, d}(t_1, t_2) & = &  \gamma_{r_{n}, d} \Re \{ x_{r_{n}}^*(t_2-t_1)\}, \label{eq:mean-multi}\\
		W_{r_{n}, d}(t_1, t_2) & = &  |t_2-t_1|^2 \gamma_{r_{n}, d}. \label{eq:var-multi}
	\end{IEEEeqnarray}For the $n$-th branch, the branch metric for a given pair of $(x_s, x_{r_{n}}), x_{r_{n}} \neq x_s$, is defined as 
	\begin{IEEEeqnarray}{rCl} \label{eq:metric-RnDlink}
		& \min \left\{ \frac{1}{  N_{r_{n},d}}|y_{r_{n},d} [k]-y_{r_{n},d} [k-1]x_s|^2,  \frac{1}{  N_{r_{n},d}}|y_{r_{n},d} [k]-y_{r_{n},d} [k-1]x_{r_{n}}|^2 + \eta_n\right\}. 
	\end{IEEEeqnarray}Solving $\frac{1}{  N_{r_{n},d}}|y_{r_{n},d} [k]-y_{r_{n},d} [k-1]x_s|^2 < \frac{1}{  N_{r_{n},d}}|y_{r_{n},d} [k]-y_{r_{n},d} [k-1]x_{r_{n}}|^2 + \eta_n$ gives $\omega_{r_{n},d}(x_s, x_{r_{n}}) < \eta_n/2$. By defining  $\tau_n \triangleq \tau_n(x_s, x_{r_{n}})=0$ if $\omega_{r_{n},d}(x_s, x_{r_{n}}) < \eta_n/2$, and $1$ otherwise, we can write \eqref{eq:metric-RnDlink} equivalently as 
	\begin{IEEEeqnarray}{rCl} \label{eq:metric-RnDlink-v2}
		\frac{1}{  N_{r_{n},d}}|y_{r_{n},d} [k]-y_{r_{n},d} [k-1](x_s+\tau_n(x_{r_{n}}-x_s))|^2 + \tau_n\eta_n.
	\end{IEEEeqnarray} 
	Based on \eqref{eq:NMLD-general} and \eqref{eq:metric-RnDlink-v2}, after some manipulations, the detection metric for any given symbol set $\mathbf{x} = [x_s, x_{r_{1}}, \dots, x_{r_{N}}]$ can be written as
	\begin{IEEEeqnarray}{rCl} \label{eq:metric-general}
		%g([x_s, x_{r_{1}}, \dots, x_{r_{N}}]) \triangleq & \frac{1}{  N_{s,d}}|y_{s,d} [k]-y_{s,d} [k-1]x_s|^2 + \sum_{n=1}^{N} \left( \frac{1}{  N_{r_{n},d}} |y_{r_{n},d} [k]-y_{r_{n},d} [k-1](x_s+\tau_n (x_{r_{n}}-x_s))|^2 + \tau_n\eta_n \right), \\
		g(\mathbf{x}) = & \| \mathbf{V}(\mathbf{y}-\mathbf{Z}(x_s+\mathbf{T}_\mathbf{x}(\mathbf{x}-x_s)^T) )\|^2 + \| \mathbf{T}_\mathbf{x} \tilde{\bm{\eta}} \|_1,
	\end{IEEEeqnarray}
	where  $		\mathbf{V}  =  \diag \Big(\frac{1}{\sqrt{N_{s,d}}}, \frac{1}{\sqrt{N_{r_{1},d}}} , \dots, \frac{1}{\sqrt{N_{r_{N},d}}}   \Big)$, $	\mathbf{y}  =  \Big[\begin{matrix}
	y_{s,d}[k] ,
	y_{r_{1},d}[k],
	\dots ,
	y_{r_{N},d}[k]
	\end{matrix} \Big]^T$, $	\mathbf{Z}  =  \diag \big(y_{s,d}[k-1], \\  y_{r_{1},d}[k-1] , \dots,  y_{r_{N},d}[k-1]   \big)$, $	\tilde{\bm{\eta}}  =  \Big[\begin{matrix}
	0 ,
	\eta_1 ,
	\dots ,
	\eta_N
	\end{matrix} \Big]^T$, and $	\mathbf{T}_\mathbf{x}  =  \diag \left(0, \tau_1 , \dots, \tau_N    \right)$.
%	\begin{IEEEeqnarray}{rCl}
%		\mathbf{V} & = & \diag \left(\frac{1}{\sqrt{N_{s,d}}}, \frac{1}{\sqrt{N_{r_{1},d}}} , \dots, \frac{1}{\sqrt{N_{r_{N},d}}}   \right), \nonumber \\
%		\mathbf{y} & = & \left[\begin{matrix}
%			y_{s,d}[k] ,
%			y_{r_{1},d}[k],
%			\dots ,
%			y_{r_{N},d}[k]
%		\end{matrix} \right]^T,\nonumber \\
%		\mathbf{Z} & = & \diag \left(y_{s,d}[k-1], y_{r_{1},d}[k-1] , \dots, y_{r_{N},d}[k-1]   \right), \nonumber \\
%		\tilde{\bm{\eta}} & = & \left[\begin{matrix}
%			0 ,
%			\eta_1 ,
%			\dots ,
%			\eta_N
%		\end{matrix} \right]^T, \nonumber \\
%		\mathbf{T}_\mathbf{x} & = & \diag \left(0, \tau_1 , \dots, \tau_N    \right). \nonumber
%	\end{IEEEeqnarray}
	Note that in these five matrices and/or vectors, only $\mathbf{T}_\mathbf{x}$ is a function of $\mathbf{x}$.
	
	For $\tilde{\Theta} \subseteq \Theta$, we define
%	\begin{IEEEeqnarray}{rCl}
	$	\omega_{\tilde{\Theta}} (\mathbf{x}^{\rm{C}}, \mathbf{x}^{\rm{E}})  \triangleq  \sum_{n \in \tilde{\Theta}} \omega_{r_{n},d}((x_1+\mathbf{T}_{\mathbf{x}^{\rm{C}}}(\mathbf{x}^{\rm{C}}-x_1)^T)[n], (x_v+\mathbf{T}_{\mathbf{x}^{\rm{E}}}(\mathbf{x}^{\rm{E}}-x_v)^T)[n]) \sim  \mathcal{N} (u_{\tilde{\Theta}} (\mathbf{x}^{\rm{C}}, \mathbf{x}^{\rm{E}}), W_{\tilde{\Theta}} (\mathbf{x}^{\rm{C}}, \mathbf{x}^{\rm{E}}))$,
%	 \nonumber 
%	\end{IEEEeqnarray}
	with $	u_{\tilde{\Theta}} (\mathbf{x}^{\rm{C}}, \mathbf{x}^{\rm{E}})  \triangleq  \sum_{n \in \tilde{\Theta}}u_{r_{n}, d}((x_1+\mathbf{T}_{\mathbf{x}^{\rm{C}}}(\mathbf{x}^{\rm{C}}-x_1)^T)[n], (x_v+\mathbf{T}_{\mathbf{x}^{\rm{E}}}(\mathbf{x}^{\rm{E}}-x_v)^T)[n])$ and $	W_{\tilde{\Theta}} (\mathbf{x}^{\rm{C}}, \mathbf{x}^{\rm{E}})  \triangleq  \sum_{n \in \tilde{\Theta}}
	W_{r_{n}, d}((x_1+\mathbf{T}_{\mathbf{x}^{\rm{C}}}(\mathbf{x}^{\rm{C}}-x_1)^T)[n], (x_v+\mathbf{T}_{\mathbf{x}^{\rm{E}}}(\mathbf{x}^{\rm{E}}-x_v)^T)[n])$. 
%	\begin{IEEEeqnarray}{rCl}
%		u_{\tilde{\Theta}} (\mathbf{x}^{\rm{C}}, \mathbf{x}^{\rm{E}}) & \triangleq & \sum_{n \in \tilde{\Theta}}u_{r_{n}, d}((x_1+\mathbf{T}_{\mathbf{x}^{\rm{C}}}(\mathbf{x}^{\rm{C}}-x_1)^T)[n], (x_v+\mathbf{T}_{\mathbf{x}^{\rm{E}}}(\mathbf{x}^{\rm{E}}-x_v)^T)[n]), \nonumber \\
%		W_{\tilde{\Theta}} (\mathbf{x}^{\rm{C}}, \mathbf{x}^{\rm{E}}) & \triangleq & \sum_{n \in \tilde{\Theta}}
%		W_{r_{n}, d}((x_1+\mathbf{T}_{\mathbf{x}^{\rm{C}}}(\mathbf{x}^{\rm{C}}-x_1)^T)[n], (x_v+\mathbf{T}_{\mathbf{x}^{\rm{E}}}(\mathbf{x}^{\rm{E}}-x_v)^T)[n]). \nonumber
%	\end{IEEEeqnarray}
Then, based on \eqref{eq:metric-general}, we have $		g(\mathbf{x}^{\rm{C}}) - g(\mathbf{x}^{\rm{E}})  =  \| \mathbf{V}(\mathbf{y}-\mathbf{Z}(x_1+\mathbf{T}_{\mathbf{x}^{\rm{C}}}(\mathbf{x}^{\rm{C}}-x_1)^T) )\|^2 + \| \mathbf{T}_{\mathbf{x}^{\rm{C}}} \tilde{\bm{\eta}} \|_1 -  (\| \mathbf{V}(\mathbf{y}-\mathbf{Z}(x_v+\mathbf{T}_{\mathbf{x}^{\rm{E}}}(\mathbf{x}^{\rm{E}}-x_v)^T) )\|^2 + \| \mathbf{T}_{\mathbf{x}^{\rm{E}}} \tilde{\bm{\eta}} \|_1)  =  \omega_{s,d}(x_1, x_v) + \omega_{\Theta} (\mathbf{x}^{\rm{C}}, \mathbf{x}^{\rm{E}})  + \| (\mathbf{T}_{\mathbf{x}^{\rm{C}}}-\mathbf{T}_{\mathbf{x}^{\rm{E}}}) \tilde{\bm{\eta}} \|_1$.
%	\begin{IEEEeqnarray}{rCl}
%		g(\mathbf{x}^{\rm{C}}) - g(\mathbf{x}^{\rm{E}}) & = & \| \mathbf{V}(\mathbf{y}-\mathbf{Z}(x_1+\mathbf{T}_{\mathbf{x}^{\rm{C}}}(\mathbf{x}^{\rm{C}}-x_1)^T) )\|^2 + \| \mathbf{T}_{\mathbf{x}^{\rm{C}}} \tilde{\bm{\eta}} \|_1 - \nonumber \\
%		& & (\| \mathbf{V}(\mathbf{y}-\mathbf{Z}(x_v+\mathbf{T}_{\mathbf{x}^{\rm{E}}}(\mathbf{x}^{\rm{E}}-x_v)^T) )\|^2 + \| \mathbf{T}_{\mathbf{x}^{\rm{E}}} \tilde{\bm{\eta}} \|_1) \nonumber \\
%		& = & \omega_{s,d}(x_1, x_v) + \omega_{\Theta} (\mathbf{x}^{\rm{C}}, \mathbf{x}^{\rm{E}})  + \| (\mathbf{T}_{\mathbf{x}^{\rm{C}}}-\mathbf{T}_{\mathbf{x}^{\rm{E}}}) \tilde{\bm{\eta}} \|_1, \nonumber
%	\end{IEEEeqnarray}
	Finally, the PEP for detecting $\mathbf{x}^{\rm{C}}$ as $\mathbf{x}^{\rm{E}}$ can be obtained as 
	$$ \Pr [g(\mathbf{x}^{\rm{C}}) > g(\mathbf{x}^{\rm{E}}) ] = 
	Q \left(
	\frac{-u_{s,d}(x_1, x_v) - u_{\Theta} (\mathbf{x}^{\rm{C}}, \mathbf{x}^{\rm{E}})  - \| (\mathbf{T}_{\mathbf{x}^{\rm{C}}}-\mathbf{T}_{\mathbf{x}^{\rm{E}}}) \tilde{\bm{\eta}} \|_1}{\sqrt{W_{s,d}(x_1, x_v) + W_{\Theta} (\mathbf{x}^{\rm{C}}, \mathbf{x}^{\rm{E}})}}
	\right).
	$$ In the following, without loss of generality, we let the relays with index belonging to sets $\Theta^{\rm{C}}$ and $\Theta^{\rm{E}}$ detect correctly and wrongly, respectively. The two sets are disjoint and $\Theta^{\rm{E}} \cup \Theta^{\rm{C}} = \Theta$. 

	%--------------------------------------------
	\subsection{SER Performance Analysis}
	%-----------------
	\subsubsection{For Error-Free Relaying}
	For the scenario of error-free relaying, there is $\Theta^{\rm{C}}=\Theta$ and $\Theta^{\rm{E}}=\emptyset$, and we have $\mathbf{x}^{\rm{C}}= x_1 \mathbf{1}_{ N+1}$, and $\mathbf{T}_{\mathbf{x}^{\rm{C}}}=\mathbf{0}_{ N+1}$. The conditional PEP for detecting $\mathbf{x}^{\rm{C}}$ as $\mathbf{x}^{\rm{E}}$ is calculated as
	\begin{IEEEeqnarray}{rCl}
	 \Pr [g(\mathbf{x}^{C}) > g(\mathbf{x}^{\rm{E}}) | \Theta^{\rm{C}}=\Theta ,\Theta^{\rm{E}}=\emptyset ] & = & \Pr[\omega_{s,d}(x_1, x_v) + \omega_{\Theta} (\mathbf{x}^{\rm{C}}, \mathbf{x}^{\rm{E}})  + \| -\mathbf{T}_{\mathbf{x}^{\rm{E}}} \tilde{\bm{\eta}} \|_1 > 0] \nonumber \\
		& = & Q \left(
		\frac{-u_{s,d}(x_1, x_v)-u_{\Theta} (\mathbf{x}^{\rm{C}}, \mathbf{x}^{\rm{E}})+\|\mathbf{T}_{\mathbf{x}^{\rm{E}}} \tilde{\bm{\eta}} \|_1 /2 }{\sqrt{W_{s,d}(x_1, x_v) + W_{\Theta} (\mathbf{x}^{\rm{C}}, \mathbf{x}^{\rm{E}})}}
		\right) . \label{eq:PEP-general-perfect}
	\end{IEEEeqnarray}	
	Similarly to the single relay case, according to \eqref{eq:mean-multi} and \eqref{eq:var-multi}, the problem of maximizing \eqref{eq:PEP-general-perfect} can be written as	
	%\begin{IEEEeqnarray}{rCl}
	%\underset{\mathbf{x}^{\rm{E}}}{\min}	\quad & \sqrt{ \Re \{1-x_1^*x_v
	%	\} \gamma_{s,d}+\sum_{n=1}^{N} 
	%	\Re \{1- x_1^*(x_v+\tau_n (x_{u_{n}}-x_v))\}\gamma_{r_{n}, d} } 
	%  + \nonumber \\
	%&   \frac{\|\mathbf{T}_{\mathbf{x}^{\rm{E}}} \tilde{\bm{\eta}} \|_1}{2  } \frac{1}{\sqrt{\Re \{1-x_1^*x_v
	%	\} \gamma_{s,d}+\sum_{n=1}^{N} 
	%	\Re \{1- x_1^*(x_v+\tau_n (x_{u_{n}}-x_v))\}\gamma_{r_{n}, d} }}
	% \\
	%\subto \quad & \mathbf{x}^{\rm{E}} = [x_v, x_{u_{1}}, \dots, x_{u_{N}}]\in \mathcal{X}^N, x_v \neq x_1. 
	%\end{IEEEeqnarray}
	\begin{IEEEeqnarray}{rCl}
		\raisebox{-1.5\normalbaselineskip}[0pt][0pt]{%
			({\bf P4})
		} \quad	 \underset{\mathbf{x}^{\rm{E}}}{\minimize}	& \quad &  
		\left( \Re \{1-x_1^*x_v
		\} \gamma_{s,d}+\sum_{n=1}^{N} 
		\Re \{1- x_1^*(x_v+\tau_n (x_{u_{n}}-x_v))\}\gamma_{r_{n}, d} \right)^{\frac{1}{2}}  + \nonumber \\
		&  & \frac{\|\mathbf{T}_{\mathbf{x}^{\rm{E}}} \tilde{\bm{\eta}} \|_1}{2  }
		\left( \Re \{1-x_1^*x_v
		\} \gamma_{s,d}+\sum_{n=1}^{N} 
		\Re \{1- x_1^*(x_v+\tau_n (x_{u_{n}}-x_v))\}\gamma_{r_{n}, d} \right)^{-\frac{1}{2}} \nonumber \\
		\subto & \quad & \mathbf{x}^{\rm{E}} = [x_v, x_{u_{1}}, \dots, x_{u_{N}}]\in \mathcal{X}^{N+1}, x_v \neq x_1.  \nonumber
	\end{IEEEeqnarray}
	A dominant case is obtained when $\mathbf{x}^{\rm{E}} = x_v \mathbf{1}_{ N}$, % $\mathbf{T}_{\mathbf{x}^{\rm{E}}}=\diag(0,0, \dots
%	,0)$, and we have 
	and problem (\textbf{P4}) becomes 
	$
	\underset{x_v}{\minimize} \
	\Re \{1- x_1^*x_v\},
	$
	of which the solution is obtained when 
	$
	x_v \in \{x_2, x_M\}.
	$ For the analysis of the other values of $\mathbf{x}^{\rm{E}}$, we follow the same logic as the single relay case. Eventually, after some mathematical calculations, the solutions are obtained as: for the $n$-th branch, if $\tau_n = 0$, then $x_v \in \{x_2, x_M\}$; if $\tau_n = 1$, then $x_v \in \{x_2, x_M\}$ and $x_{u_{n}} = x_1$.
	By summing over all dominant PEP terms, { \color{black} an average approximate SER for the scenario of error-free relaying  is obtained as
	\begin{IEEEeqnarray}{rCl} \label{eq:bound-general-C}
		\mathcal{P}^{\rm{C}}_{\rm{e}}  \triangleq & 2 \prod_{n=1}^{N}(1-\epsilon_n)  \sum_{m=0}^{N-1} \binom{N}{m} \int 
		Q \left( \sin \left( \frac{\pi}{M}\right) \sqrt{\gamma_m}  + \left(\sum_{n=m+1}^{N} \frac{\eta_n}{4\sin \left( \frac{\pi}{M}\right)} \right)  \frac{1}{\sqrt{\gamma_m}}
		\right) p(\gamma_m) \dd \gamma_m,
	\end{IEEEeqnarray}
	where $m$ denotes the number of branches with $\tau_n=0$, and it takes values from $0$ to $N-1$. $\gamma_m \triangleq  \gamma_{s,d} + \sum_{n =1}^m \gamma_{r_{n}, d}$ is gamma distributed. Hereafter, we let $ \bar{\gamma}_{s,r_{n}} = \bar{\gamma}_{r_{n},d} = \bar{\gamma}$, $\epsilon_n = \epsilon$, and $\eta_n = \eta$ for $n \in \Theta$. 	
	It is derived in Appendix \ref{proof-prop-1} that
		\begin{align}
			 \mathcal{P}^{\rm{C}}_{\rm{e}}  \approx &  2(1-\epsilon)^{N}\sum_{m=0}^{N-1} \binom{N}{m}  \frac{1}{2\Gamma(m+1)\bar{\gamma}_m^{m+1} } \frac{2^{-2m-1}((N-m) \eta)^{m+1}  \exp (-(N-m)\eta/4)  }{(2 \sin^2 \left( \frac{\pi}{M}\right) ( \sin^2 \left( \frac{\pi}{M}\right) /2+\bar{\gamma}_m^{-1}))^{\frac{m+1}{2}}} \notag \\ & K_{m+1}\Big(
			 \frac{(N-m) \sqrt{ \sin^2 \left( \frac{\pi}{M}\right) /2+\bar{\gamma}_m^{-1}} \eta}{2 \sqrt{2} \sin  \left( \frac{\pi}{M}\right) }
			 \Big) ,  \label{eq:PeC-ave}
		\end{align}
where $\bar{\gamma}_m = \frac{m\bar{\gamma}+\bar{\gamma}_{s,d}}{m+1}$ and $K_{m+1}(\cdot)$ is the modified Bessel function of the second kind \cite{book2-abramowitz1966handbook}.
Note that this expression applies to the single relay case.} We also prove in Appendix \ref{proof-prop-1} that $\mathcal{P}^{\rm{C}}_{\rm{e}}$ decays with a rate of $( \ln \bar{\gamma} )^{m+\frac{1}{2}} (\bar{\gamma})^{-(N+1)}$, and the full diversity order $N+1$ is achieved for error-free relaying.

	%--------------------------------------------
	\subsubsection{For Erroneous Relaying}
	Denote $\Theta^{\rm{E}} = \{1,2,\dots, N^{\rm{E}}\}$, $N^{\rm{E}} \geq 1$, for simplicity, and we have $\mathbf{x}^{C} = [ x_1, x_{r_{1}}, \dots,  x_{rN^{\rm{E}}},  x_1 \mathbf{1}_{N^{\rm{C}}}]$, $x_{r_{n}} \neq x_1$ for $n \in \Theta^{\rm{E}}$, and $\mathbf{T}_{\mathbf{x}^{\rm{C}}}=[0, \mathbf{1}_{ N^{\rm{E}}}, \mathbf{0}_{ N^{\rm{C}}}]$. Here, we assume that the detection of the $R_n-D$ link is correct if that of the $S-R_n$ link is wrong \cite{cmrc-wang2007high}, and then we have
	\begin{IEEEeqnarray}{rCl}
	\Pr [g(\mathbf{x}^{C}) > g(\mathbf{x}^{\rm{E}}) ] 
		& = & \Pr[\omega_{s,d}(x_1, x_v) + \omega_{\Theta^{\rm{C}}} (\mathbf{x}^{\rm{C}}, \mathbf{x}^{\rm{E}}) + \omega_{\Theta^{\rm{E}}} (\mathbf{x}^{\rm{C}}, \mathbf{x}^{\rm{E}})  + \| (\mathbf{T}_{\mathbf{x}^{\rm{C}}}-\mathbf{T}_{\mathbf{x}^{\rm{E}}}) \tilde{\bm{\eta}} \|_1 > 0] \nonumber \\
		& = & Q \left(
		\frac{-u_{s,d}(x_1, x_v)-u_{\Theta^{\rm{C}}} (\mathbf{x}^{\rm{C}}, \mathbf{x}^{\rm{E}})-u_{\Theta^{\rm{E}}} (\mathbf{x}^{\rm{C}}, \mathbf{x}^{\rm{E}})-\| (\mathbf{T}_{\mathbf{x}^{\rm{C}}}-\mathbf{T}_{\mathbf{x}^{\rm{E}}}) \tilde{\bm{\eta}} \|_1/2 }{\sqrt{W_{s,d}(x_1, x_v) + W_{\Theta^{\rm{C}}} (\mathbf{x}^{\rm{C}}, \mathbf{x}^{\rm{E}}) + W_{\Theta^{\rm{E}}} (\mathbf{x}^{\rm{C}}, \mathbf{x}^{\rm{E}})}}
		\right) . \label{eq:PEP-general-wrong}
	\end{IEEEeqnarray}
	To obtain the dominant PEP terms, we maximize \eqref{eq:PEP-general-wrong} over $\mathbf{T}_{\mathbf{x}^{\rm{E}}} \in \{0,1\}^{N+1}$ with $\mathbf{T}_{\mathbf{x}^{\rm{E}}}[1]=0$. After some involved analysis, the dominant terms are obtained when $\mathbf{T}_{\mathbf{x}^{\rm{E}}}=[0, \mathbf{0}_{ N^{\rm{E}}}, \mathbf{1}_{ N^{\rm{C}}}]$, $\mathbf{x}^{\rm{C}}=[x_1, x_v\mathbf{1}_{ N^{\rm{E}}}, x_1\mathbf{1}_{ N^{\rm{C}}} ]$ and $\mathbf{x}^{\rm{E}}=[x_v, x_v\mathbf{1}_{ N^{\rm{E}}}, x_1\mathbf{1}_{ N^{\rm{C}}} ]$.  By adding the multiplicity $\binom{N}{N^{\rm{E}}}$, 	{ \color{black} the average approximate SER for this scenario can be obtained as
	\begin{IEEEeqnarray}{rCl} \label{eq:bound-general-E}
		\mathcal{P}^{\rm{E}}_{\rm{e}} & \triangleq & \sum_{N^{\rm{E}} =1 }^N \frac{ \binom{N}{N^{\rm{E}}}}{M-1}   \epsilon^{N^{\rm{E}}}(1-\epsilon)^{N-N^{\rm{E}}}
		\sum_{\substack{x_v \in \mathcal{X} \\ x_v \neq x_1}} \int_{0}^{\infty}
		Q \Big( 
		\frac{1}{\sqrt{2}} 
		\sqrt{ \Re \{ 1-x_1^*x_v\}  \gamma_{s, d} } \ +  \frac{(N-2N^{\rm{E}}) \eta}{2\sqrt{2}\sqrt{ \Re \{ 1-x_1^*x_v\}  \gamma_{s, d} }}
		\Big) \nonumber \\ \
		& &  
		p(\gamma_{s, d})  \dd \gamma_{s, d}.
	\end{IEEEeqnarray}
	It is derived in Appendix \ref{proof-prop-2} that
	\begin{align}
		\mathcal{P}^{\rm{E}}_{\rm{e}} \approx &  \Bigg\{   \sum_{N^{\rm{E}} > N/2} \frac{ \binom{N}{N^{\rm{E}}}}{M-1}   \epsilon^{N^{\rm{E}}}(1-\epsilon)^{N-N^{\rm{E}}} 
		\sum_{\substack{x_v \in \mathcal{X} \\ x_v \neq x_1}} \bigg[ 1-\exp \left( -\frac{(2N^{\rm{E}}-N) \eta}{2 c_1\bar{\gamma}_{s,d}}\right) +   \frac{(2N^{\rm{E}}-N) (4c_1)^{-1/2} \eta}{ 2 \left(c_1/4+\bar{\gamma}_{s,d}^{-1}\right)^{1/2}\bar{\gamma}_{s,d}}  \notag \\  & 
		K_1 	\Big(
		\frac{(2N^{\rm{E}}-N)\sqrt{c_1/4+\bar{\gamma}_{s,d}^{-1}}}{2\sqrt{c_1}} \eta
		\Big) \exp \left((2N^{\rm{E}}-N)\eta /4\right)
		\bigg] \Bigg\} +  \Bigg\{   \frac{ \binom{N}{N/2}}{M-1}   \epsilon^{N/2}(1-\epsilon)^{N/2} 	\sum_{\substack{x_v \in \mathcal{X} \\ x_v \neq x_1}} \frac{2}{4+c_1\bar{\gamma}_{s,d}} \Bigg\} \notag \\ &  +   \Bigg\{   \sum_{N^{\rm{E}} < N/2} \frac{ \binom{N}{N^{\rm{E}}}}{M-1}   \epsilon^{N^{\rm{E}}}(1-\epsilon)^{N-N^{\rm{E}}} 
		\sum_{\substack{x_v \in \mathcal{X} \\ x_v \neq x_1}} \bigg[ \frac{(2N^{\rm{E}}-N) (4c_1)^{-1/2} \eta}{ 2 \left(c_1/4+\bar{\gamma}_{s,d}^{-1}\right)^{1/2}\bar{\gamma}_{s,d}} K_1 	\Big(
		\frac{(N-2N^{\rm{E}})\sqrt{c_1/4+\bar{\gamma}_{s,d}^{-1}}}{2\sqrt{c_1}} \eta
		\Big)  \notag \\ &  \exp \left((2N^{\rm{E}}-N)\eta /4\right)
		\bigg] \Bigg\} , \ 1 \leq N^{\rm{E}} \leq N, \label{eq:PeE-ave}
	\end{align}
	where $c_1 = \Re \{ 1-x_1^*x_v\} > 0$. Note that this expression applies to the single relay case. We also prove in Appendix \ref{proof-prop-2} that $\mathcal{P}^{\rm{E}}_{\rm{e}}$ decays with $ ( \ln \bar{\gamma} )^{\frac{1}{2}} (\bar{\gamma}_{s,d})^{-(N+1-N^{\rm{E}})}$ for $1 \leq N^{\rm{E}} \leq \frac{N}{2}$ and with $ ( \ln \bar{\gamma} )^{\frac{1}{2}} (\bar{\gamma}_{s,d})^{-(N^{\rm{E}}+1)} $ for $\frac{N}{2} \leq N^{\rm{E}} \leq N $, and the diversity order is  $\left \lceil \frac{N}{2} \right \rceil+1$. Considering both scenarios of error-free and erroneous relaying, we give \textbf{Proposition \ref{prop: diversity}} to describe the diversity order, where the first part summarizes how the number of erroneous relays affects the diversity order, the second part can be further obtained based on the worst case analysis, and  the third part gives the exact diversity order.
	\begin{proposition}\label{prop: diversity}
		For the relay network with $N$ parallel DF relays, when $M$-DPSK is used, the diversity order of both the proposed NMLD and the existing AMLD can be described as follows.
		\begin{itemize}
			\item In the case when $N^{\rm{E}}$ relays make errors,  $0 \leq N^{\rm{E}} \leq N$,   the diversity order is $d(N^{\rm{E}}) = N^{\rm{E}}+1$ if $\frac{N}{2} \leq N^{\rm{E}} \leq N $; is $d(N^{\rm{E}}) = N+1-N^{\rm{E}}$ if $0 \leq N^{\rm{E}} \leq \frac{N}{2}$. 
			\item With $r$ error-free relays, the diversity order is $d = \min_{ 0 \leq N^{\rm{E}} \leq N-r } \ d ( N^{\rm{E}} )$.
			\item All cases considered,  the achievable diversity order is $d = \left \lceil \frac{N}{2} \right \rceil+1$.
		\end{itemize}
		
	\end{proposition}
}
%	\begin{table}[t]
%		\caption{Our performance analysis results for the proposed NMLD and the existing AMLD in the non-coherent DF relay network with $N$ parallel relays and a direct link } % title of Table
%		\centering % used for centering table
%		\begin{tabular}{c  c  c  c  } % centered columns (4 columns)
%			\toprule %inserts double horizontal lines
%			$\#$ \textbf{of relays} & \textbf{Modulation} & \tabincell{c}{\textbf{SER expression}} &  \textbf{Diversity order} $d$  \\ % inserts table
%			%heading
%			\midrule % inserts single horizontal line
%			$N$  &
%			\tabincell{c}{$M$-DPSK} & Available &  $d = \left \lceil \frac{N}{2} \right \rceil+1$	 \\
%			\bottomrule
%		\end{tabular}
%		\label{tab: results-diversity-ours} % is used to refer this table in the text
%	\end{table}		
Since the NMLD is obtained by applying the accurate max-sum approximation to the AMLD, the average SER and the  diversity order expressions derived based on the NMLD are also applicable to the AMLD, which will be verified in Section \ref{sec:simu}. 	
%For a clear comparison with the known performance analysis results in the literature (please refer to Table \ref{tab: results-diversity}), our results are briefly summarized in Table \ref{tab: results-diversity-ours}. 
{\color{black}
\begin{remark}
%		The proposed NMLD can be extended to the multi-branch multi-hop non-coherent DF relay networks with the following modifications on the detection metric in \eqref{eq:NMLD-general}.
	 For a multi-branch multi-hop non-coherent DF relay network, assume there are $N$ parallel branches, and each branch has $U$ hops. Denote the $u$-th relay of the $n$-th branch  as $R_{n,u}$, $n \in \{1,2,\dots, N \}$, $u \in \{1,2,\dots, U-1\}$. Then, according to \cite{SER-multihop-2010}, the multi-hop branch $S-R_{n,1}-R_{n,2}-\dots-R_{n,U-1}$ can be transformed into an equivalent single-hop link $S-R_{n,U-1}$ in terms of the average SER and the diversity order. Thanks to this, the proposed NMLD detection metric in \eqref{eq:NMLD-general} can be applied to this network by replacing $y_{r_{n},d}$ with $y_{r_{n,U-1},d}$, $n \in \{1,2,\dots, N \}$, where 
	$y_{r_{n,U-1},d}$ denotes the  received signal at the destination from $R_{n,U-1}$.	 In this case, $\epsilon_n$ is the average SER at $R_{n,U-1}$ (for the equivalent single-hop link $S-R_{n,U-1}$). It can be calculated according to \cite{SER-multihop-2010}.
	Besides, the proposed performance analysis method remains valid due to the aforementioned multi-hop to single-hop equivalence transformation.
\end{remark}
}

	\section{Simulation Results}
	\label{sec:simu}
	This section presents simulation results in the cases of single- and multi-relay networks.  For comparison, {\color{black} the results in the coherent counterparts are also presented using the coherent AMLD in \cite{bhatnagar2011ml}}. 
	For this coherent counterpart, the CSI assumption is: each relay has the perfect instantaneous CSI of the corresponding source-relay link, and the destination has the perfect instantaneous CSI of the source/relay-destination links, but only the average CSI of the source-relay links. 
	Let all nodes transmit with the same power $P_s = 1$. As mentioned in Section \ref{sec:GDM}, we consider quasi-static Rayleigh fading, that is, each channel coefficient $h_{I, J}$, $({I}, {J}) \in \{(s,r),(s,d), (r,d)\}$, remains unchanged within one frame ($K$ symbols) duration, while varies independently from one frame to another. The default simulation settings are as follows unless specified otherwise. The modulation used is QPSK, $(\rho_{\rm{R}},\rho_{\rm{N}} )=(\rho_{\rm{R}}^{\star}, \rho_{\rm{N}}^{\star})$ as in \eqref{eq:opt-rho}, all links have the same average SNR $\bar{\gamma}$, i.e., $\bar{\gamma}_{s,d} = \bar{\gamma}_{s,r_{n}}  = \bar{\gamma}_{r_{n},d} = \bar{\gamma}$, and we set $N_{s,d} = N_{s,r_{n}}  = N_{r_{n},d} = N_{0} = 1$, $n \in \Theta$.
	%the detector we use is the proposed NMLD, 
	% Fig.~\ref{fig:SER-comparison-GDM}-\ref{fig:SER-comparison-power-allo} are for the single relay scenario. Fig.~\ref{fig:SER-and-bound-N=23}-\ref{fig:diversity-3relays} are for the multiple relay scenario. % with $L=1$ for the non-coherent setup.
	\subsection{Single Relay Network}
\begin{figure}[t]
	\centering
	\includegraphics[width=0.5\textwidth]{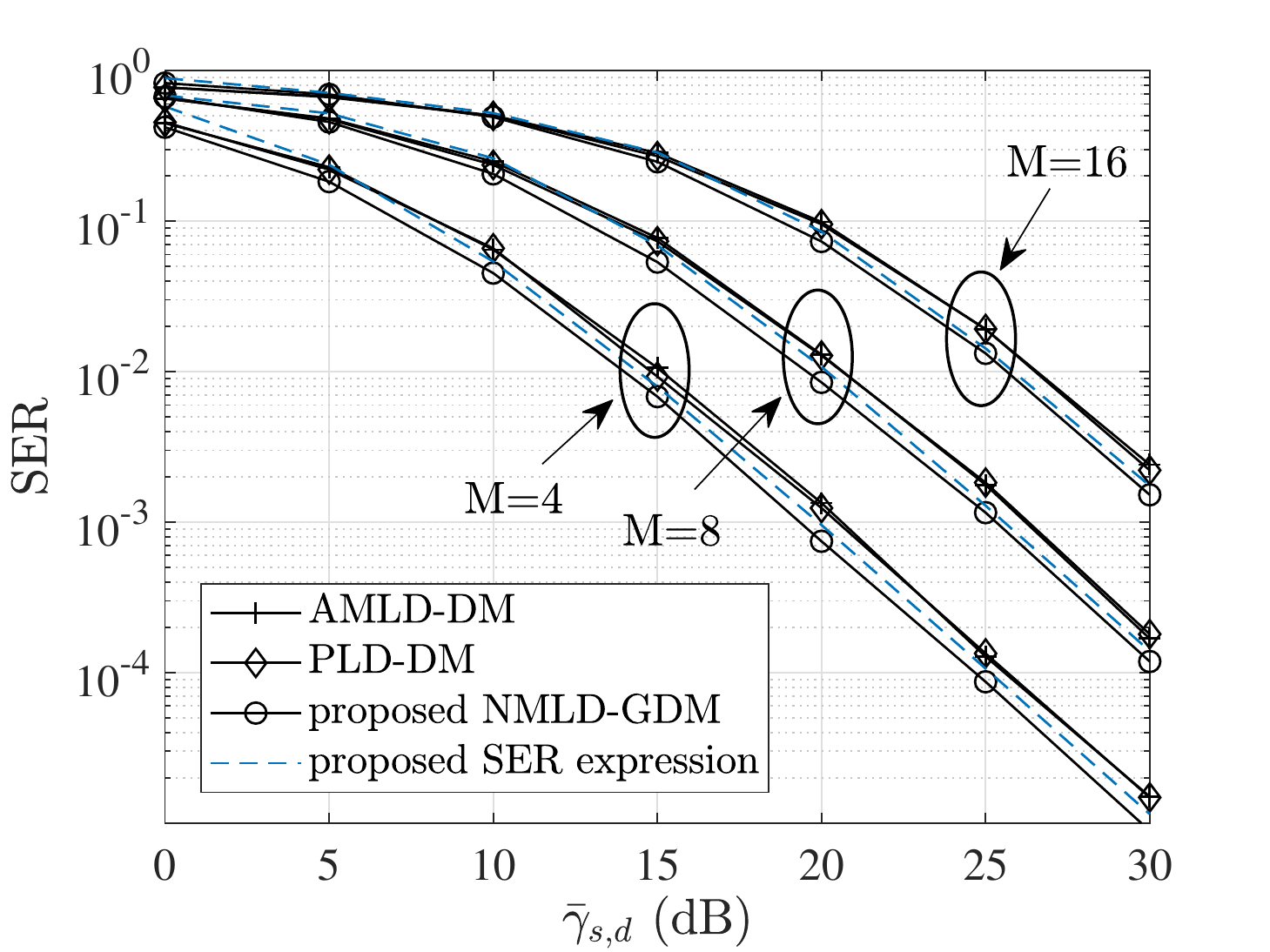}
	\caption{SER performance comparison of various schemes and the proposed SER expression using $M$-DPSK when the information frame length $K =320$ and the block length $L=16$ in the non-coherent DF single relay network under quasi-static Rayleigh fading (the fading coefficients remain fixed for the duration of one frame, while vary independently from one frame to another).}
	\label{fig:SER-comparison-GDM}
\end{figure}

\begin{figure} [t] 	 
	\centering
	\subfigure[{\color{black}Average run time in Matlab} ]{
		\label{fig:RunTime-M}
		\includegraphics[width=0.48\textwidth]{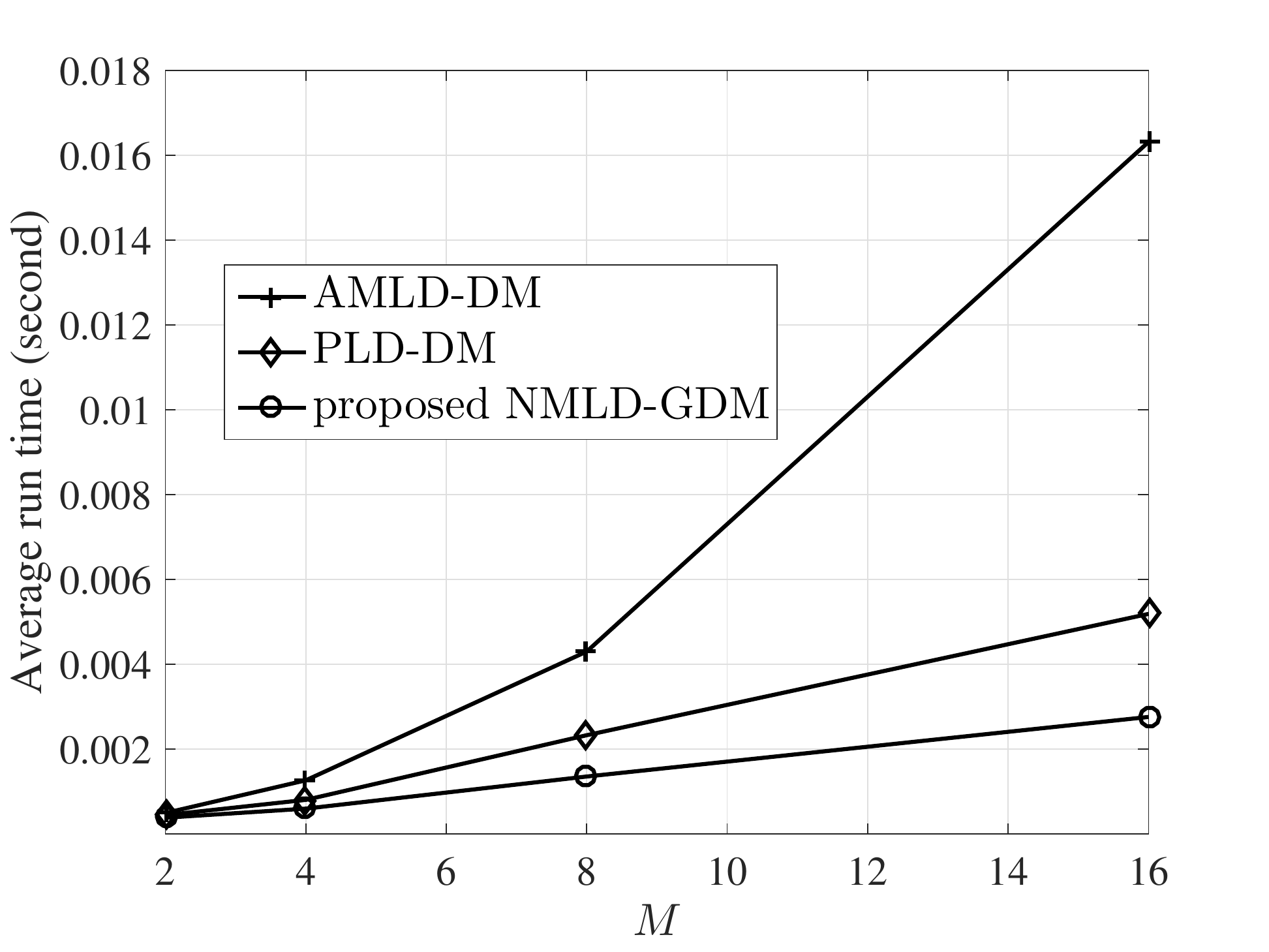}}
	\subfigure[ {\color{black}Total number of operations from Table \ref{table:metric-complexity} } ]{
		\label{fig:tab-M}
		\includegraphics[width=0.48\textwidth]{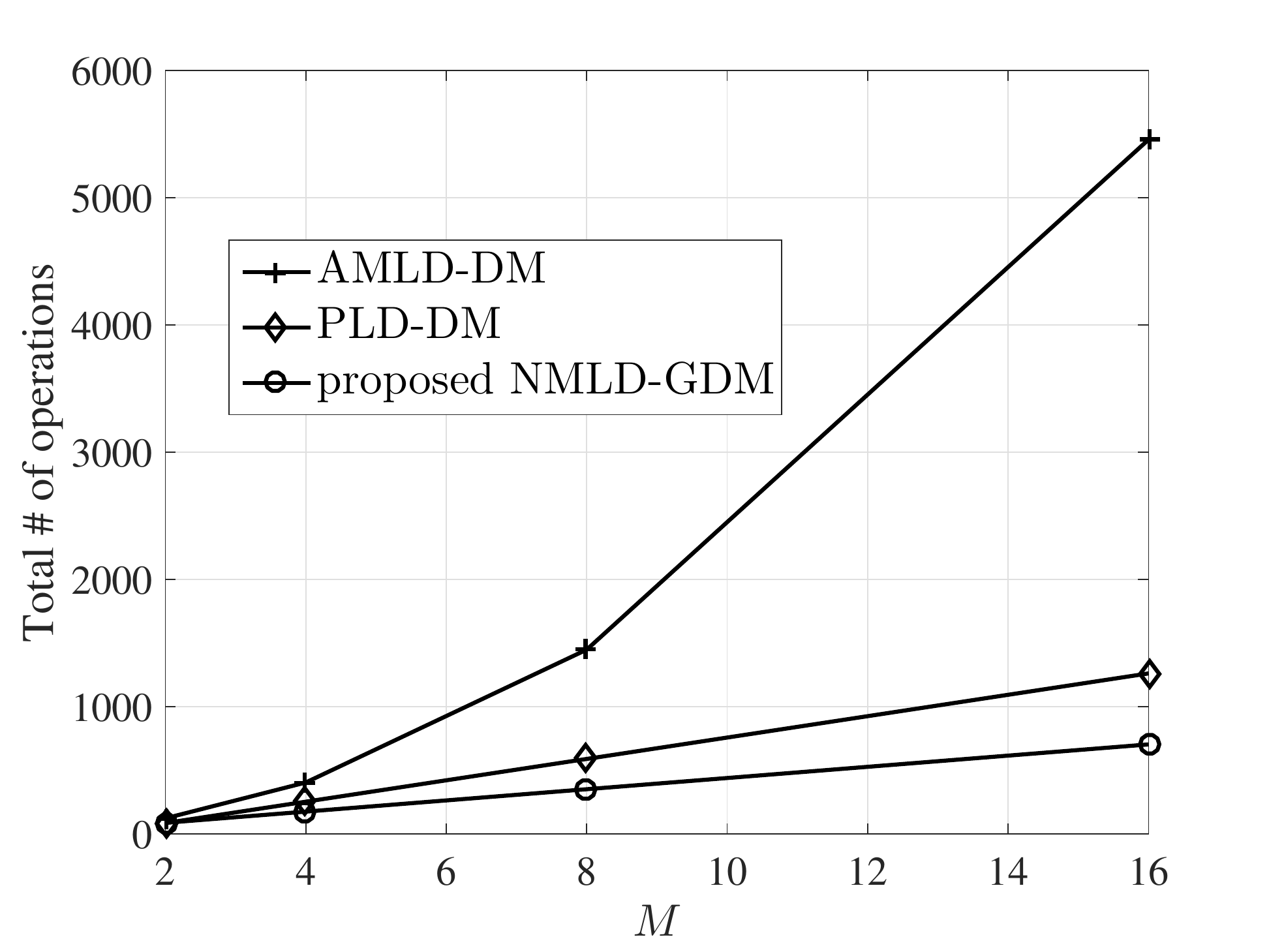}}
	%	\captionsetup{font={small}} 
	\caption{ {\color{black}Detection complexity comparison of various schemes using $M$-DPSK in the non-coherent DF single relay network under quasi-static Rayleigh fading (the fading coefficients remain fixed for the duration of one frame, while vary independently from one frame to another).}}
	\label{fig:complexity-single-relay} %% label for entire figure
\end{figure}

	%------------------	
%	\subsubsection{	{\color{black}Run Time and Performance Comparison}}
 Fig.~\ref{fig:SER-comparison-GDM} compares the SER performance of various schemes when the information frame length $K =320$ and block length $L=16$. It is seen that the proposed NMLD-GDM outperforms the AMLD-DM and PLD-DM. For example, to achieve an SER of $10^{-3} $ using $8$-PSK, the performance gap between NMLD-GDM and AMLD(PLD)-DM is around $2$ dB. In addition, the proposed SER expression is shown to be accurate for all three values of $M$ over the whole SNR range. {	{\color{black} Fig.~\ref{fig:complexity-single-relay} shows the complexity comparison of various schemes with different mudulation size $M \in \{2,4,8,16\}$.  Fig.~\ref{fig:RunTime-M} compares the average run time in Matlab, while Fig.~\ref{fig:tab-M} compares the total number of operations (c.f. Table \ref{table:metric-complexity}). For the run time comparison, the average SNR $\bar{\gamma}=20$ dB, and $10^4$ symbol detections are performed for each value of $M$ using each detector under consideration. We can see that the run time (and the total number of operations) of the proposed NMLD is linear in $M$, while those of the AMLD and PLD are quadratic and linear in $M$, respectively. This agrees with the complexity order analysis in Table \ref{table:metric-complexity}. } \color{black} Together with the SER comparison in Fig.~\ref{fig:SER-comparison-GDM}, it can be seen that the proposed NMLD-GDM outperforms the AMLD-DM with a reduced complexity order.}

%	\begin{figure}[t]
%		\centering
%		\includegraphics[width=3.4in]{comparison-SER-v2}
%		\caption{SER performance comparison of various schemes and the proposed SER bound using $M$-PSK when the frame length $K =320$ and the block length $L=16$ in the non-coherent DF single relay network under quasi-static Rayleigh fading (the fading coefficients remain fixed for the duration of one frame, while vary independently from one frame to another).}
%		\label{fig:SER-comparison-GDM}
%	\end{figure} 

		%------------------	
%	\subsubsection{	{\color{black}Performance with Different $L$ and $\rho_{\rm{N}}$}}
\begin{figure} [t] 	 
	\centering
	\subfigure[For various values of block length $L$]{
		\label{fig:SER-comparison-L}
		\includegraphics[scale=0.48]{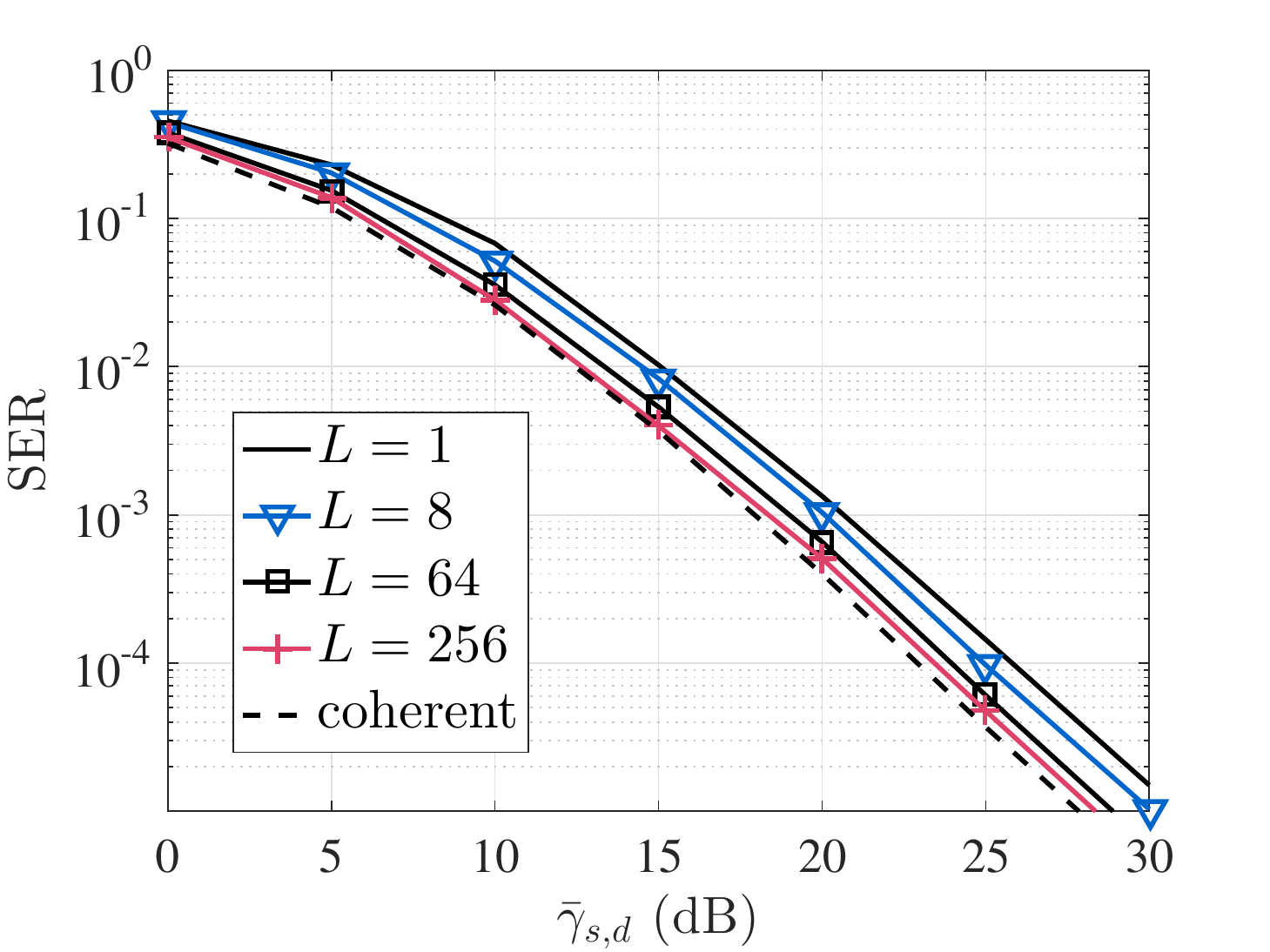}}
	\subfigure[For $\rho_{\rm{N}} \in (0,1)$ when $\bar{\gamma}=10 \text{ and }20$ dB]{
		\label{fig:SER-comparison-power-allo}
		\includegraphics[scale=0.48]{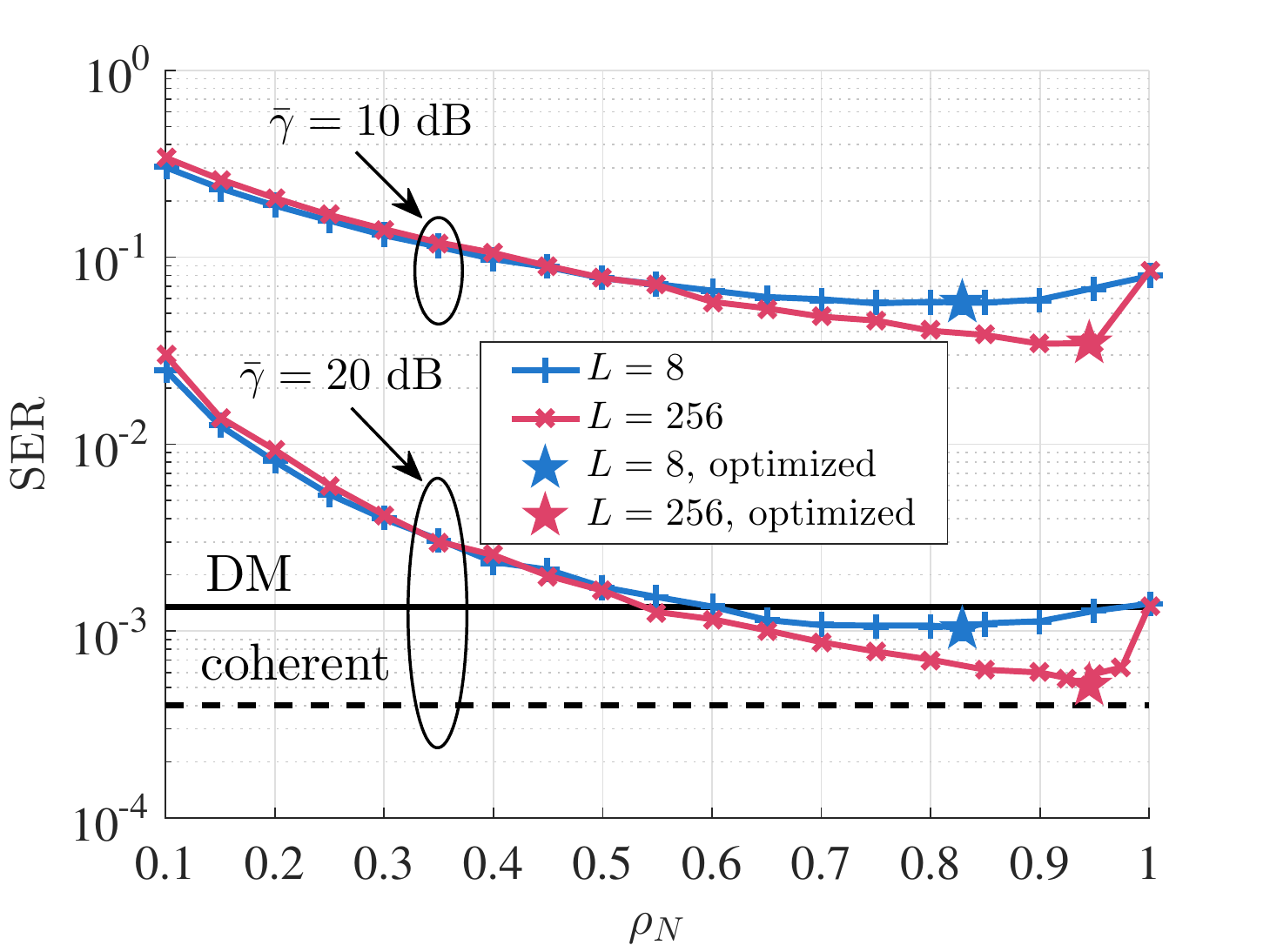}}
	%	\captionsetup{font={small}} 
	\caption{SER performance comparison using the proposed NMLD when the information frame length $K=512$ in the non-coherent DF single relay network using QPSK under quasi-static Rayleigh fading. The SER of the coherent counterpart is presented as a lower bound.}
	\label{fig:SER-comparison-L-rho} %% label for entire figure
\end{figure}
	Fig.~\ref{fig:SER-comparison-L} shows the impact of the block length $L$ on the SER performance when $K=512$. The SER of the coherent counterpart is presented as a lower bound, while that of the DM scheme (the $L=1$ case) as an upper bound. It is observed that by increasing the block length $L$, the SER performance can be significantly improved, and can be close to that of the coherent counterpart. For example, when $L=256$, the performance gap with the coherent scheme is within $0.5$ dB. Fig.~\ref{fig:SER-comparison-power-allo} shows the SER performance with respect to $\rho_{\rm{N}} \in (0,1)$ when $\bar{\gamma}=10 \text{ and } 20$ dB. According to \eqref{eq:opt-rho}, when $L=8 \text{ and } 256$, we have $\rho_{\rm{N}}^{\star} = 0.830 \text{ and } 0.945$, respectively. The simulate SERs for using these two optimized values are also shown in solid stars. It is obvious that for both cases when $\bar{\gamma}=10 \text{ and }  20$ dB, the simulate SERs of GDM with $L=8 \text{ and } 256$ show unique minimums, which are around the optimized values $\rho_{\rm{N}}^{\star}$ calculated using \eqref{eq:opt-rho}. This verifies that the solution in \eqref{eq:opt-rho} is accurate from middle to high SNR values. Let us take a closer look at the case of $\bar{\gamma}=20$ dB. It can be seen that the SERs of DM and the coherent counterpart do not change with $\rho_{\rm{N}}$, since no power allocation is exploited. With improper power allocation, such as when $\rho_{\rm{N}} < 0.5$, the SER performance for GDM is shown to be worse than that of DM. By contrast, with proper power allocation, such as when $\rho_{\rm{N}} > 0.6$, GDM is shown to outperform DM. In the case when $L=256$, the proposed scheme with $\rho_{\rm{N}}^{\star}=0.945$ (see the solid red star) performs close to the coherent counterpart, which agrees with the earlier observations in Fig.~\ref{fig:SER-comparison-L}. It is also seen that with $\rho_{\rm{N}}=1$ (the RSs and NSs are allocated the same power $P_s$), the SER of GDM is the same as that of DM, saying that the conventional DM is a special case of the GDM scheme.

	%\begin{figure}[t]
	%	\centering
	%	\includegraphics[width=3.8in]{QPSK-different-L}
	%	\caption{SER performance comparison for various values of block length $L$ when $K=512$ for in the non-coherent DF single relay network. }
	%	\label{fig:SER-comparison-L}
	%\end{figure} 
	
	% when  optimal power allocation scheme
	%\begin{figure}[t]
	%	\centering
	%	\includegraphics[width=3.8in]{opt-power-allo}
	%	\caption{SER performance comparison of various schemes for $\rho_{\rm{N}} \in (0,1)$ when $\bar{\gamma}=20$ dB for the non-coherent DF single relay network.}
	%	\label{fig:SER-comparison-power-allo}
	%\end{figure} 
	
	\subsection{Multiple Relay Networks}
	%\begin{figure}[H]
	%	\centering
	%	\includegraphics[width=3.2in]{comparison-with-amld-v2}
	%	\caption{SER performance comparison of the AMLD and the proposed NMLD for the non-coherent DF multiple relay networks with $N=2,3$ parallel relays.}
	%	\label{fig:SER-and-bound-N=23}
	%\end{figure}

	In the multi-relay networks, to focus on the SER performance of the proposed NMLD, we adopt the conventional DM scheme without exploiting power allocation. 
	
	%%-------------------
%	\subsubsection{	{\color{black} Performance Comparison with AMLD and PLD}}
	\begin{figure} [h] 	 
		\centering 
		\subfigure[ { \color{black} SER}]{
			\label{fig:simu-SERwithN}
			\includegraphics[width=0.48\textwidth]{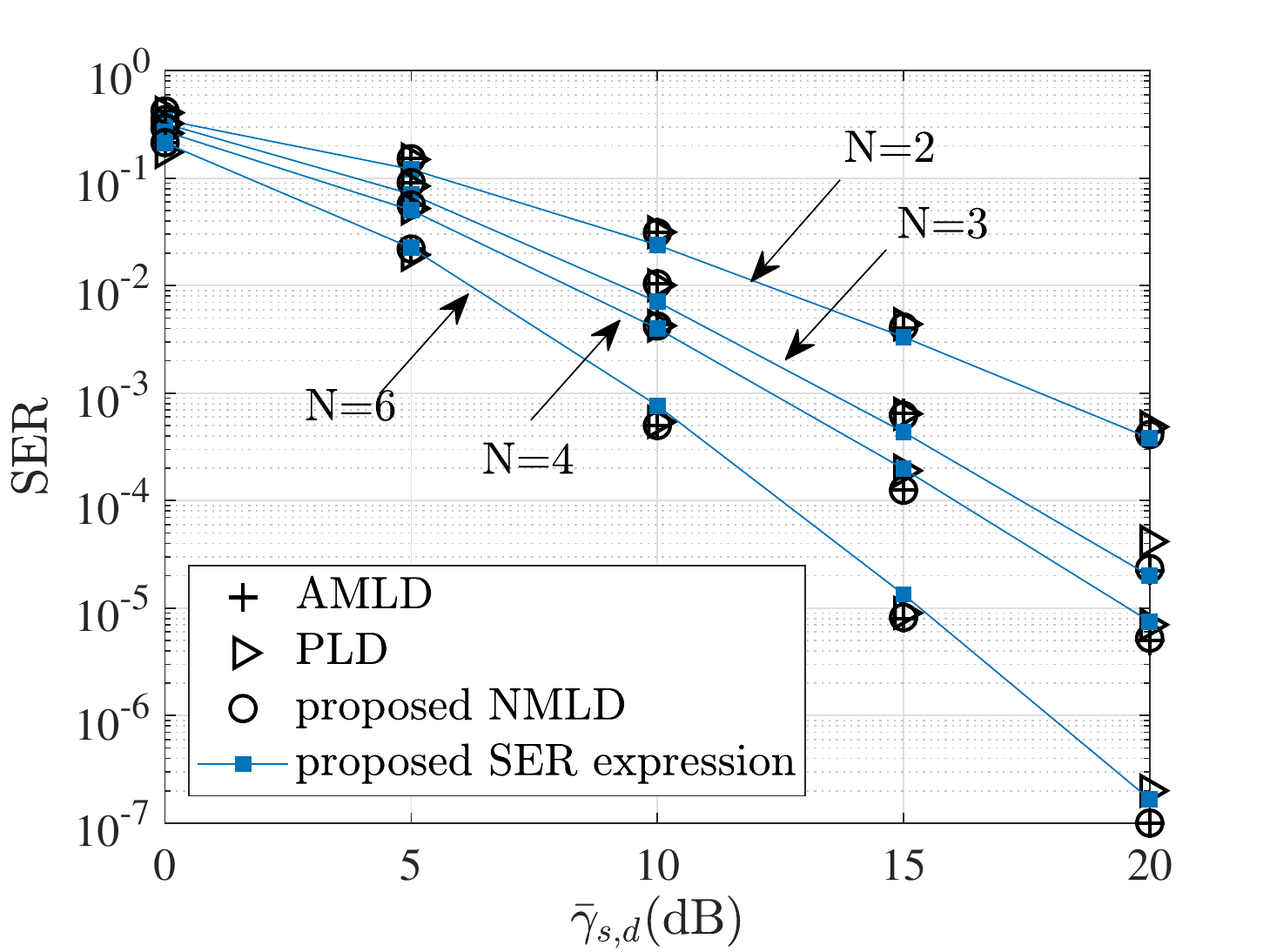}}	
		%	\captionsetup{font={small}}
		\subfigure[{ Detection complexity versus $N$}]{
		\label{fig:TableN}
		\includegraphics[width=0.48\textwidth]{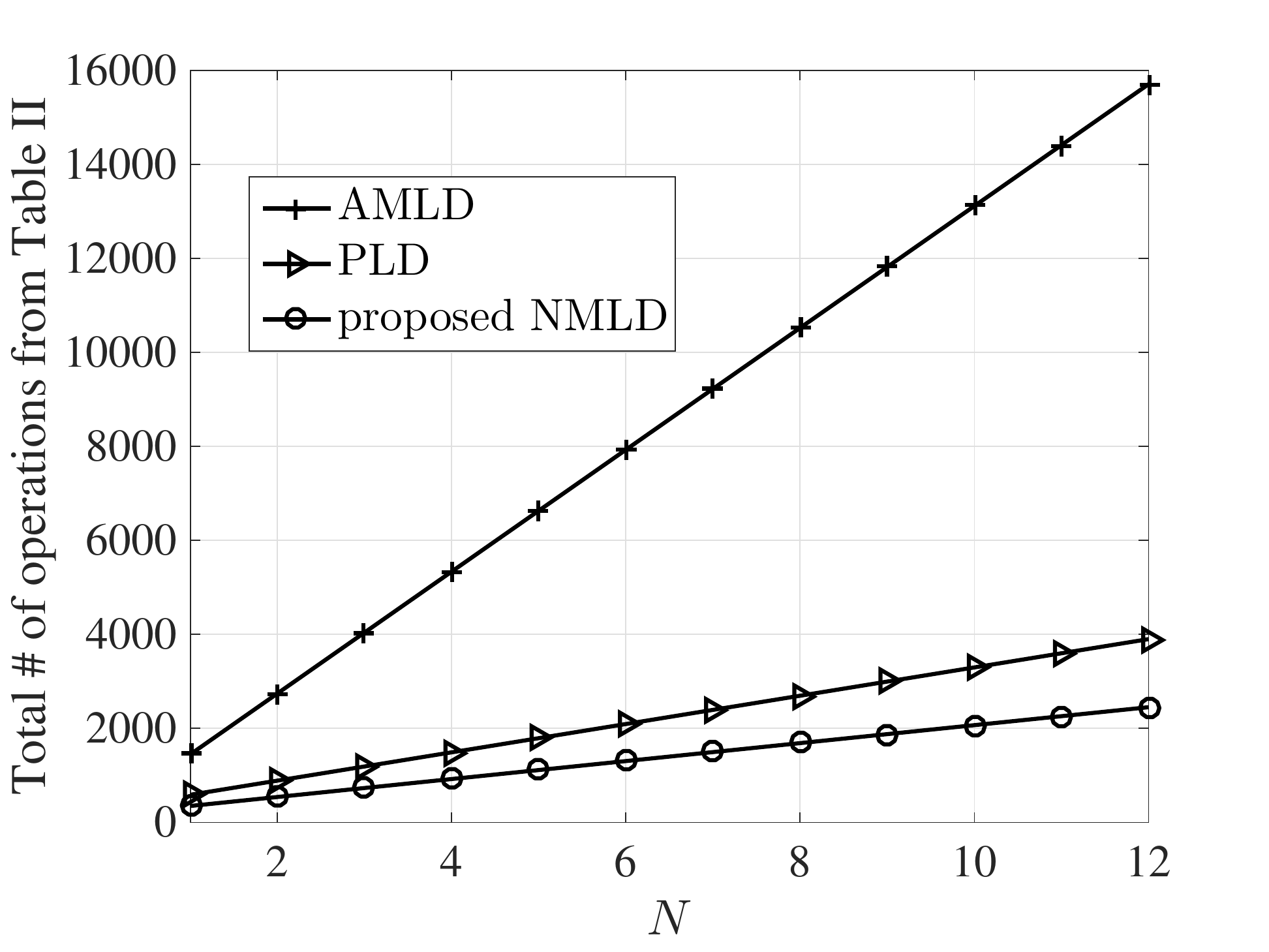}} 
		\caption{ \color{black} SER performance and detection complexity comparisons of various detectors in the non-coherent DF parallel relay network with $N$ relays using $8$-DPSK DM  for $\bar{\gamma}_{s,r_{n}}=\bar{\gamma}_{r_{n},d}=\bar{\gamma}_{s,d} + 10$ dB and $n=1,2,\cdots,N$.}
		\label{fig:SER-N-multi} %% label for entire figure
	\end{figure}
	{ \color{black} Fig.~\ref{fig:SER-N-multi} shows the performance  and  complexity comparisons with different relay number $N$ using $8$-DPSK. 
	 Fig.~\ref{fig:simu-SERwithN} shows the simulate SERs of the AMLD, the PLD and the proposed NMLD with {\color{black} $N \in \{2,3,4,6\}$} DF relays, when $\bar{\gamma}_{s,r_{n}}=\bar{\gamma}_{r_{n},d}=\bar{\gamma}_{s,d} + 10$ dB, $n=1,2,\cdots,N$, for $8$-DPSK signals. {\color{black}The proposed SER expression in \eqref{eq:PeC-ave} and \eqref{eq:PeE-ave} is also plotted to show its accuracy.  We can see that the SER performance of the NMLD is nearly the same as that of the AMLD. Besides, the SER expression is verified to be very accurate for all values of $N$.  Meanwhile, we can see that for both the AMLD and NMLD, the achievable diversity order for $N=2$ is $2$, for $N=3$ is $3$, for $N=4$ is $3$, and for $N=6$ is $4$, which are in exact agreement with the proposed diversity order expression $\left \lceil \frac{N}{2} \right \rceil+1$.} 	An interesting observation is that for infinite $N$, we have $\lim\limits_{N \to \infty} \frac{\text{achievable diversity order}}{\text{full diversity order}} = \lim\limits_{N \to \infty} \frac{\left \lceil \frac{N}{2} \right \rceil+1}{N+1} = \frac{1}{2}$, which means that asymptotically only half of the full diversity is achievable.
	 Fig.~\ref{fig:TableN} compares the total number of operations (c.f. Table \ref{table:metric-complexity}) with respect to $N$. It can be seen that all three detectors have linear complexities in $N$, but the proposed NMLD can save a considerable number of operations. For example, when $N=12$, the complexity saving by the NMLD (compared to the AMLD) is up to $\frac{15703-2451}{15703} = 84.39 \%$. Overall, the proposed NMLD can achieve similar performance as the AMLD but with a considerably reduced complexity.} 

	%When $M>2$, although the performance analysis for AMLD is not available in the literature, we can obtain some insights by simulations. 
	
\begin{figure} [t] 
	 	\centering
	 	\subfigure[\color{black} For comparison of the AMLD and the proposed NMLD when $N=2 \text{ and }3$]{
	 		\label{fig:SER-and-bound-N=23}
	 		\includegraphics[width=0.48\textwidth]{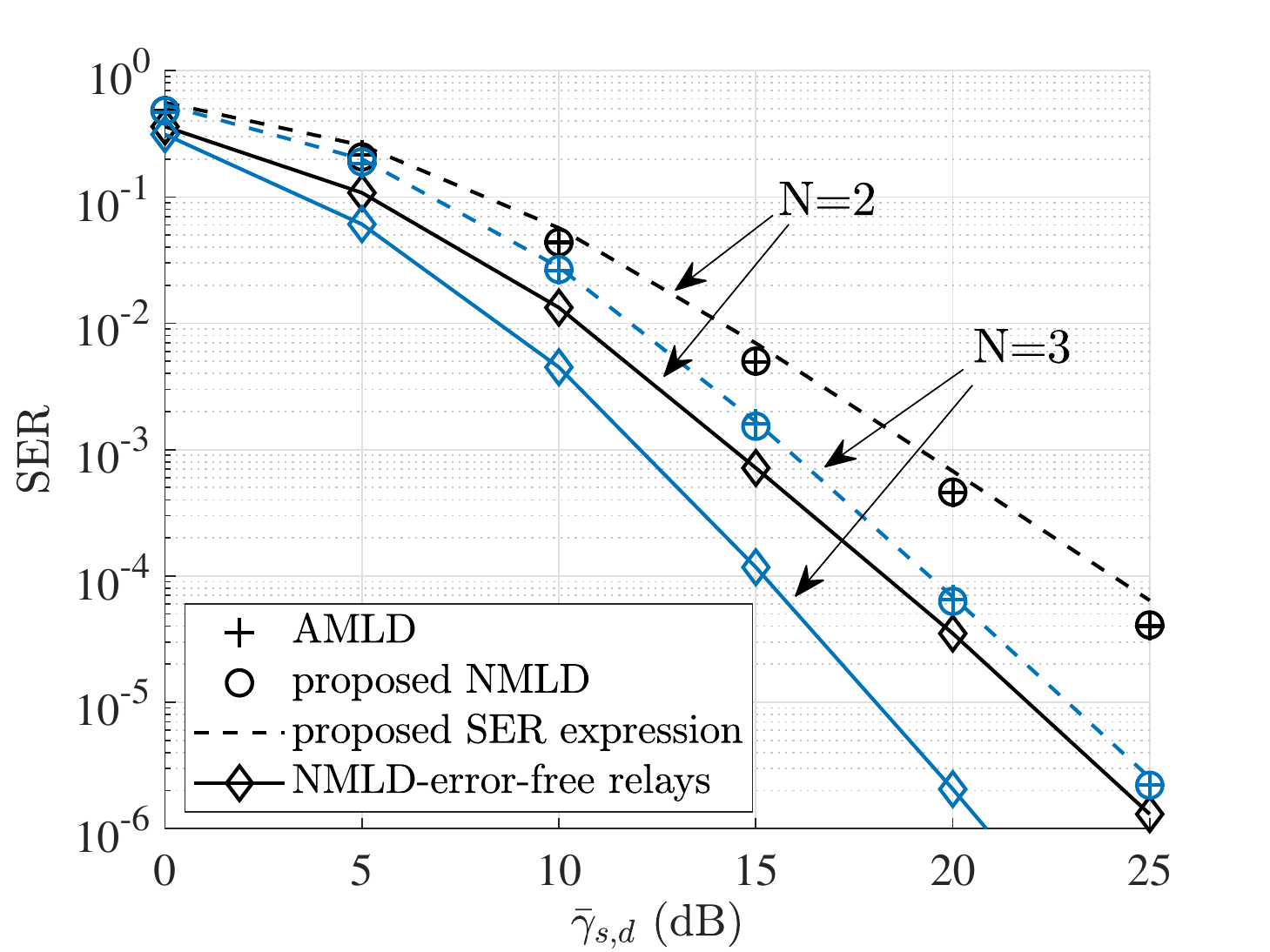}}
	 	\subfigure[For the AMLD with $r$ error-free relays when $N=3$]{
	 		\label{fig:diversity-3relays}
	 		\includegraphics[width=0.48\textwidth]{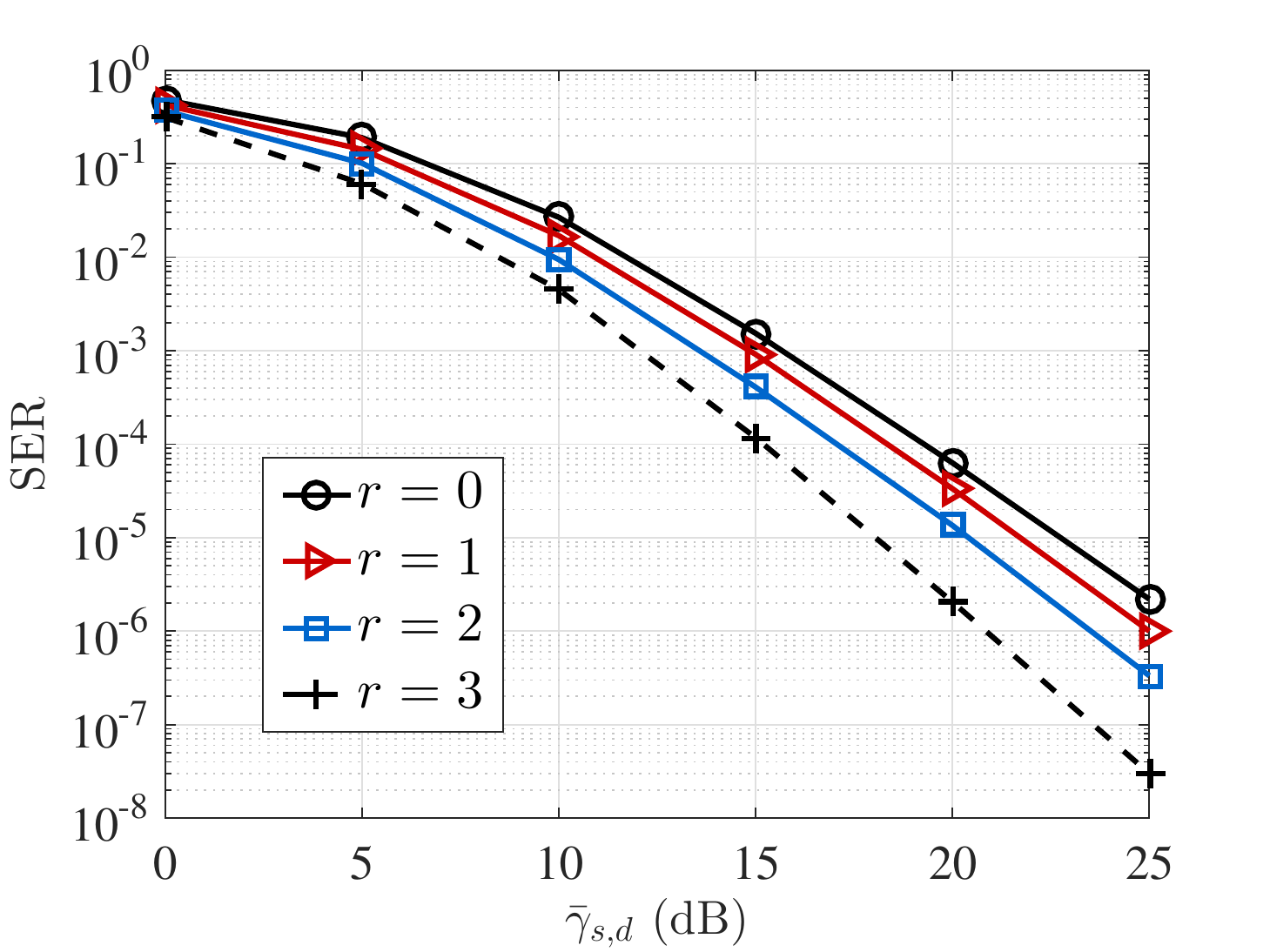}}	
	 	%	\captionsetup{font={small}} 
	 	\caption{SER performance  of various detectors for the conventional DM in the non-coherent DF multiple relay networks with $N$ parallel relays using DQPSK.}
	 	\label{fig:SER-multi-2} %% label for entire figure
	 \end{figure}

	 Fig.~\ref{fig:SER-and-bound-N=23} shows the SER performances of the AMLD and the proposed NMLD in the non-coherent DF relay networks with $N=2 \text{ and } 3$ parallel relays. It is  seen that the NMLD achieves almost the same SER performance as the AMLD, {\color{black}and the proposed SER expression calculated from \eqref{eq:PeC-ave} and \eqref{eq:PeE-ave} is accurate. Moreover, it is shown that both the NMLD and the AMLD achieve exactly the proposed diversity order  $\left \lceil \frac{N}{2} \right \rceil+1$}, while when all  relays are error-free, the NMLD achieves the full diversity order $N+1$. { \color{black} Fig.~\ref{fig:diversity-3relays} further shows the simulate SERs of the AMLD  with $r$ error-free relays when $N=3$. It is seen that for $r=0,1 \text{ and }2$, the AMLD achieves the diversity order $3$, while for $r=3$, the full diversity order $4$ can be achieved. The results are in exact agreement with the second part of \textbf{Proposition \ref{prop: diversity}}.}

	\begin{figure}[t]
		\centering
		\includegraphics[width=0.48\textwidth]{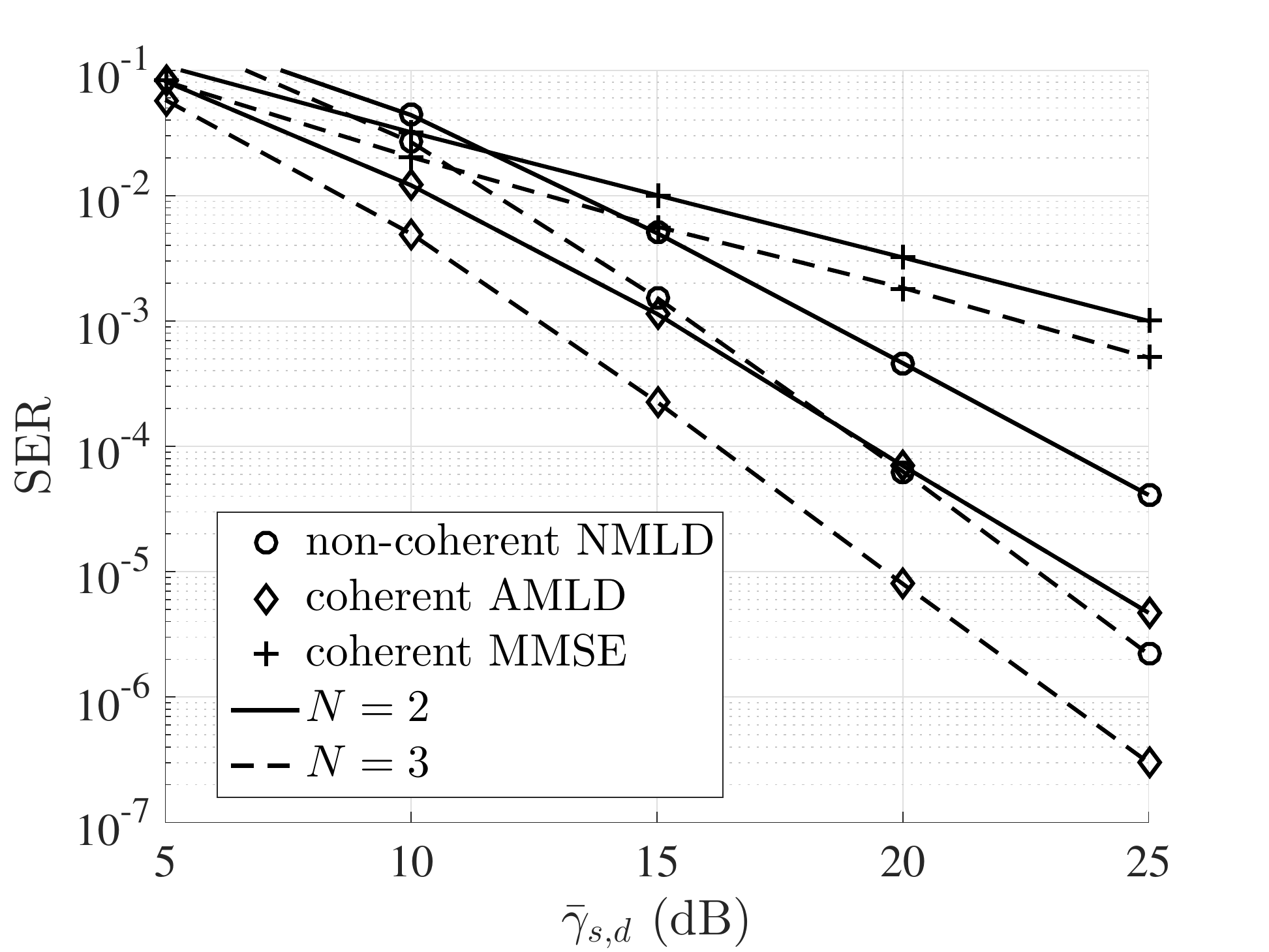}
		\caption{ \color{black} SER performance comparison of various detectors in the non-coherent (using the conventional DM) and coherent DF relay networks  with $N \in \{2,3\}$ parallel relays using QPSK. }
		\label{fig:comparison-with-coherent}
	\end{figure}

	Fig.~\ref{fig:comparison-with-coherent} compares the simulate SERs in non-coherent and coherent relay networks for $N=2 \text{ and } 3$. {\color{black} The AMLD in \cite{bhatnagar2011ml} (called the coherent AMLD) and the minimum mean squared error (MMSE) detector are used for the coherent case}, for which the CSI assumption is described at the beginning of this section. 
%	{ \color{black} Specifically, the detection rule of the MMSE detector is $ \arg\min_{x \in \mathcal{X}} \big\{  x -  h_{\text{MRC}}^{*} y_{\text{MRC}} /( |h_{\text{MRC}}^{*}|^2 + N_{0}) \big\}$, where $h_{\text{MRC}}$ and $y_{\text{MRC}}$ are the equivalent channel response and observation, respectively,  after maximal-ratio combining at the destination.} 
	{ \color{black} The MMSE detector is one of the most practical linear-complexity detector. We can see that the non-coherent NMLD largely outperforms the coherent MMSE detector at SERs below $10^{-2}$. The MMSE detector ignores the relay detection errors, which results in its poor performance.  For each value of $N$, it can be observed that the SERs of the non-coherent NMLD and coherent AMLD decay at nearly the same rate, but the former suffers about $3$ dB performance loss as compared to the later.} It was proved in \cite{cmrc-wang2007high} that if we assume the instantaneous CSI of all links to be available at the destination,  coherent detection can achieve the full diversity order $N+1$. Unlike this,  the coherent AMLD considered here assumes the availability of the average (instead of the instantaneous) CSI of the source-relay links at the destination. It is seen from Fig.~\ref{fig:comparison-with-coherent} that the coherent AMLD achieves the diversity order $\left \lceil \frac{N}{2} \right \rceil+1$.
%	 It was shown to achieve the maximum achievable diversity order $\left \lceil \frac{N}{2} \right \rceil+1$ through extensive simulation in \cite{ours-twc}. This coincides with the results in Fig.~\ref{fig:comparison-with-coherent}.
	  Moreover,  the non-coherent NMLD is shown to  achieve the same diversity order $\left \lceil \frac{N}{2} \right \rceil+1$ as the coherent AMLD. 
%	The results suggest that the coherent and non-coherent setups will cause difference in terms of the coding gain but not the diversity order, and that the diversity order loss is possibly due to the common system assumption of the average CSI of the source-relay links. 
	%The results of the diversity order analysis in Proposition \ref{prop: diversity-perfect-relays} and \ref{prop: diversity-error-relays} for the non-coherent network can be transferred to the coherent network since the SER performance of these two types of networks may differ mainly in coding gain. 

	%--------------------------------------------
	%--------------------------------------------

	%%--------------------------------------------------------------------------------------------------------
	%%--------------------------------------------------------------------------------------------------------

	\section{  Conclusion}
	% $\tilde{y}$
	%\textcolor{red}{till here}
	In this work, we proposed a new detector, i.e., the NMLD, for the non-coherent relay network with $N$ parallel DF relays using GDM based on $M$-PSK. An accurate SER expression was then derived, and based on which an optimized transmission power allocation scheme was designed. Simulation results showed that the proposed non-coherent scheme can greatly reduce the performance gap between  the conventional non-coherent DM scheme and the coherent counterpart. 
	For example, for a block length of $256$ symbols at SER $10^{-5}$ using DQPSK in the single relay case, the gap was reduced from approximately $3$ dB to within $0.5$ dB. Moreover, by analyzing the behavior of the SER expression at high SNR, it was proved that the diversity order of both the proposed NMLD and the state-of-the-art AMLD is exactly $\left \lceil \frac{N}{2} \right \rceil+1$. The accuracy of the diversity order was further confirmed by extensive simulation, revealing that the full diversity order $N+1$ is not achievable for $N>1$. 	
%	We have also observed from simulation that the proposed non-coherent scheme can achieve the same diversity order $\left \lceil \frac{N}{2} \right \rceil+1$ as its coherent counterpart in various relay channel scenarios, such as {\color{black}with $N \in \{1,2,3,4,6 \}$ relays}. 

%	{\color{black} As future research directions, we plan to extend our results for the coherent DF relay networks,  and some promising preliminary results have been reported in \cite{ours-icassp}.  In this case, quadrature amplitude modulation (QAM) is more generally used than PSK considered here. However, performance analysis for QAM signals is much more challenging especially with multiple relays, and this will be left for our future investigation. }
	
% SWIPT-enabled relay networks

 %%--------------------------------------------------------------------------------------------------------
	%%--------------------------------------------------------------------------------------------------------
	\appendix
	\subsection{Expression of $\epsilon$}
	\label{app:SER-SRlink}
	%Based on \eqref{eq:receive-p2p}, the instantaneous receive SNR can be approximated as $\frac{(\rho_{\rm{T}}/\rho_{\rm{R}}) |y_{{I}, {J}}[k_2]|^2}{(1+\rho_{\rm{T}}/\rho_{\rm{R}}) N_{{I}, {J}}}
	%\approx  \phi_{\rm{T}} \gamma_{{I}, {J}}$, where $\gamma_{{I}, {J}} \triangleq \frac{|h_{{I}, {J}}|^2 }{N_{{I}, {J}}}$ is the instantaneous link SNR. We denote $\bar{\gamma}_{{I}, {J}} = \mathbb{E}_{ |h_{{I}, {J}}| } [\gamma_{{I}, {J}}]$ as the average link SNR. 
	An expression of the average SER for $M$-PSK using coherent detection over a Rayleigh fading channel is obtained in \cite[eq. (8.112)]{simon2005digital} as 	$	\mathcal{P}^{\rm{co}}_{\rm{e}} (\bar{\gamma})=  \frac{M-1}{M} -\sqrt{\frac{\bar{\gamma} \sin^2(\frac{\pi}{M})}{1+\bar{\gamma} \sin^2(\frac{\pi}{M}) }}  \times \Big[
	\frac{1}{2} + \frac{1}{\pi} \taninv \Big(
	\sqrt{\frac{\bar{\gamma} \sin^2(\frac{\pi}{M}) }{1+\bar{\gamma} \sin^2(\frac{\pi}{M}) }} \\ \cot \frac{\pi}{M}
	\Big)
	\Big]$, 
%	\begin{IEEEeqnarray}{rCl}
%		\mathcal{P}^{\rm{co}}_{\rm{e}} (\bar{\gamma})=  \frac{M-1}{M} -\sqrt{\frac{\bar{\gamma} \sin^2(\frac{\pi}{M})}{1+\bar{\gamma} \sin^2(\frac{\pi}{M}) }}  \times \left[
%		\frac{1}{2} + \frac{1}{\pi} \taninv \left(
%		\sqrt{\frac{\bar{\gamma} \sin^2(\frac{\pi}{M}) }{1+\bar{\gamma} \sin^2(\frac{\pi}{M}) }} \cot \frac{\pi}{M}
%		\right)
%		\right]	\nonumber
%	\end{IEEEeqnarray}
	with $\bar{\gamma}$ denoting the average SNR. By using {\color{black} the average receive SNR } $  \phi_{\rm{T}} \bar{\gamma}_{s,r}= \frac{1}{1/\rho_{\rm{T}} + 1/\rho_{\rm{R}}}\bar{\gamma}_{s,r}$, $\rm{T} \in \{\rm{R},\rm{N}\}$, {\color{black} obtained in \eqref{eq:approx-high-noise} }to replace $\bar{\gamma}$ in this expression, we obtain
%	$\epsilon =	\epsilon (\rho_{\rm{R}}, \rho_{\rm{N}}) = 
%	(1/L)\mathcal{P}^e (1, \rho_{\rm{R}}) + (1-1/L) \mathcal{P}^e \left( \sqrt{\rho_{\rm{N}}/\rho_{\rm{R}} },\rho_{\rm{R}}  \right),$
	\begin{IEEEeqnarray}{rCl} \label{eq:ser-relay}
		\epsilon =	 
		(1/L)\mathcal{P}^{\rm{co}}_{\rm{e}} (\phi_{R}\bar{\gamma}_{s,r}) + (1-1/L) \mathcal{P}^{\rm{co}}_{\rm{e}} \left( \phi_{\rm{N}}\bar{\gamma}_{s,r} \right).
	\end{IEEEeqnarray}
%	\begin{IEEEeqnarray}{rCl} % \label{eq:ser-relay}
%		\mathcal{P}^e (\sqrt{\rho_{\rm{T}}/\rho_{\rm{R}}}, \rho_{\rm{R}})= & \frac{M-1}{M} -\sqrt{\frac{\phi_{\rm{T}} \bar{\gamma}_{s,r} \sin^2(\frac{\pi}{M})}{1+ \phi_{\rm{T}} \bar{\gamma}_{s,r} \sin^2(\frac{\pi}{M}) }}  \times \left[
%		\frac{1}{2} + \frac{1}{\pi} \taninv \left(
%		\sqrt{\frac{ \phi_{\rm{T}} \bar{\gamma}_{s,r} \sin^2(\frac{\pi}{M}) }{1+ \phi_{\rm{T}} \bar{\gamma}_{s,r} \sin^2(\frac{\pi}{M}) }} \cot \frac{\pi}{M}
%		\right)
%		\right]	. \nonumber
%	\end{IEEEeqnarray} 
%
%\vspace{-14mm}	%%-----------------------------------------
	\subsection{Monotonicity of the Second Term in \eqref{eq:PE}}
	\label{app-mono-PE}
%		$\frac{\partial  \log \epsilon(z)}{\partial z} = \frac{1}{\epsilon} \frac{\partial  \epsilon(z)}{\partial z} < 0$
%		
%		
%	$(\rho_{\rm{T}}/\rho_{\rm{R}}) = \frac{\rho_{\rm{N}}}{\rho_{\rm{R}}}$, $\epsilon$ decreasing, 	$\eta \triangleq \left(1+\rho_{\rm{T}}/\rho_{\rm{R}}\right)\log\frac{(1-\epsilon)(M-1)}{\epsilon}$, to prove $\tilde{\mathcal{P}}^{\rm{E}}(\gamma_{s,d})$ decreasing.
	
%	Denote $z = \sin \left( \frac{\pi}{M}\right) \sqrt{2 \phi_{\rm{N}}} > 0$,
Denote $z = \sin \left( \frac{\pi}{M}\right) \sqrt{2 \phi_{\rm{N}}} > 0$ and $q(z) = z
\sqrt{ \gamma_{s,d}} - \frac{ \log \frac{(1-\epsilon)(M-1)}{\epsilon} }{2  z \sqrt{  \gamma_{s,d}} } $. The second term in \eqref{eq:PE} can be written as $\tilde{\mathcal{P}}_{\rm{N}}^{\rm{E}}(z) = \frac{2 \epsilon}{M-1} Q(q(z))$ for NSs.\footnote{For presentational convenience, $M>2$ is considered, but the analysis of $M=2$ is only slightly different.} For $\tilde{\mathcal{P}}_{\rm{N}}^{\rm{E}}(z)$ to be decreasing in $z$ (equivalently, in $\phi_{\rm{N}}$), by taking derivative, a sufficient and necessary condition is obtained as 
%\begin{IEEEeqnarray}{rCl}
%	\frac{\partial \tilde{\mathcal{P}}^{\rm{E}}(z) }{ \frac{2}{M-1} \partial z} =  \frac{\partial \epsilon}{\partial z}  Q \left(q(z)
%	\right) + \epsilon \partial  Q ( q(z)
%	)/  \partial z < 0 . \label{eq:app-B-cd-1}
%\end{IEEEeqnarray}
$	\frac{\partial \tilde{\mathcal{P}}_{\rm{N}}^{\rm{E}}(z) }{ \frac{2}{M-1} \partial z} =  \frac{\partial \epsilon}{\partial z}  Q \left(q(z)
\right) + \epsilon \partial  Q ( q(z)
)/  \partial z < 0$. 
Since $\frac{ \log \frac{(1-\epsilon)(M-1)}{\epsilon}  }{2  z^2 \gamma_{s,d} } \to \infty \implies \frac{ \log \frac{(1-\epsilon)(M-1)}{\epsilon}  }{2  z^2 \gamma_{s,d} } \gg 1 \iff   \frac{ \log \frac{(1-\epsilon)(M-1)}{\epsilon} }{2  z \sqrt{  \gamma_{s,d}} } \gg z \sqrt{ \gamma_{s,d}}$, we have 
$q(z) \approx - \frac{ \log \frac{(1-\epsilon)(M-1)}{\epsilon} }{2 z  \sqrt{  \gamma_{s,d}} } < 0 $. By applying this to the derivative $\frac{\partial \tilde{\mathcal{P}}_{\rm{N}}^{\rm{E}}(z) }{ \frac{2}{M-1} \partial z}$ and then using $Q(x) \approx \frac{1}{2}e^{-\frac{x^2}{2}}$, $x > 0$, the sufficient and necessary condition can be written as % and after some calculations,   
 \begin{IEEEeqnarray}{rCl}
% 	& - \frac{\partial  \epsilon}{\partial z} + \frac{  - 2   \log \frac{(1-\epsilon)(M-1)}{\epsilon}  \frac{\partial \epsilon}{\partial z} z - 2 \epsilon \log^2 \frac{(1-\epsilon)(M-1)}{\epsilon}   }{8z^3\gamma_{s,d}} + \nonumber \\ &  \frac{\partial \epsilon}{\partial z} / ( \frac{1}{2} \exp ( - \log^2 \frac{(1-\epsilon)(M-1)}{\epsilon} / (8z^2\gamma_{s,d})  )  )  < 0 \label{eq:app-B-1} \\
 	& -\frac{\epsilon \log^2 \frac{(1-\epsilon)(M-1)}{\epsilon}}{4z^3\gamma_{s,d}} +  \big[  2 \exp \big(
 	\frac{\log^2 \frac{(1-\epsilon)(M-1)}{\epsilon}}{8z^2\gamma_{s,d}}
 	\big) - \frac{    \log \frac{(1-\epsilon)(M-1)}{\epsilon}  }{ 4 z^2 \gamma_{s,d} } -1
 	\big] \frac{\partial  \epsilon}{\partial z} < 0 \label{eq:app-B-1} .
 \end{IEEEeqnarray}
 Note that $\frac{\partial  \epsilon}{\partial z} < 0 $. By ignoring the first  summation term (which is negative) in the left hand side (LHS) of \eqref{eq:app-B-1} and rearranging the terms, a sufficient condition is obtained as $\sqrt{M-1} ^ {-\frac{z_0}{2 }  } \Big(1 + \frac{z_0}{2 }   \Big) <  2 \epsilon^{-\frac{z_0}{4 }  }$, % after some calculations, 
% \begin{IEEEeqnarray}{rCl}
% 	\sqrt{M-1} ^ {-\frac{    \log \frac{(1-\epsilon)(M-1)}{\epsilon}  }{ 4 z^2 \gamma_{s,d} }  } (1 + \frac{    \log \frac{(1-\epsilon)(M-1)}{\epsilon}  }{ 4 z^2 \gamma_{s,d} }  ) <  2 \epsilon^{-\frac{    \log \frac{(1-\epsilon)(M-1)}{\epsilon}  }{ 8 z^2 \gamma_{s,d} }  }. \label{eq:app-B-2}
% \end{IEEEeqnarray}
%\begin{IEEEeqnarray}{rCl}
%	\sqrt{M-1} ^ {-\frac{z_0}{2 }  } \Big(1 + \frac{z_0}{2 }   \Big) <  2 \epsilon^{-\frac{z_0}{4 }  }  \label{eq:app-B-2},
%\end{IEEEeqnarray}
where $z_0=\frac{ \log \frac{(1-\epsilon)(M-1)}{\epsilon}  }{2  z^2 \gamma_{s,d} } $ (for notational convenience). Next, we prove that this sufficient condition holds true when  $z_0 \to \infty $. For the LHS, we have $\lim\limits_{z_0 \to \infty} \sqrt{M-1} ^ {-\frac{z_0}{2 }  } \Big(1 + \frac{z_0}{2 }   \Big)  = 0$. For the right hand side, since $0 < \epsilon < 1$ and $z_0 > 0$, we have $ 2 \epsilon^{-\frac{z_0}{4 }  } > 2$. Therefore, the sufficient condition holds true when  $z_0 \to \infty $. Overall, provided that $\frac{ \log \frac{(1-\epsilon)(M-1)}{\epsilon}  }{2  z^2 \gamma_{s,d} } \to \infty $ holds true, $\tilde{\mathcal{P}}_{\rm{N}}^{\rm{E}}(z)$ is monotonically decreasing in $z$, and equivalently, in $\phi_{\rm{N}}$. 
%	However, since $\gamma_{s,d}$ is not known by any node, we are motivated to introduce a more practical sufficient condition without $\gamma_{s,d}$.
	
%	Assume $\gamma_{s,d}$ has a finite value, then $z_1 \approx - (\log \epsilon)/z^2$, based on which \eqref{eq:PE-condi} can be approximated as $ 1 - (\log \epsilon)/z^2  <   \frac{2}{\sqrt{\frac{M-1}{\epsilon}}^{ (\log \epsilon)/z^2 }} = $
	
%	$\frac{\partial z_0}{\partial z} = \frac{  2   \log \frac{M-1}{\epsilon} (- \frac{1}{\epsilon} \frac{\partial \epsilon}{\partial z}) (8z^2\gamma_{s,d}) - \log^2 \frac{M-1}{\epsilon}  16 z \gamma_{s,d} }{(8z^2\gamma_{s,d})^2}$, \\
%	$\exp(-z_0) = \exp (- \log^2 \frac{M-1}{\epsilon} / (8z^2\gamma_{s,d})) <  \exp (- \log \frac{M-1}{\epsilon} / (8z^2\gamma_{s,d})) = (  \frac{M-1}{\epsilon}  ) ^{ - \frac{1}{8z^2\gamma_{s,d}}  }$, \\
%	$\exp(-z_0) = \exp (- \log^2 \frac{M-1}{\epsilon} / (8z^2\gamma_{s,d})) = (  \frac{M-1}{\epsilon}  ) ^{ -\log \frac{M-1}{\epsilon} \frac{1}{8z^2\gamma_{s,d}}  } $ 
% \vspace{-3mm}
	%--------------------------------------
	\subsection{Relationship between the NMLD and the PLD}
	\label{app:NMLD and PLD}
	Note that for a clear comparison, we follow the notations and definitions in \cite{ref1-bhatnagar2012decode} as closely as possible. Let $x_p$ and $x_q$ denote a given pair of any two different symbols belonging to $\mathcal{X}$. By reducing the candidate set of the relay symbols from $\mathcal{X}^N$ to $\{x_p, x_q\}^N$ for $M>2$, the NMLD detection metric in \eqref{eq:NMLD-general} can be approximated using $\beta^{I,J}$, $I \in \{ p,q\}$, $J = \{ p,q\} \setminus I$, as  $	\beta^{I,J} = \frac{1}{N_{s,d}}|y_{s,d}[k]-y_{s,d}[k-1]x_I|^2 + \sum_{n=1}^{N} \min \left\{
	\frac{1}{N_{r_{n},d}}|y_{r_{n},d}[k]-y_{r_{n},d}[k-1]x_I|^2,  \right.  \left. \min_{ \substack{x_{r_{n}} \in \{x_p, x_q\} }} \frac{1}{N_{r_{n},d}} |y_{r_{n},d}[k]-y_{r_{n},d}[k-1]x_{r_{n}}|^2 +\eta_n 
	\right\}$
%	\vspace{-6mm}
	%  (or equivalently, by only considering the error in decoding of $x_p$ as $x_q$ at the relay)
%	\begin{IEEEeqnarray}{rCl}	
%	\small	
%	\beta^{I,J} 
%	& = &  \frac{1}{N_{s,d}}|y_{s,d}[k]-y_{s,d}[k-1]x_I|^2 + \sum_{n=1}^{N} \min \left\{
%	\frac{1}{N_{r_{n},d}}|y_{r_{n},d}[k]-y_{r_{n},d}[k-1]x_I|^2,  \right.  \nonumber  \\
%	& & \left. \min_{ \substack{x_{r_{n}} \in \{x_p, x_q\} }} \frac{1}{N_{r_{n},d}} |y_{r_{n},d}[k]-y_{r_{n},d}[k-1]x_{r_{n}}|^2 +\eta_n % 2\sigma_{s,d}^2
%	\right\}
%	\nonumber  
%	\\
%	& = & \frac{1}{N_{s,d}}|y_{s,d}[k]-y_{s,d}[k-1]x_I|^2 + \sum_{n=1}^{N} \min \left\{
%	\frac{1}{N_{r_{n},d}}|y_{r_{n},d}[k]-y_{r_{n},d}[k-1]x_I|^2,  \right.  \nonumber \\
%	& & \left.  \frac{1}{N_{r_{n},d}}|y_{r_{n},d}[k]-y_{r_{n},d}[k-1]x_J|^2 +\eta_n % 2\sigma_{s,d}^2
%	\right\}
%	\nonumber . 
%	\end{IEEEeqnarray}
	After some algebra, we have
	\begin{IEEEeqnarray}{rCl} \label{eq:gxp}
	\beta^{p,q} =  & \frac{1}{N_{s,d}}|y_{s,d}[k]-y_{s,d}[k-1]x_p|^2+ \sum_{n=1}^{N}\begin{cases}
	\frac{1}{N_{r_{n},d}}|y_{r_{n},d}[k]-y_{r_{n},d}[k-1]x_p|^2, & \text{ if }  t_n \geq - \frac{\eta_n}{2}  , \IEEEeqnarraynumspace \\
	\frac{1}{N_{r_{n},d}}|y_{r_{n},d}[k]-y_{r_{n},d}[k-1]x_q|^2 +\eta_n , & \text{ if }   t_n < - \frac{\eta_n}{2}, \IEEEeqnarraynumspace
	\end{cases} 
	\end{IEEEeqnarray}
	and
	\begin{IEEEeqnarray}{rCl} \label{eq:gxq}
	\beta^{q,p} 
	= & \frac{1}{N_{s,d}}|y_{s,d}[k]-y_{s,d}[k-1]x_q|^2+\sum_{n=1}^{N}\begin{cases}
	\frac{1}{N_{r_{n},d}}|y_{r_{n},d}[k]-y_{r_{n},d}[k-1]x_q|^2, & \text{ if }  t_n \leq  \frac{\eta_n}{2} , \IEEEeqnarraynumspace \\
	\frac{1}{N_{r_{n},d}}|y_{r_{n},d}[k]-y_{r_{n},d}[k-1]x_p|^2 +\eta_n , & \text{ if }   t_n > \frac{\eta_n}{2} , \IEEEeqnarraynumspace
	\end{cases}
	\end{IEEEeqnarray}
	where $t_n = \frac{1}{N_{r_{n},d}}\Re \{ y_{r_{n},d}^*[k] y_{r_{n},d}[k-1] (x_p-x_q) \}$.
	Then, based on \eqref{eq:gxp} and \eqref{eq:gxq}, the detection metric of the PLD is obtained as
	% same detection metric as that of PLD in \cite[eq. (26)]{bhatnagar2011ml} \hl{confirm citation} as % \hl{no inline for cases }
	\begin{IEEEeqnarray}{rCl}
	\small
	\Lambda^{p,q} \triangleq \beta^{q,p}- \beta^{p,q}=  t_0 + \sum_{n=1}^{N} \begin{cases}
	- \frac{\eta_n}{2}, & \text{ if }  t_n < - \frac{\eta_n}{2} , \\
	t_n  & \text{ if }   - \frac{\eta_n}{2} \leq t_n \leq \frac{\eta_n}{2} , \\
	\frac{\eta_n}{2} , & \text{ if }   t_n > \frac{\eta_n}{2} ,
	\end{cases} \nonumber
	\end{IEEEeqnarray} 
	where $t_0 = \frac{1}{N_{s,d}}\Re \{ y_{s,d}^*[k] y_{s,d}[k-1] (x_p-x_q) \}$ and $\Lambda^{p,q}$ is the log-likelihood ratio between $x_p$ and $x_q$. Note that when $M=2$, the detection metrics of the PLD and the NMLD are exact the same.  
	%------------------------------
	\subsection{\color{black} Approximate Average SER and Diversity Order Analysis for Error-Free Relaying}
	\label{proof-prop-1}	
	 $\mathcal{P}^{\rm{C}}_{\rm{e}}$ in \eqref{eq:bound-general-C} can be written  as  $		 2(1-\epsilon)^{N}\sum_{m=0}^{N-1} \binom{N}{m} \int_0^{\infty} 
	Q \left( \sin \left( \frac{\pi}{M}\right) \sqrt{\gamma_m} + \frac{(N-m) \eta}{4\sin \left( \frac{\pi}{M}\right)} \frac{1}{\sqrt{\gamma_m}}
	\right)  p (\gamma_m )  \dd \gamma_m $,
%	\begin{IEEEeqnarray}{rCl}
%		 2(1-\epsilon)^{N}\sum_{m=0}^{N} \binom{N}{m} \int_0^{\infty} 
%		Q \left( \sin \left( \frac{\pi}{M}\right) \sqrt{\gamma_m} + \frac{(N-m) \eta}{4\sin \left( \frac{\pi}{M}\right)} \frac{1}{\sqrt{\gamma_m}}
%		\right)  p (\gamma_m )  \dd \gamma_m , \label{eq:appA-1}
%	\end{IEEEeqnarray}
	with $p (\gamma_m ) =  \frac{(\gamma_m)^{m}}{\Gamma(m+1)\bar{\gamma}_m^{m+1} } e^{-\frac{\gamma_m}{\bar{\gamma}_m}}.$
	By applying $Q(x) \approx \frac{1}{2}e^{-\frac{x^2}{2}}$ to the integral term in $\mathcal{P}^{\rm{C}}_{\rm{e}}$, we have
	\begin{IEEEeqnarray}{rCl} 
		& & \int_0^{\infty} Q \left(
		\sin \left( \frac{\pi}{M}\right) \sqrt{\gamma_m} + \frac{(N-m) \eta}{4\sin \left( \frac{\pi}{M}\right)} \frac{1}{\sqrt{\gamma_m}}
		\right)  p (\gamma_m )  \dd \gamma_m \nonumber \\
		& \approx & \frac{1}{2\Gamma(m+1)\bar{\gamma}^{m+1} } \int_0^{\infty}
		\exp 	\Big(-
		\frac{(2 \sin^2 \left( \frac{\pi}{M}\right) \gamma_m  + (N-m) \eta/2)^2}{ 8 \sin^2 \left( \frac{\pi}{M}\right) \gamma_m}
		\Big) (\gamma_m)^{m} e^{-\frac{\gamma_m}{\bar{\gamma}_m}} \dd \gamma_m. \label{eq:appA-3}
	\end{IEEEeqnarray}
	For the integral term in \eqref{eq:appA-3} we have
	
	\begin{IEEEeqnarray}{rCl}
%	\begin{equation}
%	\begin{aligned}
	&	& \int_0^{\infty}
		\exp 	\Big(-
		\frac{(2 \sin^2 \left( \frac{\pi}{M}\right) \gamma_m  + (N-m) \eta/2)^2}{ 8 \sin^2 \left( \frac{\pi}{M}\right) \gamma_m}
		\Big) (\gamma_m)^{m} e^{-\frac{\gamma_m}{\bar{\gamma}_m}} \dd \gamma_m \nonumber \\
	&	= & \frac{2^{-2m-1}((N-m) \eta)^{m+1}  \exp (-(N-m)\eta/4)  }{(2 \sin^2 \left( \frac{\pi}{M}\right) ( \sin^2 \left( \frac{\pi}{M}\right) /2+\bar{\gamma}_m^{-1}))^{\frac{m+1}{2}}}K_{m+1}\Big(
		\frac{(N-m) \sqrt{ \sin^2 \left( \frac{\pi}{M}\right) /2+\bar{\gamma}_m^{-1}} \eta}{2 \sqrt{2} \sin  \left( \frac{\pi}{M}\right) }
		\Big) 
		\label{eq:prop1-1} \\
	&	\approx & \sqrt{\pi}2^{-2m} \sin^{-m-\frac{1}{2}} \left( \frac{\pi}{M}\right) ((N-m) \eta)^{m+\frac{1}{2}}( \sin^2 \left( \frac{\pi}{M}\right) /2+\bar{\gamma}_m^{-1})^{-\frac{m}{2}-\frac{3}{4}} \exp (-(N-m)\eta/4) \nonumber \\
	&	& \exp \big(-
		(N-m) \sqrt{ \sin^2 \left( \pi/M\right) /2+\bar{\gamma}_m^{-1}}\eta/(2 \sqrt{2} \sin  \left( \pi/M\right) )
		\big)   \label{eq:prop1-2}\\
%		\propto &  \eta^{m+\frac{1}{2}} \exp \left( -(N-m)\eta /4 
%		\right) \exp (-(N-m)\eta/4) \nonumber \\
	&	\propto & ( \ln \bar{\gamma}_m )^{m+\frac{1}{2}}
		(\bar{\gamma}_m)^{-(N-m)}. \label{appA-2}
%	\end{aligned}
%	\end{equation}
	\end{IEEEeqnarray}
	\eqref{eq:prop1-1} is obtained based on \cite{book1-gradshteyn2014table}, { and we obtain \eqref{eq:PeC-ave}}. Next, we show how to obtain \eqref{eq:prop1-2}. By applying the binomial theorem, we have
	$ \lim\limits_{\bar{\gamma}_m \to \infty}( \sin^2 \left( \frac{\pi}{M}\right) /2+\bar{\gamma}_m^{-1})^{1/2} = ( \sin^2 \left( \frac{\pi}{M}\right) /2)^{1/2} + \frac{1}{2}( \sin^2 \left( \frac{\pi}{M}\right) /2)^{-1/2}\bar{\gamma}_m^{-1} $. Then for the term inside the Bessel function $K_{m+1}(\cdot)$, we have $\lim\limits_{\bar{\gamma}_m \to \infty} (N-m)\eta ( \sin^2 \left( \frac{\pi}{M}\right) /2+\bar{\gamma}_m^{-1})^{1/2} \\ \sin^{-1} \left( \frac{\pi}{M}\right)  = \lim\limits_{\bar{\gamma}_m \to \infty} c_0  (( \sin^2 \left( \frac{\pi}{M}\right)  /2)^{1/2}  + \frac{1}{2}( \sin^2 \left( \frac{\pi}{M}\right) /2)^{-1/2}\bar{\gamma}_m^{-1} ) \ln \bar{\gamma}_m = c_0 ( \sin^2 \left( \frac{\pi}{M}\right) /2)^{1/2} \ln \bar{\gamma}_m \gg 0$ with $c_0$ denoting some positive constant. Therefore based on $K_1(x) \approx (\pi/2)^{1/2} x^{-1/2} \exp(-x)$ when $x \gg 0$, we can approximate the $K_{m+1}(\cdot)$ term and obtain \eqref{eq:prop1-2}. Finally, accordingly to  $\mathcal{P}^{\rm{C}}_{\rm{e}}$ and \eqref{appA-2}, it is concluded that for error-free relaying, $\mathcal{P}^{\rm{C}}_{\rm{e}}$ decays with $ ( \ln \bar{\gamma}_m )^{m+\frac{1}{2}}
	(\bar{\gamma}_m)^{-(N-m)}  (\bar{\gamma}_m)^{-(m+1)}=( \ln \bar{\gamma}_m )^{m+\frac{1}{2}} (\bar{\gamma}_m)^{-(N+1)}$, and the proposed NMLD achieves the full diversity order $N+1$.
	
	%  \vspace{-6.1mm}
	\subsection{\color{black} Approximate Average SER and Diversity Order Analysis for  Erroneous Relaying}
	\label{proof-prop-2}                  
	First, we analyze the case when $2N^{\rm{E}}-N > 0$, i.e., the number of erroneous relays is larger than that of error-free ones. By applying  $Q(x) \approx \frac{1}{2}e^{-\frac{x^2}{2}}$, we have
	\begin{IEEEeqnarray}{rCl}
		& & \frac{1}{\bar{\gamma}_{s,d}} \int_{0}^{\infty} Q \left(
		\sqrt{\frac{c_1}{2}} \sqrt{\gamma_{s, d}}  - \frac{(2N^{\rm{E}}-N) \eta}{ 2 \sqrt{2c_1}} \frac{1}{\sqrt{\gamma_{s, d}}}
		\right) 
		\exp\left(-\frac{\gamma_{s, d}  }{\bar{\gamma}_{s,d}}\right) \dd \gamma_{s, d} \nonumber \\
		& \approx & \frac{1}{\bar{\gamma}_{s,d}} \int_{0}^{\frac{(2N^{\rm{E}}-N) \eta}{2c_1}\nonumber }  
		\exp\left(-\frac{\gamma_{s, d}  }{\bar{\gamma}_{s,d}}\right)  \dd \gamma_{s, d} +  \frac{1}{2\bar{\gamma}_{s,d}} \int_{\frac{(2N^{\rm{E}}-N)  }{2c_1} \eta}^{\infty}  
		\exp \left(-
		\frac{ ( c_1 \gamma_{s, d} - \frac{2N^{\rm{E}}-N}{2} \eta)^2}{4 c_1  \gamma_{s, d}  }
		\right)
		\exp	\Big(-\frac{\gamma_{s, d}  }{\bar{\gamma}_{s,d}}	\Big)  \dd \gamma_{s, d} \nonumber
%		 \\
%		& = & Z_1+ Z_2, \nonumber
	\end{IEEEeqnarray}
 where $c_1 = \Re \{ 1-x_1^*x_v\} > 0$. $Z_1$ and $Z_2$ are used to represent the first and second summation terms, respectively. {	{We have $	Z_1 = 1 -\exp \left( -\frac{(2N^{\rm{E}}-N) \eta}{2 c_1\bar{\gamma}_{s,d}}\right) $,} and 
	$ \lim\limits_{\bar{\gamma}_{s,d}  = \bar{\gamma}  \to \infty}
	Z_1 =  1-\lim\limits_{\bar{\gamma}   \to \infty}\exp \left( -\frac{(2N^{\rm{E}}-N) \eta}{2 c_1\bar{\gamma}}\right) = 1-\exp \left( -c_2\lim\limits_{\bar{\gamma}   \to \infty} \frac{\ln \bar{\gamma} }{\bar{\gamma}}\right) =  0
	$ with $c_2$ denoting some positive constant. Now we deal with $Z_2$, which can be written as
	\begin{IEEEeqnarray}{rCl}
		Z_2 
		& \approx &  \frac{1}{2\bar{\gamma}_{s,d}} \int_{0}^{\infty} 
		\exp 	\Big(-
		\frac{ ( c_1 \gamma_{s,d} - \frac{2N^{\rm{E}}-N}{2} \eta)^2}{4 c_1 \gamma_{s,d} }
			\Big)
		\exp	\Big(-\frac{\gamma_{s,d}  }{\bar{\gamma}_{s,d}}	\Big)  \dd \gamma_{s,d} \label{eq:prop2-1}\\
		& = &  \frac{(2N^{\rm{E}}-N) (4c_1)^{-1/2} \eta}{ 2 \left(c_1/4+\bar{\gamma}_{s,d}^{-1}\right)^{1/2}\bar{\gamma}_{s,d}} K_1 	\Big(
		\frac{(2N^{\rm{E}}-N)\sqrt{c_1/4+\bar{\gamma}_{s,d}^{-1}}}{2\sqrt{c_1}} \eta
			\Big) \exp \left((2N^{\rm{E}}-N)\eta /4\right) \label{eq:W2-1}  \\
		& = & V_1 V_2 V_3, \nonumber
	\end{IEEEeqnarray}
	where $V_1 \triangleq  \frac{(2N^{\rm{E}}-N) (4c_1)^{-1/2} \eta}{ 2 \left(c_1/4+\bar{\gamma}_{s,d}^{-1}\right)^{1/2}\bar{\gamma}_{s,d}}, V_2 \triangleq K_1(\cdot) \text{ and } V_3 \triangleq \exp \left((2N^{\rm{E}}-N)\eta /4\right) $. \eqref{eq:prop2-1} is obtained by approximating $Z_2$ using an upper bound. It can be numerically shown that the value of \eqref{eq:prop2-1} is very close to $Z_2$, and therefore it is an accurate approximation. Similarly to \eqref{eq:prop1-1}, \eqref{eq:W2-1} is obtained based on \cite{book1-gradshteyn2014table}, { and we obtain the expression in the first curly bracket of \eqref{eq:PeE-ave}}. Now we analyze the diversity order. As $\bar{\gamma}_{s,d} \to \infty$, based on $K_1(z) \approx (\pi/2)^{1/2} z^{-1/2} \exp(-z)$ when $z \gg 0$, 
	we can get
	%\begin{IEEEeqnarray}{rCl}
	%	V_2 \triangleq & K_1 \left(
	%	2B\eta (4c_1)^{-1/2} \left( c_1/4+\bar{\gamma}^{-1} \right)^{1/2}
	%	\right) \\
	%	\approx  &  (\pi/4)^{1/2} (B\eta)^{-1/2}C^{1/4} \left( c_1/4+\bar{\gamma}^{-1} \right)^{-1/4} \exp \left(
	%	-2B\eta (4c_1)^{-1/2}(c_1/4+\bar{\gamma}^{-1})^{1/2}
	%	\right)\\
	%	\approx  & (\pi/2)^{1/2} (4B \ln \bar{\gamma}  (1/4+(2A)^{-1} \bar{\gamma}^{-1}))^{-1/2} \exp (-4B \ln \bar{\gamma}  (1/4+(2A)^{-1} \bar{\gamma}^{-1})) \\
	%	\approx & (\pi/2)^{1/2}(2A^{-1}B \frac{\ln \bar{\gamma}}{ \bar{\gamma}})^{-1/2} \bar{\gamma}^{2A^{-1}B\bar{\gamma}^{-1}} 
	%\end{IEEEeqnarray}
	\begin{IEEEeqnarray}{rCl}
		V_2 
		& \approx  &  (\pi/4)^{1/2} \left((2N^{\rm{E}}-N)\eta/2 \right)^{-1/2}(4 c_1)^{1/4} \left( c_1/4+\bar{\gamma}_{s,d}^{-1} \right)^{-1/4} \exp 	\Big( -
		\frac{(2N^{\rm{E}}-N)\sqrt{c_1/4+\bar{\gamma}_{s,d}^{-1}}}{2\sqrt{c_1}} \eta
			\Big), \nonumber \\
		V_1V_2 
%		& \approx & (\pi/4)^{1/2} \left((2N^{\rm{E}}-N)\eta/2 \right)^{1/2} (4 c_1)^{-1/4} \left( c_1/4+\bar{\gamma}^{-1} \right)^{-3/4} \bar{\gamma}^{-1} \exp \left( -
%		\frac{(2N^{\rm{E}}-N)\sqrt{c_1/4+\bar{\gamma}^{-1}}}{2\sqrt{c_1}} \eta
%		\right) \nonumber \\	
		& \propto & ( \ln \bar{\gamma} )^{\frac{1}{2}} \left( c_1/4+\bar{\gamma}_{s,d}^{-1} \right)^{-3/4} \bar{\gamma}_{s,d}^{-1} (\bar{\gamma}_{s,d})^{ 
			- 2(2N^{\rm{E}}-N) (4c_1)^{-1/2}(c_1/4+\bar{\gamma}_{s,d}^{-1})^{1/2} } \nonumber \\
%		\propto & ( \ln \bar{\gamma} )^{\frac{1}{2}} \left( c_1/4 \right)^{-3/4} \bar{\gamma}^{-1} (\bar{\gamma})^{ 
%			-2(2N^{\rm{E}}-N)  (4c_1)^{-1/2}(c_1/4+\bar{\gamma}^{-1})^{1/2} } \nonumber \\
		& \propto & ( \ln \bar{\gamma} )^{\frac{1}{2}} \bar{\gamma}_{s,d}^{-1} (\bar{\gamma}_{s,d})^{-\frac{2N^{\rm{E}}-N}{2}}. \nonumber
	\end{IEEEeqnarray}
	It is straightforward to show that 
	$
	V_3  \propto  (\bar{\gamma}_{s,d})^{\frac{2N^{\rm{E}}-N}{2}}
	$, and then we have $ V_1V_2V_3 \propto ( \ln \bar{\gamma} )^{\frac{1}{2}} \bar{\gamma}_{s,d}^{-1}$. Therefore $\mathcal{P}^{\rm{E}}_{\rm{e}}$ decays with $ ( \ln \bar{\gamma} )^{\frac{1}{2}} (\bar{\gamma}_{s,d})^{-(N^{\rm{E}}+1)} $. { Considering the constraint that $2N^{\rm{E}}-N > 0$, the diversity order is determined by the worst case when $N^{\rm{E}}= \left \lfloor \frac{N}{2} \right \rfloor + 1$.} So it is  $\frac{N}{2}+2$ and $\frac{N+1}{2}+1$, respectively, for even and odd $N$.

	{ When $2N^{\rm{E}}-N = 0$, similarly we can obtain $	Z_1 = 0$ and $Z_2 = \frac{2}{4+c_1\bar{\gamma}_{s,d}}$. Then the expression in the second  curly bracket of \eqref{eq:PeE-ave} can be obtained}.	
	When $2N^{\rm{E}}-N < 0$, similarly to the error-free relaying case, { the  expression in the third curly bracket of \eqref{eq:PeE-ave} can be obtained}. After some calculation, it can be shown that $\mathcal{P}^{\rm{E}}_{\rm{e}}$ decays with $ ( \ln \bar{\gamma} )^{\frac{1}{2}} (\bar{\gamma}_{s,d})^{-(N+1-N^{\rm{E}})}$ for $2N^{\rm{E}}-N \leq 0$, and therefore the diversity order is $N+1-N^{\rm{E}}$. {Considering $2N^{\rm{E}}-N \leq 0$, the diversity order is determined by the worst case when $N^{\rm{E}} = \left \lfloor \frac{N}{2} \right \rfloor$. } So it is  $\frac{N}{2}+1$ and $\frac{N+1}{2}+1$, respectively, for even and odd $N$. Overall speaking, for erroneous relaying, we obtain the diversity as $\frac{N}{2}+1$ and $\frac{N+1}{2}+1$, for even and odd $N$, respectively.  

	\bibliographystyle{IEEEtran} 
	\bibliography{ref_relay}
	\ifCLASSOPTIONcaptionsoff
	\newpage
	\fi

\end{document}